\documentclass{article}
\usepackage{graphicx,epsfig}
\usepackage{graphicx,subfigure}
\usepackage{arxiv}
\usepackage{amsmath}
\usepackage[utf8]{inputenc} 
\usepackage[T1]{fontenc}    
\usepackage{hyperref}       
\usepackage{url}            
\usepackage{booktabs}       
\usepackage{amsfonts,bm}       
\usepackage{microtype}      
\usepackage[mathscr]{euscript}
\usepackage[dvipsnames]{xcolor}
\usepackage[numbers, compress]{natbib}
\title{Surface wave scattering by multiple flexible fishing cage system}

\author{
  Siluvai Antony Selvan\\
  School of Mathematics and Statistics\\
  The University of Melbourne\\
   Parkville VIC 3010, Australia\\
  \texttt{antony.selvan22@gmail.com} \\
   \And
 R. Gayathri\\
  Department of Mathematics\\
  SRM Institute of Science and Technology\\
  Kattankulathur-603203, India\\
  \texttt{gayathri.ramachandran08@gmail.com} \\
    \And
  Harekrushna Behera*\\
  Department of Mathematics\\
  SRM Institute of Science and Technology\\
  Kattankulathur-603203, India\\
  \texttt{hkb.math@gmail.com} \\
       \And
M H Meylan\\
  School of Mathematical and Physical Sciences\\
  University of Newcastle\\
 NSW 2308, Australia.\\
  \texttt{mike.meylan@newcastle.edu.au} \\   
}

\begin{document}
\maketitle

\begin{abstract}
A study of the wave dynamics around a multiple fishing cage system is carried out under the assumption of the linear water wave theory and small-amplitude wave response. The Fourier--Bessel series expansion of the velocity potential is derived for regions enclosed under the open-water and cage systems and in the immediate vicinity. Further, the scattering between the cages is accounted for by employing Graf's addition theorem. The porous flexible cage system is modelled using Darcy's law and the three-dimensional membrane equation. The edges of the cages are moored along its circumferences to balance its position in the deep sea. The unknown coefficients in the potentials are obtained by employing the matched eigenfunction method in conjunction with the least-squares approximation method. In addition, the far-field scattering coefficients for the entire system are obtained by expanding the Bessel and Hankel functions in the plane wave representation form. Numerical results such as the hydrodynamic forces, scattering coefficients, and power dissipation are investigated for various cage and wave parameters. The wave loading on the cage system can be significantly damped by the spatial arrangement, membrane tension, and porous-effect parameter. Moreover, the far-field results suggest that the cage system can also be used as a breakwater.

\end{abstract}


\section{Introduction}

There is a huge demand on seafood owing to the increasing population (\citet{duarte2009will} and \citet{golden2016nutrition}). Simultaneously, there is a need to save the endangered species in the ocean due to global warming and ocean acidification (\citet{mccormick2013ocean}). Both can be made possible using fish farming, which was widely employed in the local ponds and lakes in the past decades (\citet{ helal2017water} and \citet{zhang2011emergy}). In order to implement the same in the ocean, there are a series of challenges posed by the marine environment. Hence, suitable modeling needs to be carried out for explaining the wave response of such a system. The fish cage can be modeled as the cylindrical perforated membrane-like structure with perforated bottom moored along its circumference (\citet{chan2001wave} and \citet{mandal2016gravity}). 
Although wave attenuation by a single flexible porous cage has been well studied, however, there are no studies in the literature regarding the wave response of the multiple flexible perforated fishing cage system, which forms the ideal model for the fish farm. Therefore, the authors are motivated to study the hydrodynamic response of a system of multiple fishing cages. In the case of such a cage system, the parameters such as porosity, flexibility and geometry play a major role in explaining the wave response of a single/multiple cage systems. The significance of such parameters have been investigated by many researchers in different models for various applications.

In general, the presence of porosity in any off-shore structures can be modeled using Darcy's law (\citet{chwang1983porous} and \citet{yu1994wave}), which damps the high amplitude wave by dissipating its energy. Several studies have carried to study the wave dissipation by the rigid cylindrical porous structures (\citet{williams2000water,fredriksson2003fish} and \citet{sarkar2019hydrodynamic}). The permeability, size and location of the perforation significantly control the hydrodynamic loads acting on the porous cylinder (\citet{williams2000water}). The theoretical and experimental studies on a porous cylinder were carried out by \citet{zhao2011theoretical}, which suggests that there exist additional components corresponding to the porous effect effects along with the wave-radiating damping. In the case of a surface piercing cylinder, the variation of draft, porosity and radii influences the wave run-up and hydrodynamic loading on the porous cylinder (\citet{sarkar2019water}). \citet{sarkar2019hydrodynamic} investigated the wave interaction with the surface-piercing bottom-mounted porous cylinder, where the resonance effect was observed at a particular wavenumber in the presence of porosity. Later, \citet{behera2020wave} studied an influence of truncated concentric porous cylinder on the inner rigid platform, where the optimum number of porous structures were estimated for damping the wave amplitude. Recently, the cnoidal wave diffraction from the dual concentric cylinders in the presence of an arc-shaped outer cylinder was analyzed by \citet{zhai2020hydrodynamic}.

In addition to the porosity, the inclusion of flexibility into the off-shore structures reduce the damage experienced due to the wave loading. Further, it also proves to be cost-effective and easily portable (\citet{lee1990wave, williams1991flexible, meylan1994flexible}). The wave passage through the vertical flexible porous membrane was analyzed by \citet{chan2001wave}, where the tensile force was incorporated in the standard two-dimensional Euler--Bernoulli's equation. They observed that the deformation of structure increases for larger flexibility and the presence of perforation substantially decreases the deformation. Other than this, the significance of both the porosity and flexibility have also investigated in the case of horizontal floating/submerged two-/three- dimensional structures (\citet{behera2015hydroelastic, meylan2017water, behera2018wave,selvan2019reduction} and \citet{zheng2020wave}). In the case of cylindrical structures, \citet{mandal2013hydroelastic} studied the effect of a flexible porous cylinder on the inner truncated rigid platform, where the hydrodynamic wave loads were analyzed for various flow and structural parameters. The dynamics of the flexible porous cylinder mounted on the truncated rigid bottom was investigated by \citet{su2015analysis}. Recently, the gravity wave interaction with the cylindrical fish cage having bottom perforated membrane was analyzed by \citet{mandal2016gravity} and found that the proper selection of membrane tension along with the cage radius consequently reduces the deflection of the cylindrical side-wall.

In the previous study of wave scattering from the single fishing cage (\citet{su2015analysis} and \citet{mandal2016gravity}), the bottom of the fish cage was either mounted to a rigid bed or clamped. However, in the physical \textcolor{black}{viewpoint}, the fishing cage can be moored along its circumferences, which was not addressed in the previous studies of \citet{su2015analysis} and \citet{mandal2016gravity}. To the best of authors' knowledge, the topic of fish farm constituting of multiple fishing cages haven't addressed in any of the previous studies in the literature. In the present study, the surface wave interaction with the multiple fishing cage system has investigated. Moreover, the fishing cages are moored along its cylindrical circumferences using the frictionless linear spring. In order to consider the scattering between the cages, Graff's addition theorem \textcolor{black}{has} employed. This theorem forms the basis of modeling the wave scattering from any multiple floating structures (\citet{park2012numerical, park2017water, zheng2020hydroelastic} and \citet{zheng2020water}). Further, the quantities like power dissipation reflected and transmitted wave amplitudes at the far-field have analyzed in the present study to check the usage of multiple fishing cage system as a breakwater.

The present manuscript is organized as follows: Section 2 contains the mathematical formulation, where the details of the governing equation and the associated boundary conditions for the physical problem are discussed; Section 3 showcases the method of solution, in which the procedure for solving the physical problem by employing the method of an eigenfunction in conjunction with the technique of least-squares approximation are explained in detail. Further, the brief description about the derivation of far-field amplitude function and the power dissipation is given; Section 4 contains the result and discussion, which further subdivided into the dual, triple and multiple fishing cage systems along with the time-domain simulation. In the case of the dual and triple fishing cages, the hydrodynamic wave forces, far-field amplitude functions and power dissipation are investigated. Moreover, for multiple fishing cage systems, far-field amplitude functions, power dissipation and flow distributions are analyzed. The technique of time-domain simulation is given in the last subsection of results and discussion. The supplementary files are attached to show the surface wave scattering by the single and multiple fishing cage system; In the last Section (i.e. Section 5), the detailed summary and significance of the present work are highlighted.

\section{Mathematical formulation}\label{mathform}
\begin{figure}[ht!]
 	\begin{center}
 		\subfigure[]{\includegraphics*[width=8cm]{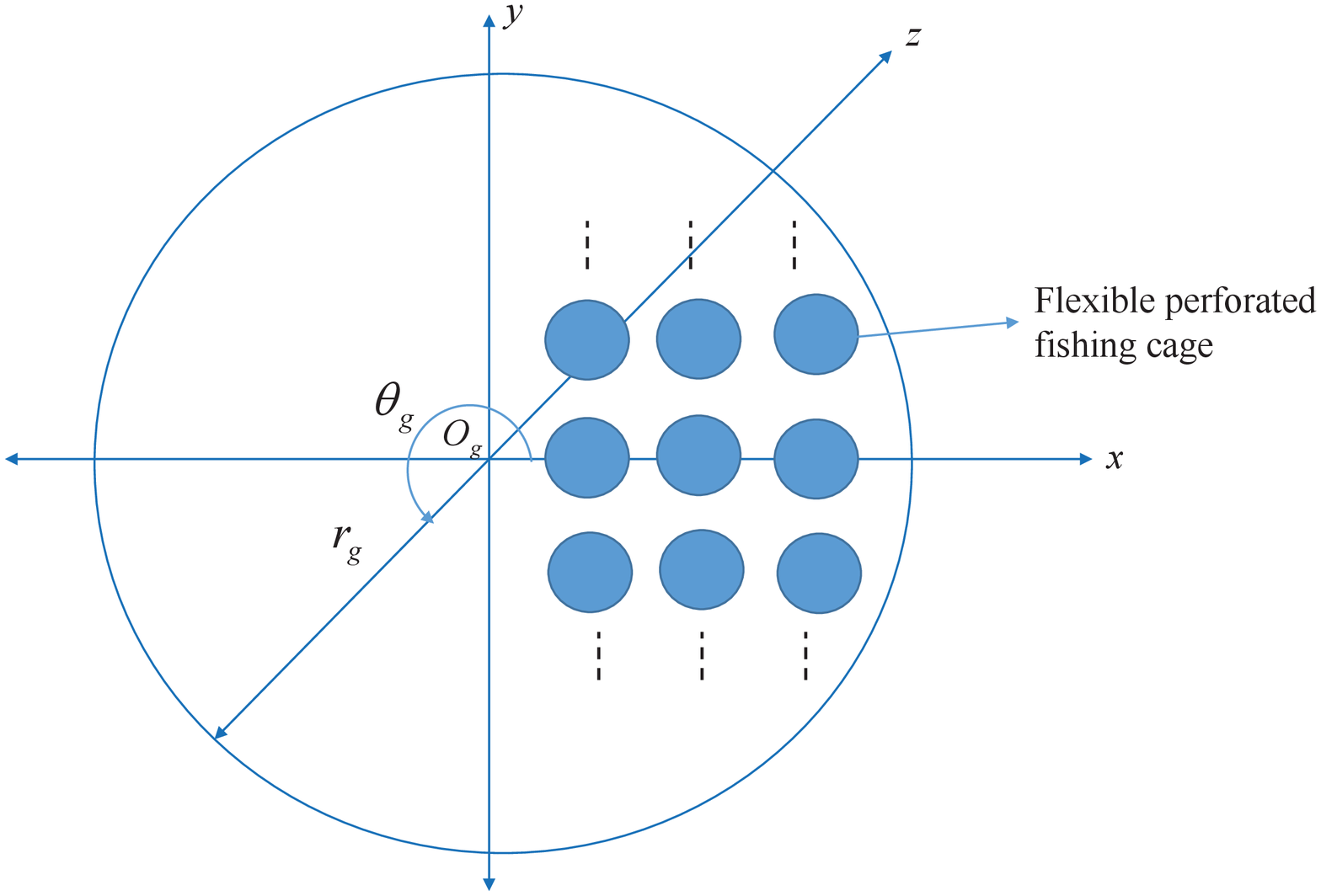}}
 		\subfigure[]{\includegraphics*[width=8.4cm]{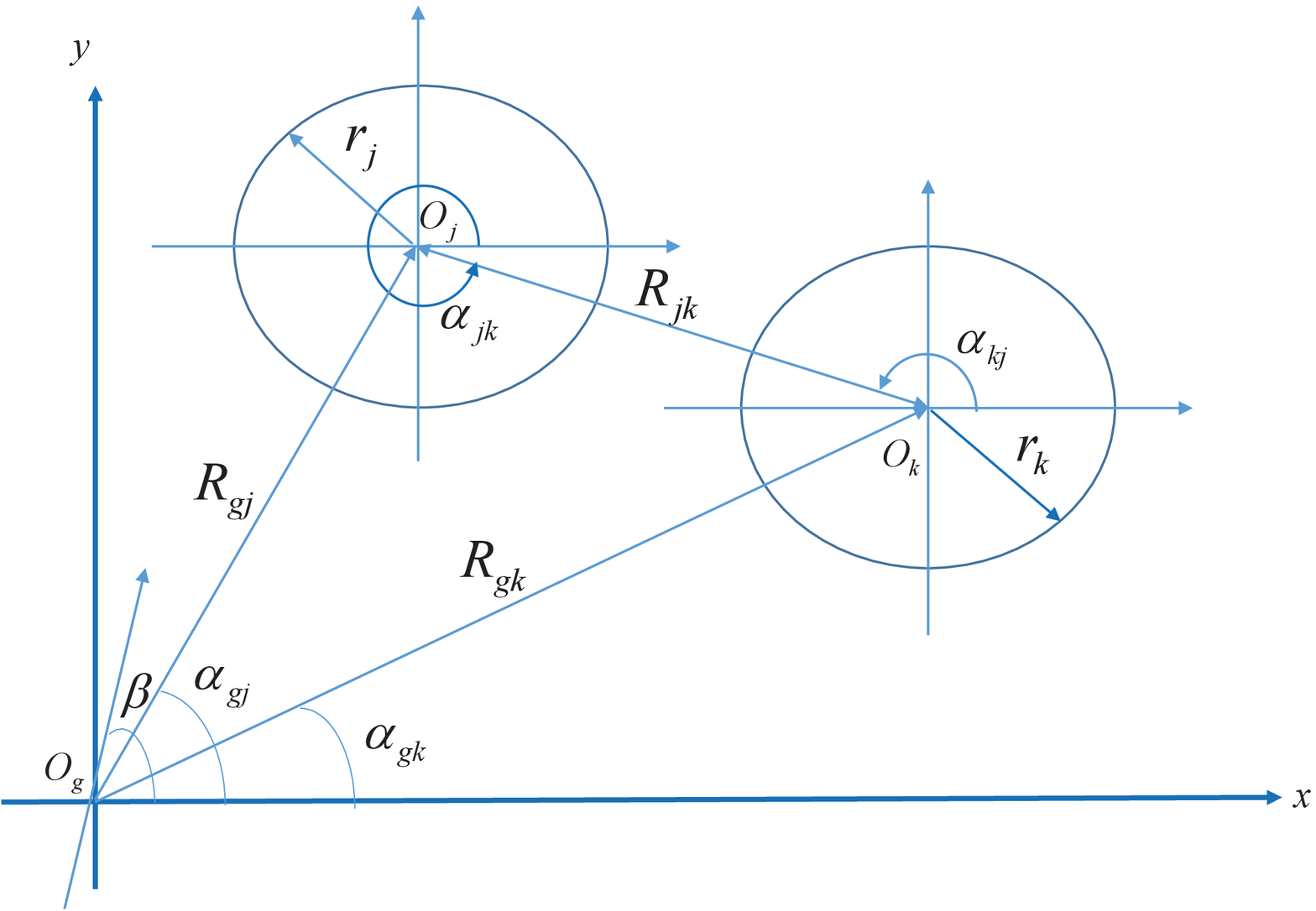}}
 		\subfigure[]{\includegraphics*[width=8.4cm]{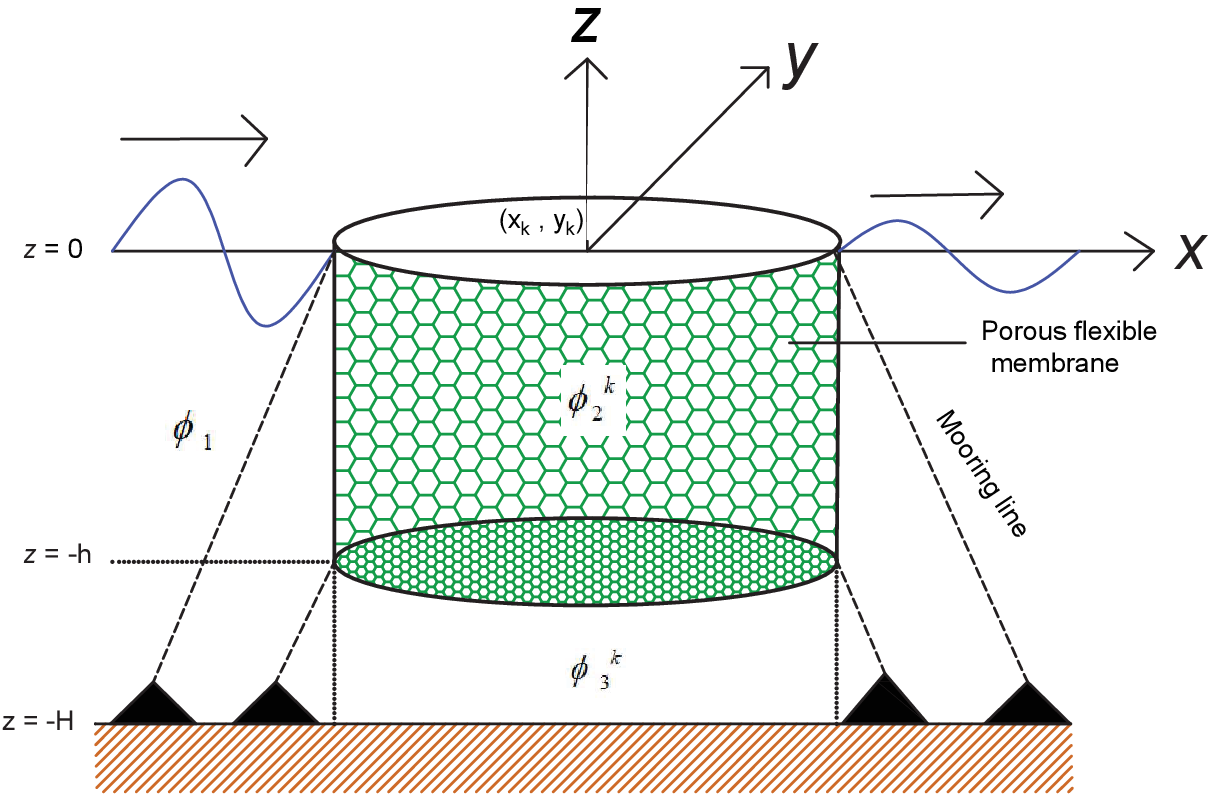}}
 		\subfigure[]{\includegraphics*[width=15.4cm]{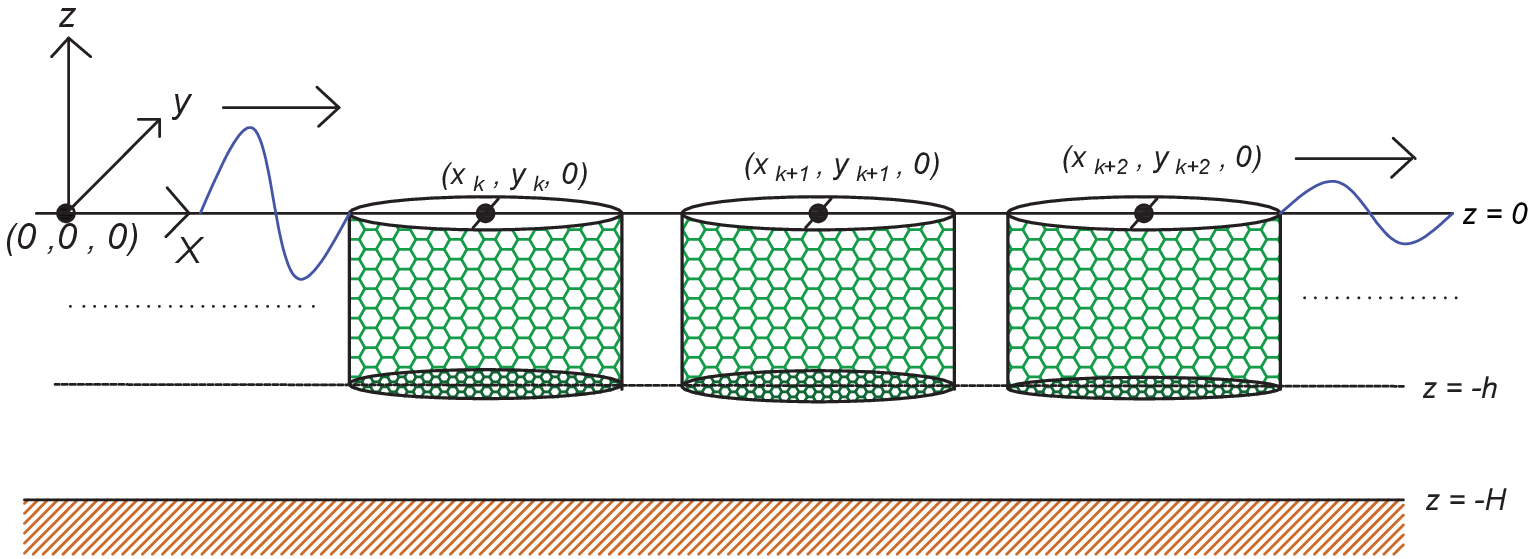}}
 		\caption{Schematic sketch of the defined physical problem; (a) Top view showing the global Cylindrical coordinate ($O_gr_g\theta_gz$), (b) Top view showing the $k$\textsuperscript{th}- and $j$\textsuperscript{th}-cylinders, (c) Lateral view of a flexible perforated moored fishing cage and (d) Lateral view showing the array of multiple fishing  cages.}\label{fig1} 
 	\end{center}		
 \end{figure}

 The dynamics of surface waves around the system of $\mbox{N}$ multiple flexible fishing cages [Fig.~\ref{fig1}(a)] is studied in the global Cartesian coordinates $(x,y,z)$ with origin overlap with the still water having the final depth $H$ and $z$-axis pointed upward direction. The position of arbitrary $j$\textsuperscript{th} and $k$\textsuperscript{th} fishing cages are marked as $(x_j, y_j)$ and $(x_k, y_k)$ in the global co-ordinate system, respectively [Fig.~\ref{fig1}(b)], which form the origins of the local polar coordinate systems $(r_j,\theta_j,z)$ and $(r_k,\theta_k,z)$ with $k=1,2,\dots,\text{N}$ and $j=1,2,\dots,\text{N}$. The thicknesses of entire cylindrical fishing cages are assumed as $d$, which is assumed very small on comparing with the wavelength of an incident monochromatic wave ($\lambda$) as a consequence of small amplitude wave theory. Further, the radii of an array of floating fishing cages are considered as $b_k$ with $k=1,2,\dots,\text{N}$ [Fig.~\ref{fig1}(d)].
 The incompressible and inviscid fluid is assumed to be having an irrotational and \textcolor{black}{time-harmonic} motion of the form $\mathrm{exp}({-\mathrm{i}\omega t})$. The domain enclosed between the still water and a rigid bottom is divided into an exterior and $\text{N}$ interior regions.
 The general velocity potential potential is defined as $\Phi(r,\theta,z,t)=\Re\{\phi(r,\theta,z)\,\mathrm{e}^{-\mathrm{i}\omega t}\}$ with $\phi(r,\theta,z)$ be the spatial velocity potential of  $k$\textsuperscript{th} region where $\omega$ be the angular frequency and $\Re$ denotes the real part. This general potential $\phi(r,\theta,z)$ satisfying all the assumption of fluids obeys the Laplace equation, which can be expressed as
 \begin{align}
 \frac{\partial^2\,\phi}{\partial r^2}+\frac{1}{r}\frac{\partial\,\phi}{\partial r}+ \frac{1}{r^2} \frac{\partial^2\,\phi}{\partial\,\theta^2}+\frac{\partial^2\,\phi}{\partial\,z^2}=0, \label{1}
 \end{align}
 Further, the spatial velocity potential in the $k$\textsuperscript{th} region is given as $\phi_2^k(r_k,\theta_k,z)$ and $\phi_3^k(r_k,\theta_k,z)$ with $k=1,2,\dots,\text{N}$. In an outer region, it is denoted by $\phi_1(r_k,\theta_k,z)$.
 The fluid depth enclosing the fishing cage is subdivided into the barrier region $\varLambda_B:= \{\,z\,|\,-h\leq\,z\leq\,0\}$ and gap region $\varLambda_G:= \{\,z\,|\,-H\leq\,z\leq\,-h\}$. The normal velocity becomes zero along the rigid bed in both the interior and outer regions, which is given by
 \begin{align}
 &\frac{\partial \phi_1}{\partial z}=0, \quad \text{at} \quad z=-H,\label{2}
 \\
 &\frac{\partial \phi^k_3}{\partial z}=0, \quad \text{at} \quad z=-H,\label{3}
 \end{align}
 where $k=1,2,\dots\text{N}$. The linear form of the free surface boundary condition at both the open-water and cage covered regions are given as
 \begin{align}
 & \frac{\partial \phi_1}{\partial z}-K\,\phi_1=0, \quad \text{at} \quad z=0, \label{4}
 \\
 & \frac{\partial \phi^k_2}{\partial z}-K\,\phi^k_2=0, \quad \text{at} \quad z=0, \label{5}
 \end{align}
 where $K=w^2/g$ and $g$ is the acceleration due to gravity. Using Darcy's law, the proportionality between the hydrodynamic pressure and normal velocity along the cage system is obtained, which are given as
 \begin{align}
 & \frac{\partial\,\phi^k_2}{\partial\,z}= \mathrm{i}k_0G_v\,\{\phi^k_2-\phi_1\}+\mathrm{i}\omega\eta_k, \quad \text{at} \quad r_k=b_k, \quad z\in\varLambda_B, \label{6}
 \\
 & \frac{\partial\,\phi^k_2}{\partial\,z}= \mathrm{i}k_0G_h\,\{\phi^k_2-\phi^k_3\}+\mathsf{i}\omega\zeta_k, \quad \text{at} \quad 0<r_k<a_k, \quad z=-h, \label{7}
 \end{align}
 where $G_v$ and $G_h$ be the complex porous-effect parameters of cylindrical and circular membrane, respectively. The deflection of $k$\textsuperscript{th} cylindrical  and submerged membranes are denoted as $\eta_k(r_k,\theta_k,z)$ and $\zeta_k(r_k,\theta_k,z)$, respectively. The linear form of dynamical boundary condition corresponding to the bottom circular membrane  reads as
 \begin{align}
 &\left(Q\,\frac{\partial^3 \phi^k_3}{\partial\,z^3}-m_m\,\frac{\partial \phi^k_3}{\partial z}-K\,\phi^k_3 \right)-\mathrm{i}k_0G_h\left(Q\frac{\partial^2 \phi^k_3}{\partial z^2}-m_m\right)\bigg(\phi^k_2-\phi^k_3\bigg)+K\phi^k_2=0,\, \label{8} 
 \end{align}
 where $Q=T_1/\rho g$ and $m_m=\rho_md/\rho g$, with $T_1$ and $\rho_m$ being the tensile force and density of bottom perforated circular membrane, respectively. The deflection of $k$\textsuperscript{th} cylindrical perforated membrane
 $\eta_k(r_k,\theta_k,z)$ is governed by
 \begin{align}
 & T_2 \frac{d^2\eta_k}{dz^2}+m_s\omega^2\eta_k=2\mathrm{i}\omega\rho b_k\int_{0}^{\pi} \bigg(\phi_1-\phi^k_2\bigg)\,\mathrm{e}^{\mathrm{i}\,(\pi-\theta_k)} d\theta_k \,\,\text{at}\,\,z\in\varLambda_B\,\text{and}\,\,r=b_k, \label{9}
 \end{align}
 where $k=1,2,\dots\text{N}$. Further, $T_2$ and $m_s$ are the tensile force and uniform mass per unit length of the cylindrical perforated membrane, respectively. The matching of velocity and pressure along the interface of $k$\textsuperscript{th} fish cage can be expressed as
 \begin{align}
 &\frac{\partial\,\phi_1}{\partial r}= \frac{\partial\,\phi^k_2}{\partial r} \quad \text{at} \quad z\in \varLambda_B \quad \text{for} \quad r_k= b_k, \label{10}
 \\
 &\frac{\partial\,\phi_1}{\partial r}= \frac{\partial\,\phi^k_3}{\partial r} \quad \text{at} \quad z\in \varLambda_G \quad \text{for} \quad r_k= b_k, \label{11}
 \end{align}
\begin{align}
& \phi_1 = \phi^k_3, \quad \text{at} \quad z\in \varLambda_G \quad \text{for} \quad r_k= b_k, \label{12}\\
& \frac{\partial \phi^k_2}{\partial z}= \frac{\partial \phi^k_3}{\partial z}, \quad \text{at} \quad r_k=b_k \quad \text{and} \quad z=-h, \label{13}
\end{align}
 Moreover, the condition for mooring along the circumference at both the ends of the cylindrical perforated membrane is given by
 \begin{align}
 T_2\frac{d\eta_k}{dz}-2\mu\sin^2\gamma_u\eta_k=0,\,\quad\,\text{at} \quad z=-h,\,\, z=0 \quad \text{and} \quad r_k=b_k, \label{14}
 \\
 T_2\frac{d\eta_k}{dz}-2\mu\sin^2\gamma_d\eta_k=0,\,\quad\,\text{at} \quad z=-h,\,\, z=0 \quad \text{and} \quad r_k=b_k, \label{15}
 \end{align}
 where $\mu$, $\gamma_u$ and $\gamma_d$ are the mooring spring constant, the angle extended by the upper and lower mooring lines with respect to negative $z$-axis, respectively. 

\section{Method of solution}\label{Solution}
In the outer region, the velocity potential $\phi_1(r_k,\theta_k,z)$ satisfying the Laplace equation~\eqref{1} with the associated bottom boundary condition~\eqref{2} is derived as
\begin{align}
&\phi_1= \sum_{m=-\infty}^{\infty}\left(\mathcal{I}_k J_m(k_0\,r_k) \mathrm{e}^{\mathrm{i}m(\pi/2-\beta)} g_0(z)+\sum_{k=1}^{\text{N}}\sum_{q=0}^{\infty} A^k_{mq}\,H^{(1)}_m(k_qr_k)\,g_q(z)\right)\mathrm{e}^{\mathrm{i}m\theta_k} \label{16} 
\end{align}
where $\displaystyle\, \mathcal{I}_k=(\mathsf{i}g/\omega)\mathrm{e}^{\mathrm{i}k_0(x_k\cos\beta+y_k\sin\beta)}$ is the phase factor with $\beta$ being the phase angle. Here, $J_m$ and $H^{(1)}_m$ indicate the Bessel and first kind Hankel functions of order $m$. The unknown coefficients of open-water region for $k$\textsuperscript{th} fishing cage is denoted by $A^k_{mq}$. Further, the above Eq.~\eqref{16} satisfies the far-field radiation condition
\begin{align}
\lim_{r_k \to \infty} \phi_1= \sum_{m=-\infty}^{\infty}\left(\mathcal{I}_k J_m(k_0\,r_k) \mathrm{e}^{\mathrm{i}m(\pi/2-\beta)} g_0(z)\right),\label{17}
\end{align}
where the scattered portion of the Eq.~\eqref{16} becomes zero. By employing Graff's addition theorem as in \citet{abramowitz1948handbook} and \citet{zheng2020water}, the Hankel function in the scattered portion of Eq.~\eqref{16} can be rewritten as
\begin{align}
&\phi_1= \sum_{m=-\infty}^{\infty}\bigg[\mathcal{I}_k J_m(k_0\,r_k) \mathrm{e}^{\mathrm{i}m(\pi/2-\beta)}\,\mathrm{e}^{\mathrm{i}m\theta_k}\,g_0(z)+\sum_{q=0}^{\infty}\bigg(A^k_{mq}\,H^{(1)}_m(k_qr_k)\mathrm{e}^{\mathrm{i}m\theta_k}+\sum_{{\substack{j=1 \\ j\neq k}}}^{\text{N}}A^j_{mq}\,\sum_{n=-\infty}^{\infty}\,(-1)^n\, \nonumber
\\
&\hspace{7cm}H^{(1)}_{m-n}(k_qR_{jk})\,\mathrm{e}^{\mathrm{i}(m\alpha_{kj}-n\alpha_{jk})}\,J_n(k_qr_k)\mathrm{e}^{\mathrm{i}n\theta_k}\bigg)g_q(z)\bigg], \label{18}
\end{align}
where the eigenfunction $g_q(z)$ associated with the outer region is given by
\begin{align}
g_q(z)=\frac{\cosh k_q(z+H)}{\cosh (k_qH)}, \label{19}    
\end{align}
in which, the eigenvalues $k_q$ satisfies the following dispersion relation
\begin{align}
\omega^2=gk_q\tanh(k_qH), \label{20}    
\end{align}
which can be solved numerically to obtain the one real root $(k_0)$ and infinite number of imaginary roots $(k_1,k_2,k_3,\dots)$. Similarly, the velocity potentials 
corresponding to the interior region of $k$\textsuperscript{th} fishing cage after solving the equation~\eqref{1} along with Eq.~\eqref{3} are given by 
\begin{align}
&\phi^k_2 = \sum_{m=-\infty}^{\infty} \left(\sum_{q=0}^{\infty} B^k_{mq} I_m(p_qr_k) M_q(z)\right)\,\mathrm{e}^{\mathrm{i}m\theta_k}, \label{21}
\\
&\phi^k_3 = \sum_{m=-\infty}^{\infty} \left(\sum_{q=0}^{\infty} B^k_{mq} I_m(p_qr_k) N_q(z)\right)\,\mathrm{e}^{\mathrm{i}m\theta_k}, \label{22}	\end{align}
where $I_m$ is the modified Bessel function of first kind having the order $m$, and the unknown coefficients of the interior region of $k$\textsuperscript{th} fishing cage is denoted as $B^k_{mq}$. The associated eigenfunctions are given as
\begin{align}
&M_q(z)= \frac{\cosh p_q(z+H)-E_q \sinh p_q(z+H)}{\cosh (p_qH)-E_q \sinh (p_qH)},\label{23}
\\
&N_q(z)= \frac{\{\tanh p_q(H-h)-E_q\} \{\cosh p_q(z+H)\}}{\{\tanh p_q(H-h)\} \{\cosh (p_qH)-E_q\sinh(p_qH)\}},\label{24}
\\ 
& E_q=\frac{V_qp_q\tanh^2p_q(H-h)}{V_qp_q\tanh p_q(H-h)-\{1-\tanh^2p_q(H-h)\}\{K-\mathrm{i}k_0G_2V_q\}},\label{25}
\end{align}
where $V_q=T_1p_q^2-m_sw^2$. Further, the eigenvalues $p_q$ satisfies the following dispersion relation in the interior cage region, which is expressed as
\begin{align}
K-p_q\tanh(p_qH)=E_q(K\tanh(p_qH)-p_q).\label{26}
\end{align}
Solving the Eq.~\eqref{26} numerically yields infinitely many complex roots $(p_0, p_1, p_2,\dots)$ for $G\neq 0$. On the other hand, in the case of $G=0$, the one real roots $(p_0)$ and infinitely many imaginary roots $(p_0, p_1, p_2,\dots)$ are obtained. The deflection of $k$\textsuperscript{th} cylindrical membrane obtained by solving Eq.~\eqref{9} can be represented as
\begin{align}
&\eta_k= D^k_{m1} \frac{\cos \lambda z}{\cos \lambda_1H} +D^k_{m2} \frac{\sin \lambda z}{\sin \lambda_2H}+\frac{\mathrm{i}\omega\pi\rho b_k}{T_2(k_0^2+\lambda^2)}\bigg[\sum_{q=0}^{\infty}B^k_{mq}\,I_m(p_qb_k)\,M_q(z)-\mathcal{I}_kJ_m(k_0b_k)\nonumber
\\
&\hspace{0.2cm}e^{\mathrm{i}m(\pi/2-\beta)}\,g_0(z)-\sum_{q=0}^{\infty}\bigg[A^k_{mq}H_m^{(1)}(k_qb_k)+\sum_{{\substack{j=1 \\ j\neq k}}}^{\text{N}}\sum_{n=-\infty}^{\infty}A^j_{nq}H^{(1)}_{m-n}(k_qR_{jk})\mathrm{e}^{\mathrm{i}(m\alpha_{kj}-n\alpha_{jk})} J_m(k_qb_k) \bigg]g_q(z)\bigg],\label{27}
\end{align}
where $\displaystyle\lambda=\sqrt{\frac{m_s}{T_1}}\omega$. 
By employing the velocity continuity as in Eqs.~\eqref{10} and \eqref{11}, the following the linear system of equation is obtained
\begin{align}
&\mathcal{I}_k J'_m(k_0b_k) Y_{lq}\,\mathrm{e}^{\mathrm{i}m(pi/2-\beta)} \delta_{lq} +\sum_{q=0}^{\infty} \bigg[A^k_{mq} H'^{(1)}_{m}(k_qb_{k}) Y_{lq} \delta_{lq}+\sum_{{\substack{j=1 \\ j\neq k}}}^{N} \sum_{n=-\infty}^{\infty} A^j_{nq} (-1)^m H^{(1)}_{n-m}(k_qR_{jk}) \nonumber
\\
&\hspace{5.5cm}\mathrm{e}^{\mathrm{i}(n\alpha_{kj}-m\alpha_{jk})} J'_m(k_qb_k) Y_{lq} \delta_{lq}-B^k_{mq} I'_m(p_q b_k)\big[M_{ql}+N_{ql}\big]\bigg]=0,\label{28}
\end{align}
where $m=0,1,2,\dots$, $l=0,1,\dots$ and $q=0,1,2,\dots$. The orthogonal relations $Y_{l}$, $M_{ql}$ and $N_{ql}$ with respect to open water region are given by
\begin{align}
&Y_{lq}=\int_{-H}^{0}g_l(z) g_q(z)\,dz,\quad
M_{ql}=\int_{-h}^{0}M_q(z)\,g_l(z)\,dz,\quad \mbox{and} \quad N_{ql}=\int_{-H}^{-h}N_q(z)\,g_l(z)\,dz, \label{29}
\end{align}
with $q=0,1,\dots$ and $l=0,1,\dots$. Using Eqs.~\eqref{6} and \eqref{12}, the coupled relation is obtained for $k=1,2,\dots,\text{N}$, which is given by
\begin{align}
X^k(z)=\sum_{q=0}^{\mathbf{n}_0}\bigg[A^k_{mq}\,R^k_{mq}(z)+\sum_{{\substack{j=1 \\ j\neq k}}}^{N} \sum_{n=0}^{M} A^j_{nq} R^j_{nq}(z)\bigg]+ \sum_{q=0}^{\mathbf{n}_0} B^k_{mq} S^k_{mq}(z) +\sum_{i=1}^{2} D^k_{mi} W^k_{mi}(z)+L^k_{m0}(z), \label{30}
\end{align}
where
\begin{align*}
&R^k_{mq}(z)=
\begin{cases}
H^{(1)}_m(k_qb_k) g_q(z) \quad z\in \varLambda_G,
\\[1ex]
\left(\frac{b_k\omega^2\rho\pi}{T_2(k_0^2+\lambda^2)}-\mathrm{i}k_0G_v\right) H^{(1)}_m(k_qb_k) g_q(z) \quad z\in \varLambda_B,
\end{cases}
\\[1ex]
&R^j_{nq}(z)=
\begin{cases}
H^{(1)}_{n-m}(k_qR_{jk}) \mathrm{e}^{\mathrm{i}(n\alpha_{kj}-m\alpha_{jk})} J_m(k_qb_k) g_q(z) \quad z \in \varLambda_G
\\[1ex]
\left(\frac{b_k\omega^2\rho\pi}{T_2(k_q^2+\lambda^2)}-\mathrm{i}k_0G_v\right) H^{(1)}_{n-m}(k_qR_{jk}) \mathrm{e}^{\mathrm{i}(n\alpha_{kj}-m\alpha_{jk})} J_m(k_qb_k) g_q(z) \quad z \in \varLambda_B
\end{cases}
\end{align*}
\begin{align*}
&S^k_{mq}(z)=
\begin{cases}
-I_m(p_q b_k) N_q(z) \quad z \in \varLambda_G,
\\[1ex]
\left(\left(-\frac{b_k\omega^2\rho\pi}{T_2(p_q^2+\lambda^2)}-\mathrm{i}k_0G_v\right) I_m(p_qb_k)-I'_m(p_q b_k)\right) M_q(z) \quad z \in \varLambda_B
\end{cases}
\\[1ex]
&L^k_{m0}(z)=
\begin{cases}
\mathcal{I}_k J_m(k_0b_k)\mathrm{e}^{\mathrm{i}m(\pi/2-\beta)} g_q(z) \quad z \in \varLambda_G,
\\
\left(\frac{b_k\omega^2\rho\pi}{T_2(k_0^2+\lambda^2)}-\mathrm{i}k_0G_v\right) \mathcal{I}_k J_m(k_0b_k) \mathrm{e}^{\mathrm{i}m(\pi/2-\beta)} g_q(z) \quad z \in \varLambda_B,
\end{cases}
\end{align*}

\begin{align*}
&W^k_{m1}(z)=
\begin{cases}
0 \quad z \in \varLambda_G,
\\[1ex]
\frac{\mathrm{i}\omega\cosh (\lambda z)}{\cosh (\lambda H)} \quad z \in \varLambda_B,
\end{cases}
\hspace{4cm}
W^k_{m2}(z)=
\begin{cases}
0 \quad z \in \varLambda_G,
\\[1ex]
\frac{\mathrm{i}\omega\sinh (\lambda z)}{\cosh (\lambda H)} \quad z \in \varLambda_B
\end{cases}
\end{align*}
The above relation Eq.\eqref{30} satisfies the least square approximation method as given by,
\begin{align}
&\int_{-H}^{0}|X^k|^2 dz =\text{minimum},\label{31}\\
&\int_{-H}^{0} X^k \frac{\partial X^{k*}}{\partial A^{k*}_{mq}} dz =0,
\end{align}
where $^{*}$ denotes the complex conjugate.
The series expansion for eigenvalues are truncated for $q=\mathbf{n}_0$ roots for both the open water and submerged membrane covered regions for $k$\textsuperscript{th} fishing cage. Similarly, the Bessel series are truncated for $m=M$ for $k$\textsuperscript{th} fishing cage. This leads to $2NM(\mathbf{n}_0+1)$ system of equations (Eqs.~\eqref{28} and \eqref{30} along with the edge conditions~\eqref{14} and \eqref{15}), which can be solved for all values of $m$, $q$ and $k$ simultaneously.

\subsection{Far-field amplitude functions and power dissipation}

As $r_k\to \infty$, there exists only the progressive wave mode in the scattered part of the total potential Eq.~\eqref{16} corresponding to the exterior region, which can be expressed in the following asymptotic form
\begin{align}
\phi^s_1=\sum_{k=1}^{N}\sum_{m=-\infty}^{\infty} A^k_{m0} \sqrt{\frac{2}{\pi k_0 r_k}} e^{-\mathsf{i}m\pi/4} e^{\mathsf{i}k_0r_k}  e^{\mathsf{i}m(\theta_k-\pi/2)} g_0(z). \label{32}
\end{align}
Further, the above expression Eq.~\eqref{32} is transformed from the local cylindrical co-ordinate $(r_k, \theta_k, z)$ to the global cylindrical coordinate $(r_g, \theta_g, z)$, which can be rewritten as
\begin{align}
\phi^s_1=\sum_{k=1}^{N}\sum_{m=-\infty}^{\infty} A^k_{m0} \sqrt{\frac{2}{\pi k_0 r_k}} e^{-\mathsf{i}m\pi/4} e^{\mathsf{i}k_0R_{gk}\cos(\alpha_{gk}-\theta_g)}  e^{\mathsf{i}m(\theta_g-\pi/2)} g_0(z). \label{33}
\end{align}
Similarly, the incident portion of the total potential Eq.~\eqref{16} is expressed in the asymptotic form and extended to the global cylindrical coordinate, which is given as
\begin{align}
\phi^{in}_1=\sum_{m=-\infty}^{\infty} \mathcal{I}_k e^{-\mathsf{i}k_0R_{gk}\cos(\alpha_{gk}-\theta_g)} e^{\mathsf{i}m(\pi/2-\beta)} g_0(z). \label{34}
\end{align} 
From the Eqs.~\eqref{33} and \eqref{34}, the scattering and incident amplitude functions are denoted as
\begin{align}
\mathbf{D}(\theta_g)=\frac{\mathsf{i}\omega}{g\pi} \sum_{k=1}^{N}\sum_{m=-\infty}^{\infty} A^k_{m0} e^{-\mathsf{i}k_0R_{gk}\cos(\alpha_{gk}-\theta_g)} e^{\mathsf{i}m(\theta_g-\pi/2)}~~~\mbox{and}~~~\mathbf{A}(\theta_g)=\frac{A}{2\pi}\sum_{m=-\infty}^{\infty}e^{\mathsf{i}m(\theta_g-\beta)}.
\end{align}
The reflected and transmitted wave amplitude functions are defined as
\begin{align}
R(\theta_g)=\mathbf{D}(\beta+\theta_g+\pi)~~~\mbox{and}~~~T(\theta_g)=\mathbf{D}(\beta+\theta_g)+\mathbf{A}(\beta+\theta_g)~~~\mbox{provided}~~-\pi/2<\theta_g<\pi/2.
\end{align}
Further, the power dissipation is obtained directly from the hydrodynamic pressure acting on the vertical and submerged porous membrane, which is given as
\begin{align}
\mathcal{P}_D=\frac{k_0\rho\omega}{2}\left[ (G_v+G_v^*)\sum_{k=1}^{N}\iint_{\Omega_h}|\phi^{k}_2-\phi^{k}_3|^2 dS+(G_h+G_h^*)\sum_{k=1}^{N}\iint_{\Omega_v}|\phi^{k}_2-\phi_1|^2 dS\right],\label{e37}
\end{align}
where $\Omega_v=\{(\theta,z): 0\leq \theta_k \leq 2\pi~~\mbox{and}~~-h \leq z\leq 0 \}$ and $\Omega_h=\{(r,\theta): 0\leq r\leq b_k~~\mbox{and}~~0\leq \theta_k \leq 2\pi\}$ are the domains of integration, and $^*$ denotes the complex conjugate. The incident power per unit width of wave front is given as
\begin{align}
\mathcal{P}_{I}=\frac{\rho \omega g A^2}{4k_0}\bigg(1+\frac{2k_0H}{\sinh(2k_0H)}\bigg),    
\end{align}
which can be used for scaling the power dissipation $\mathcal{P}_D$. Hence, the non-dimensional power dissipation is denoted as $E=k_0\mathcal{P}_D/\mathcal{P}_{I}$.

\section{Results and Discussion}\label{results}

The MATLAB program have developed to solve the system of equations for determining the unknown coefficients. For numerical discussion, the following physical parameters such as water depth $H=30$m, density of water $\rho=1025$ kgm\textsuperscript{-3}, wave amplitude $A=1$m, height of fish cage $h/H=0.5$, density of submerged membrane $\rho_m=100$ kgm\textsuperscript{-3}, non-dimensional tensile force $T_1/\rho gH^2=T_2/\rho gH^2=T $ with $T=0.4$, non-dimensional mass of cylindrical cage $\alpha=0.01$,
the porous-effect parameter of cage system $G_h=G_v=G=3+3\mathsf{i}$, the mooring spring constant for mooring lines connected to both the ends of cage  
$\mu=10^{12}$N/m, the mooring angles $\gamma_u=60^\circ$ and $\gamma_d=30^\circ$, and the phase angle $\beta=0^\circ$ are fixed, unless it is highlighted in the appropriate figure's caption. The hydrodynamic wave load acting on a fishing cage in vertical direction $\mathcal{S}_v$, can be expressed as
\begin{align}
&\mathcal{S}_v = \mathsf{i}\rho\omega\,\int_{0}^{2\pi}\int_{0}^{b_k}\,\big[\phi_3^{k}(r_k,\theta_k,-h)-\phi_2^{k}(r_k,\theta_k,-h)\big] r_k dr_k \cos(\pi-\theta_k) d\theta_k.
\end{align}
Further, the non-dimensional form of vertical wave force can be represented as
\begin{align*}
C_v= \bigg|\frac{\mathcal{S}_{v}}{\rho g b_k h H}\bigg|.
\end{align*}
\subsection{Single fishing cage}
\begin{figure}[h!]
	\begin{center}
		\subfigure[Absence of cage ($k_0h=1.25$)]{\includegraphics*[width=8cm]{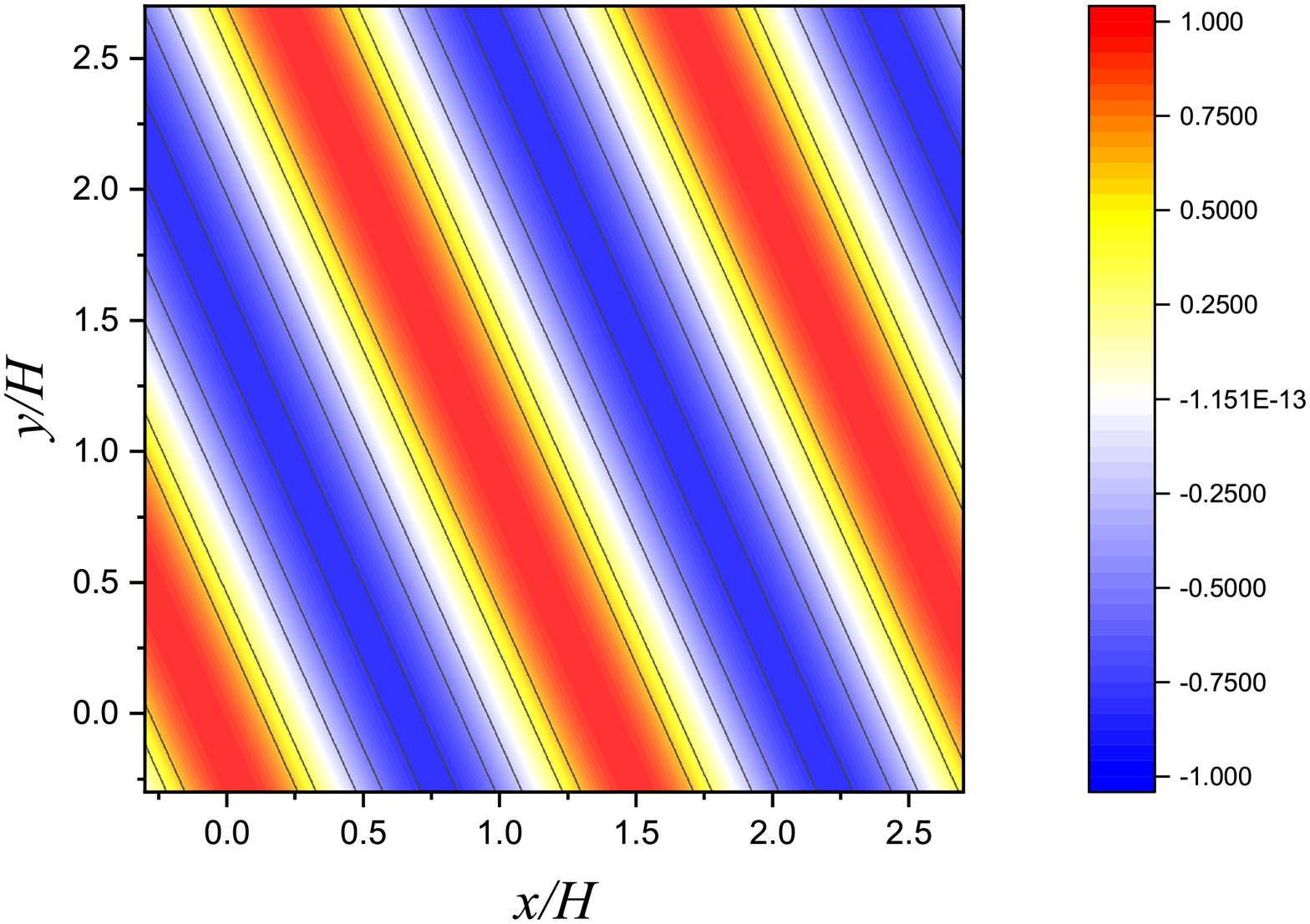}}
		\subfigure[Presence of cage ($k_0h=1$)]{\includegraphics*[width=8cm]{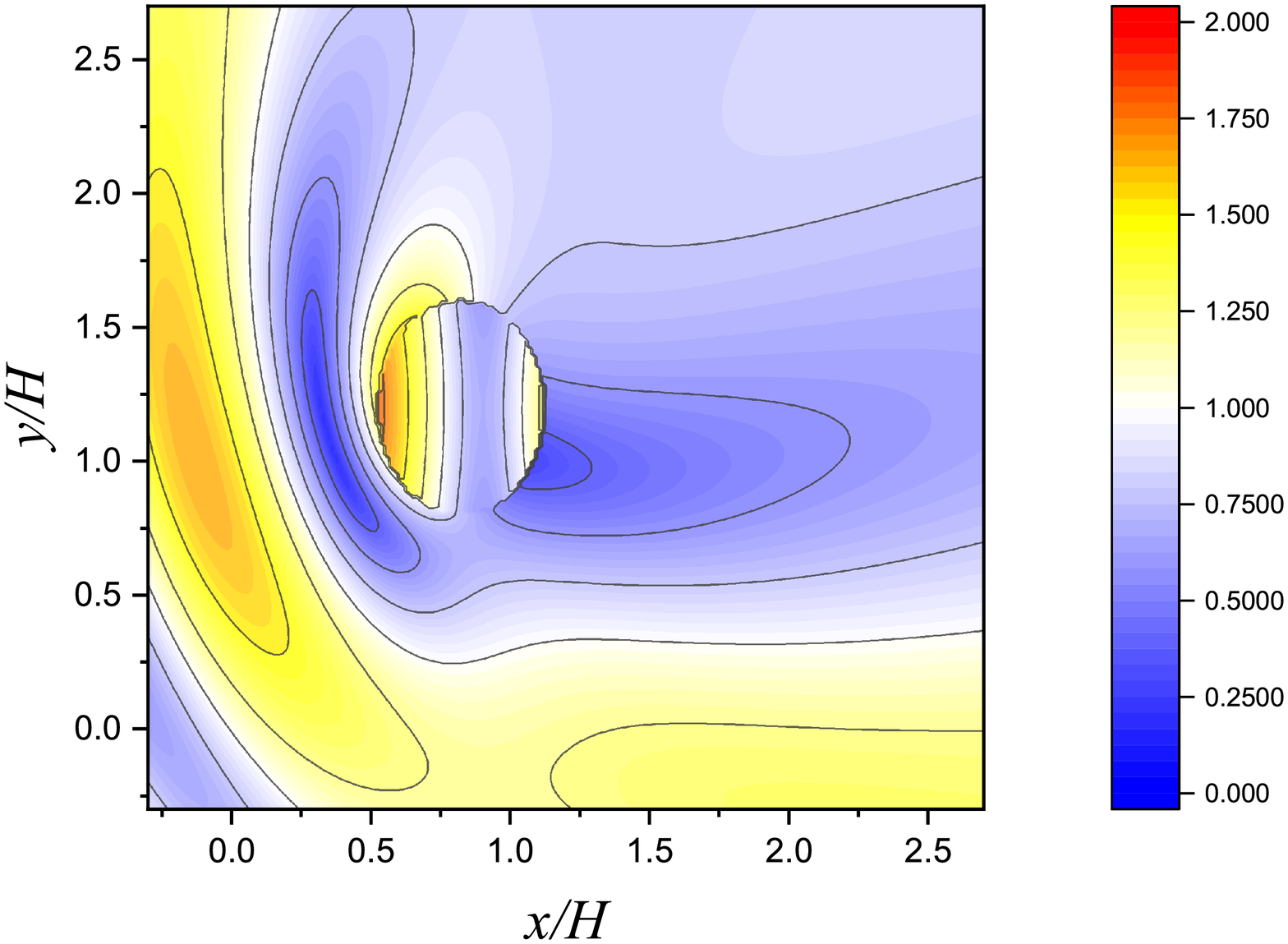}}
		\subfigure[Presence of cage ($k_0h=1.25$)]{\includegraphics*[width=8cm]{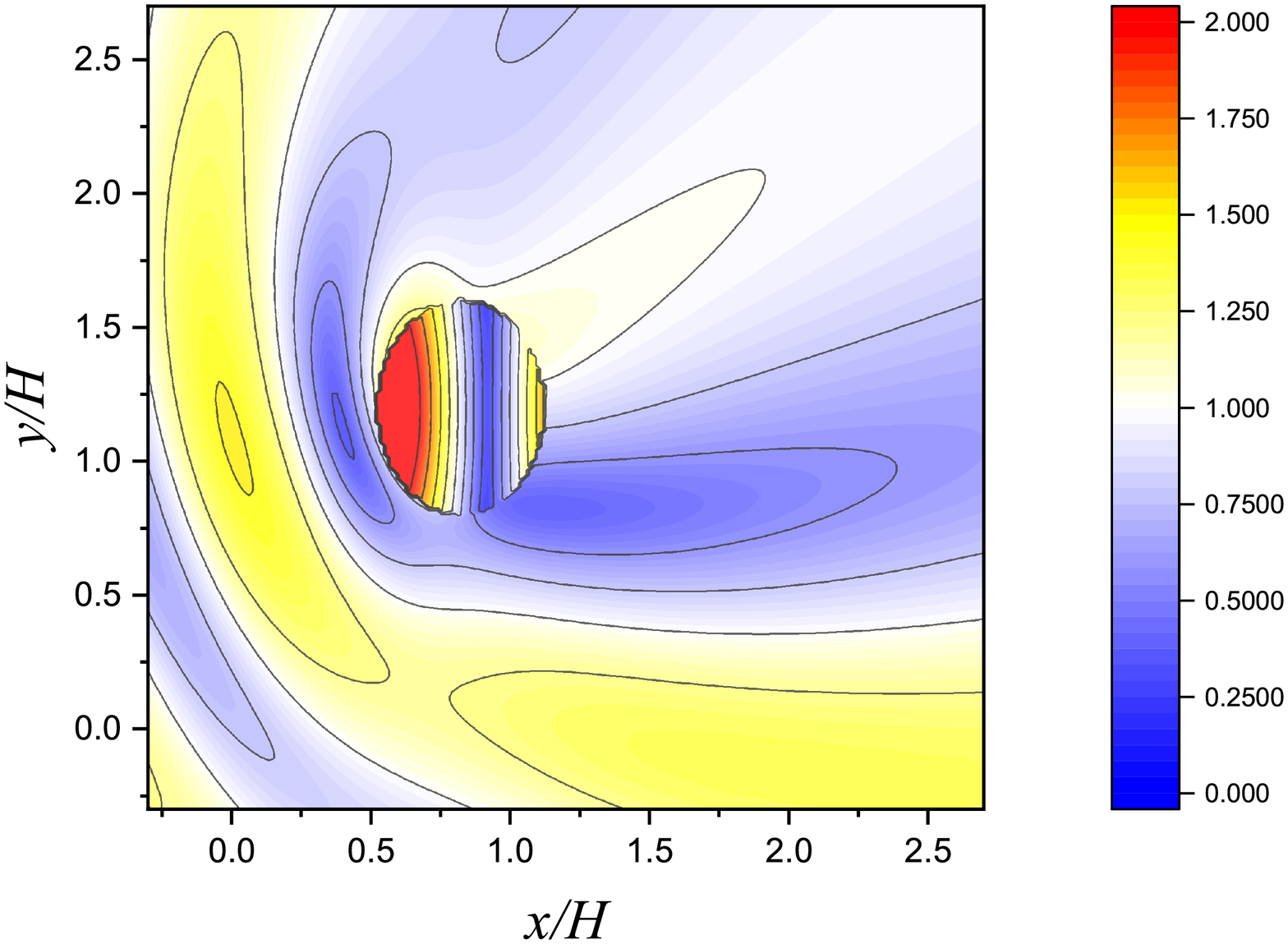}}
		\subfigure[Presence of cage ($k_0h=1.5$)]{\includegraphics*[width=8cm]{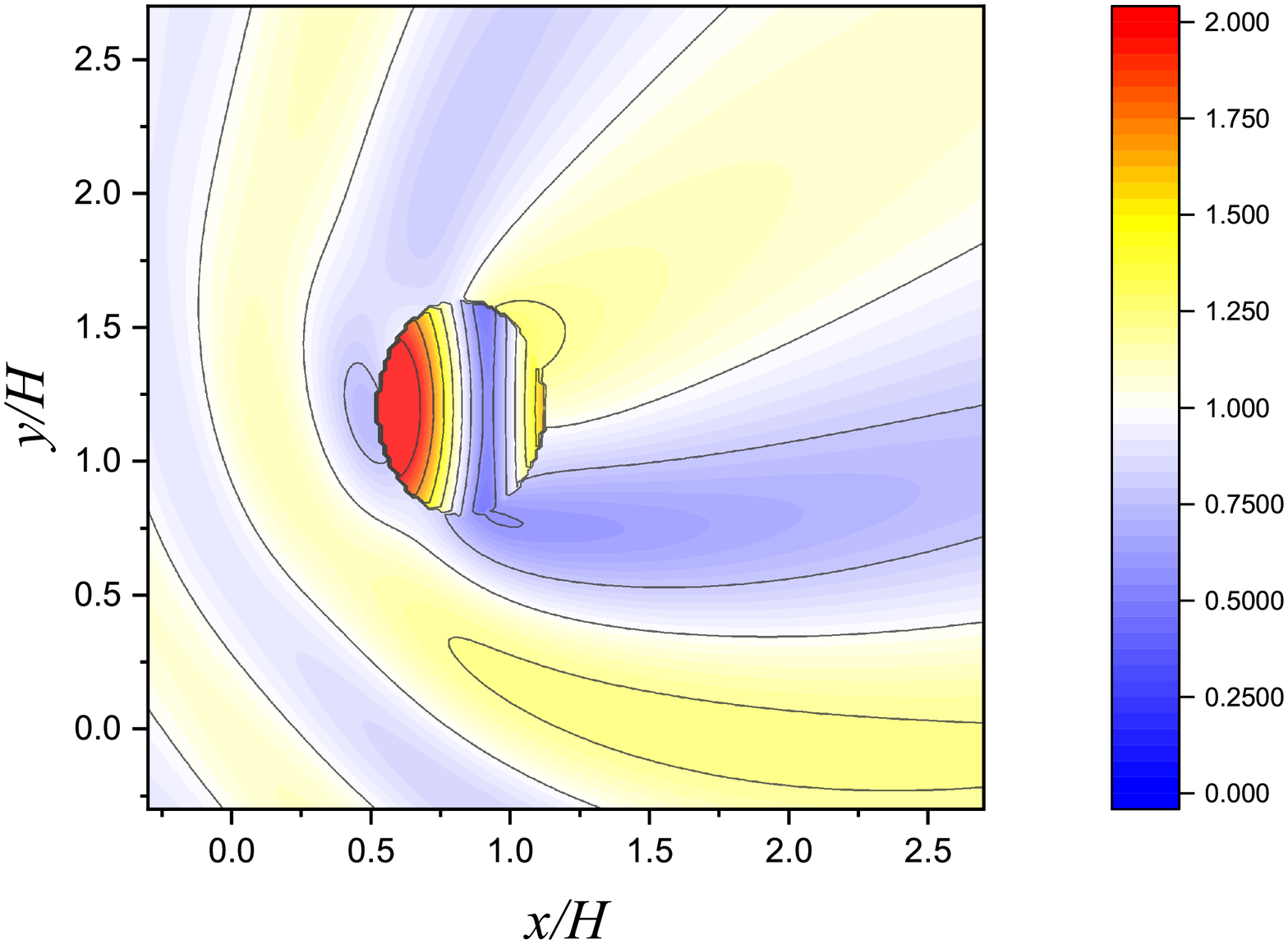}}
		\caption{Surface amplitude distributions $|\eta|$ for the (a) absence  and (b)--(d) presence of fishing cage at different non-dimensional wavenumber. The other parameters are $G=3+3\mathsf{i}$, $T=0.4$, $h/H=0.4$, $\beta=30^\circ$ and $\mu=10^6$N/m.} \label{f2} 
	\end{center}
\end{figure}

Fig.~\ref{f2} portrays the amplitude of wave-field in both the presence and absence of a fishing cage. In the first window [Fig.~\ref{f2}(a)], the incident wave propagating with the phase angle $\beta=30^\circ$ and unit peak-to-peak amplitude is illustrated. When it interacts with the single fishing cage as seen in Figs.~\ref{f2}(b)--(d), the waves get scattered, and the portion of waves is dissipated due to the presence of porosity in the cage. Further, the amplitude of scattered wave-field is less than the absolute value of incident amplitude [i.e. $|A|<2$]. On increasing the non-dimensional wavenumber $k_0h$, the wave energy decreases in the deep water. Thus, in the presence of a fishing cage, more energy dissipation occurs at the larger values of $k_0h$ as compared to that of the smaller values. It is clear from Fig.~\ref{f2}(a) that the waves with the smaller wavenumbers are damped more efficiently as compared to the waves having larger wavenumbers [Figs.~\ref{f2}(b)--(d)]. 

\begin{figure}[ht!]
	\begin{center}
		\subfigure[$k_0h=1$]{\includegraphics*[width=5.4cm,height=4.4cm]{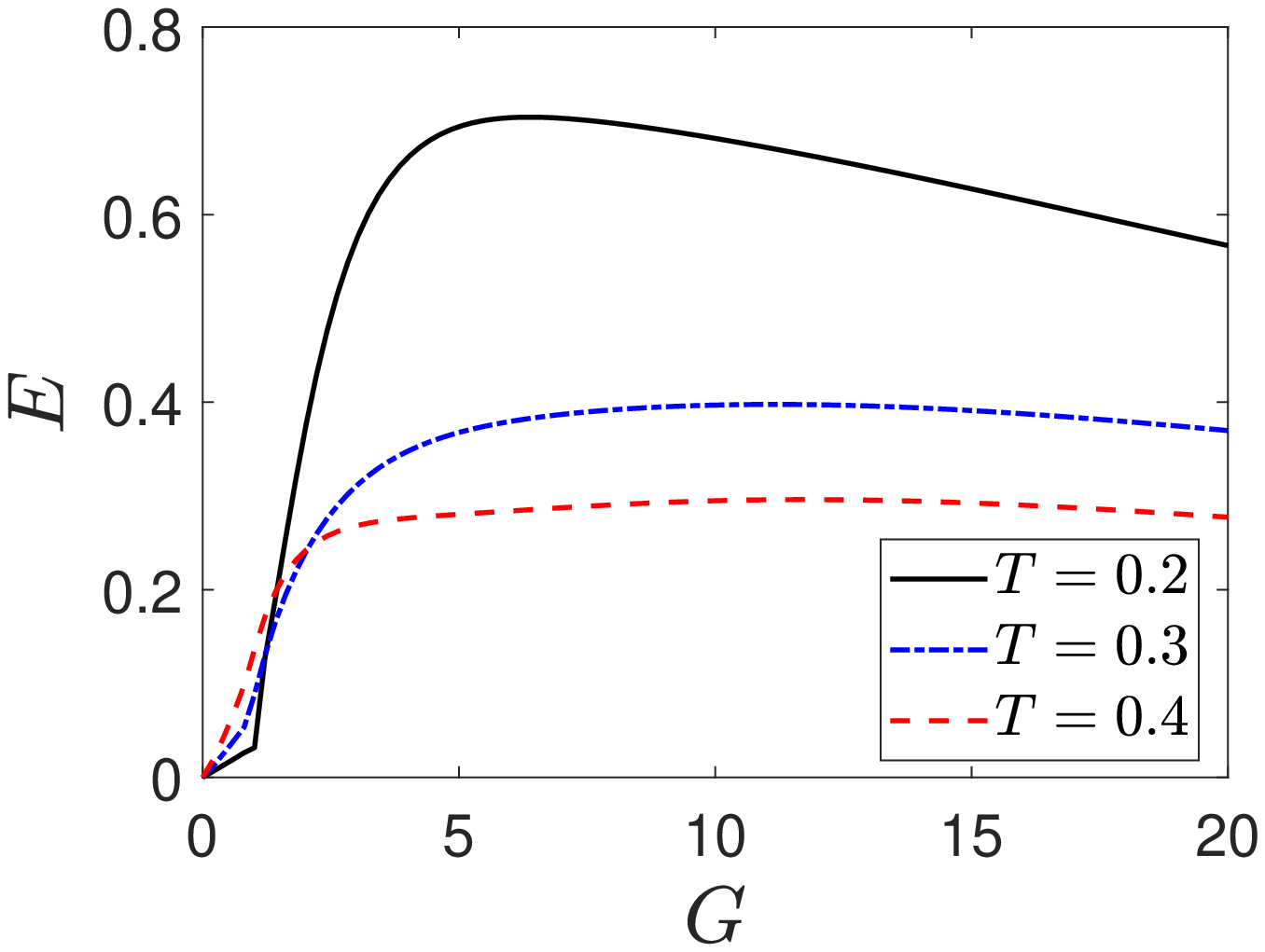}}
		\subfigure[$k_0h=1.25$]{\includegraphics*[width=5.4cm,height=4.4cm]{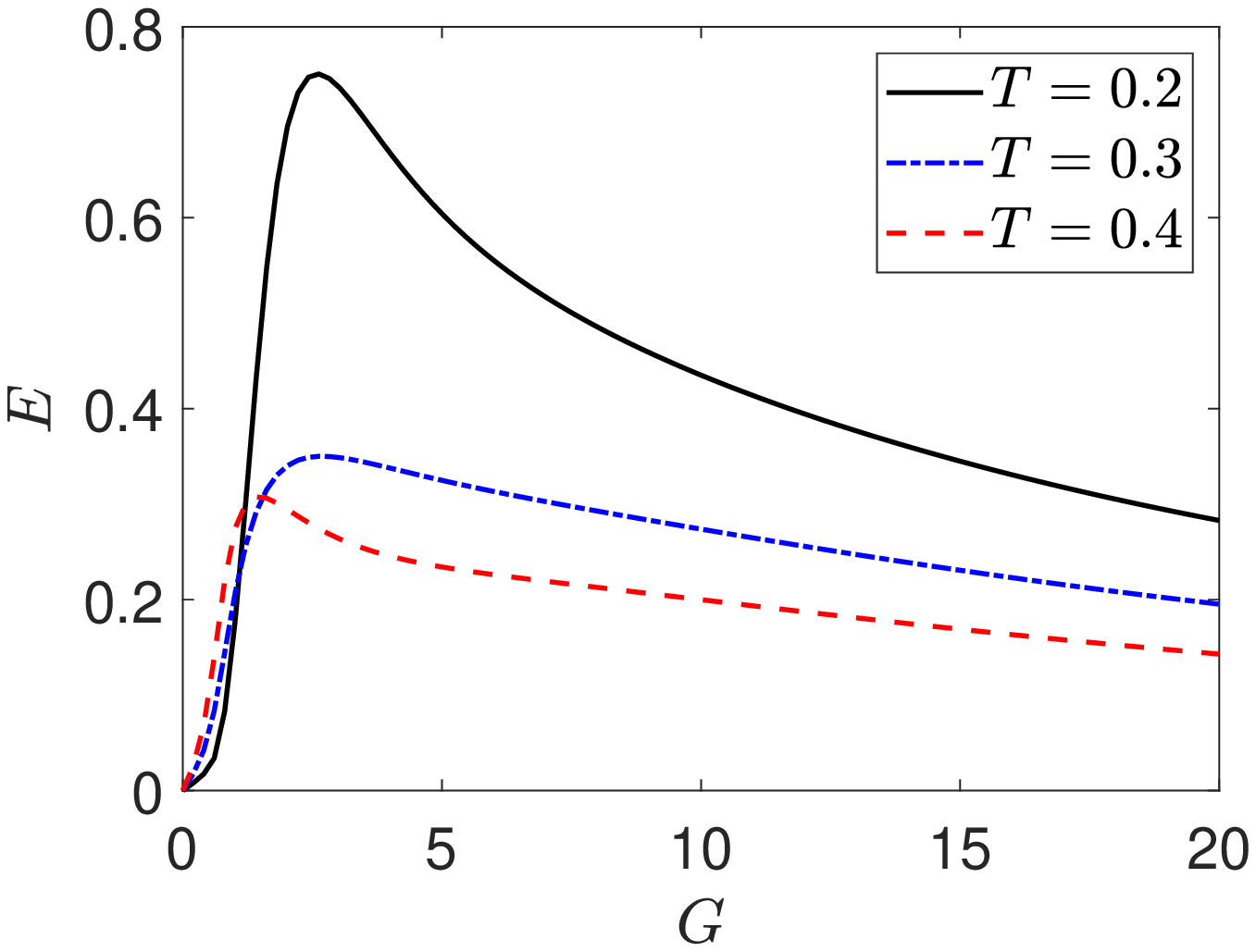}}
		\subfigure[$k_0h=1.5$]{\includegraphics*[width=5.4cm,height=4.4cm]{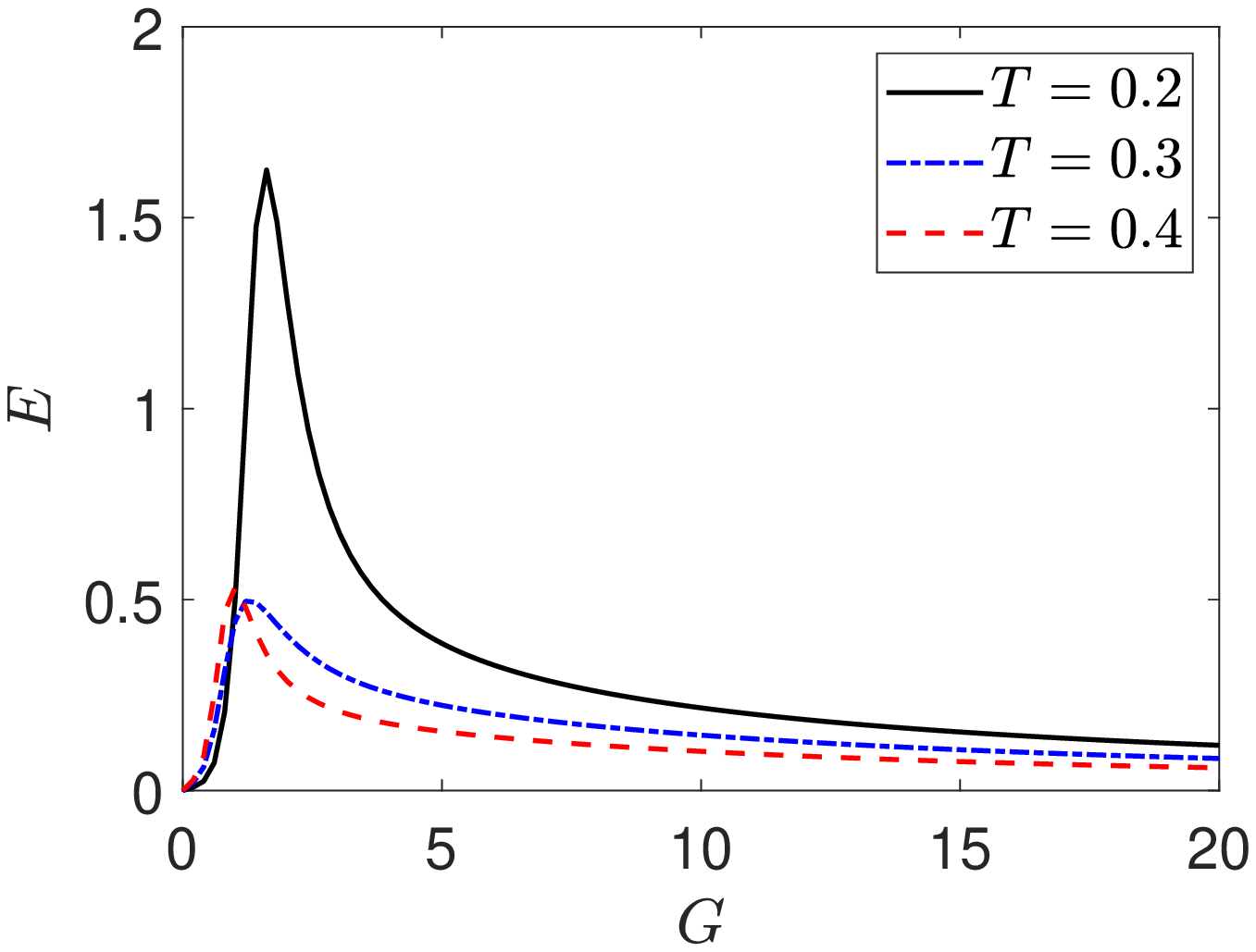}}
		\caption{Power dissipation $E$ against the real part of $G$ for different values of $T$ at various nondimensional wavenumber $k_0h$. The other parameters are $G=3+3\mathsf{i}$, $h/H=0.4$, $\beta=30^\circ$ and $\mu=10^6$N/m.} \label{f3} 
	\end{center}
\end{figure}

Figs.~\ref{f3}(a)--(c) illustrate the energy dissipation as a function of real porous-effect parameters (where the inertial effects [i.e. $Im(G)=0$ are neglected] for different values of $T$ at various wavenumbers. In general, the energy dissipation increases initially and attains maximum, then decreases for larger values of $G$. There is a reducing energy dissipation for larger values of $G$ due to the transparency of the cage to the incoming wave, where the dissipation does not occur. With an increase in the values of $T$, the dissipation follows an increasing pattern for smaller $G$, and it follows the reverse pattern for moderate and larger $G$. When the porous-effect parameter is less, the increasing tensile force develops more restoring force against the wave interacting with the cage. Thus, it dissipates more energy. On the other hand, for moderate and larger $G$, the structure gradually becomes transparent, and the waves are more likely reflected than dissipated. This is a reason behind the decreasing energy dissipation for moderate and larger values of $G$. For the waves with larger wavenumber [Fig.~\ref{f3}(c)], the energy dissipation increases for smaller and moderate values of $G$, whereas it falls significantly for larger values of $G$.

\begin{figure}[h!]
	\begin{center}
		\subfigure[$\mu=10^5$N/m]{\includegraphics*[width=5.4cm,height=4.4cm]{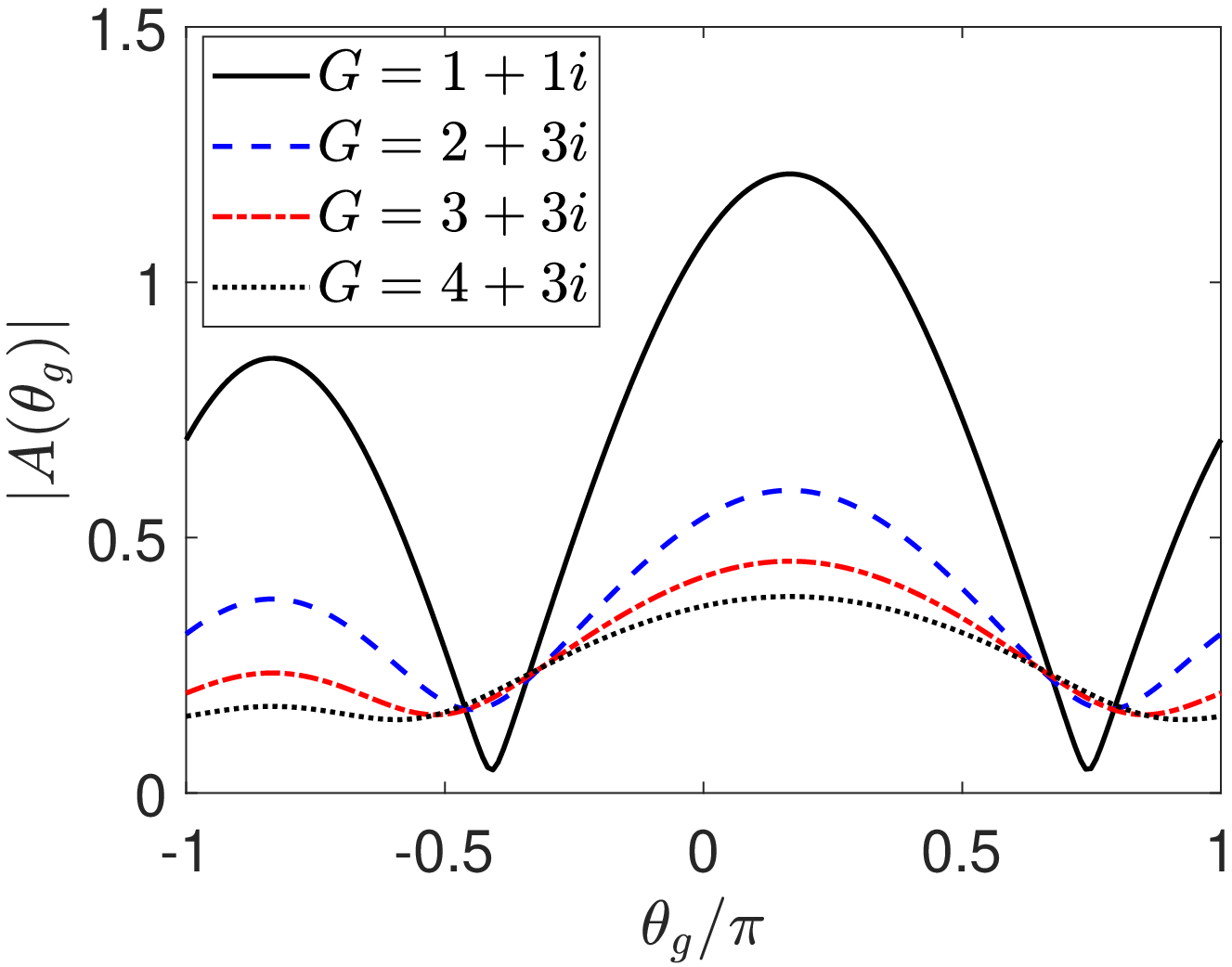}}
		\subfigure[$\mu=10^6$N/m]{\includegraphics*[width=5.4cm,height=4.4cm]{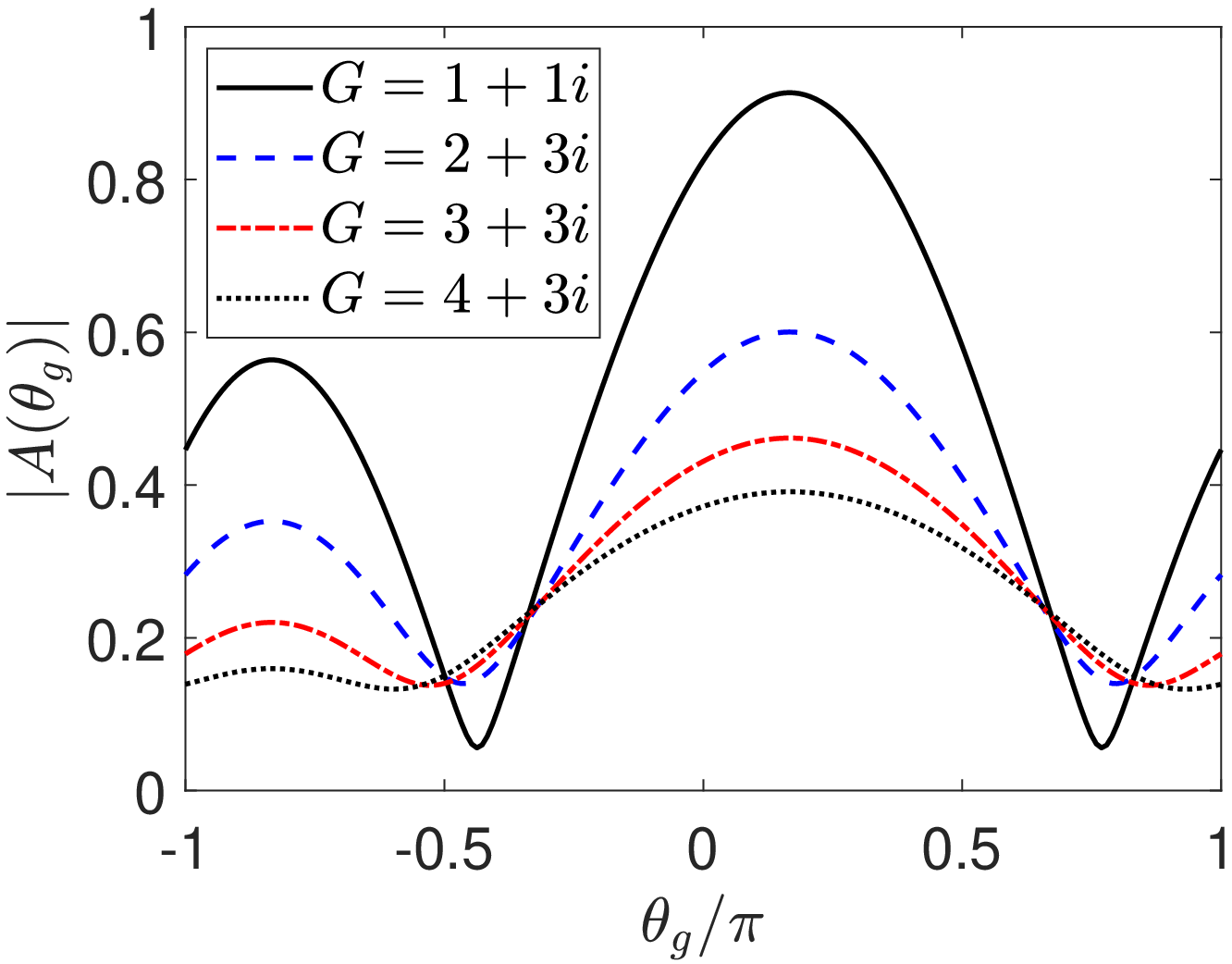}}
		\subfigure[$\mu=10^{7}$N/m]{\includegraphics*[width=5.4cm,height=4.4cm]{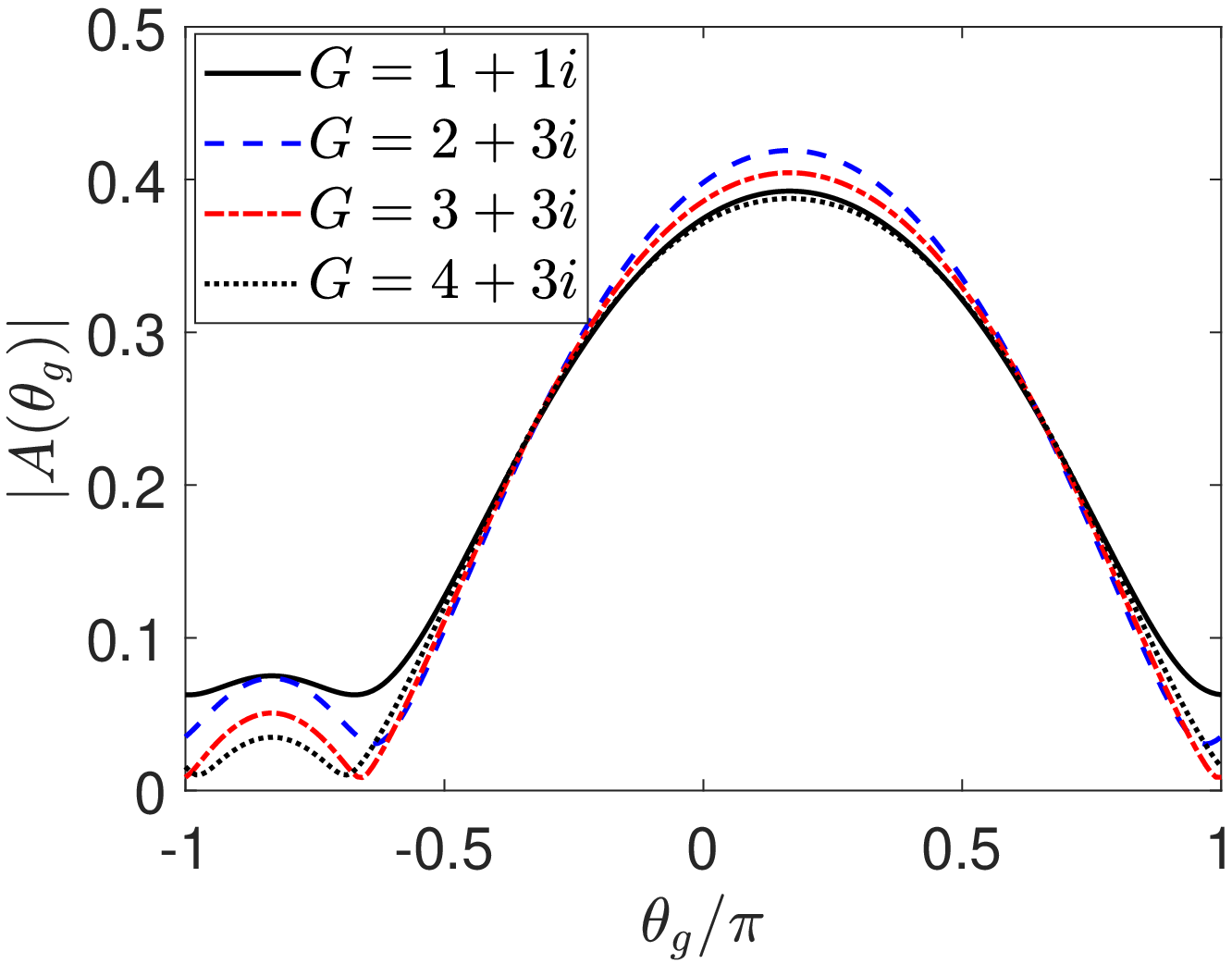}}
		\caption{Modulus of scattering coefficient $|A(\theta_g)|$ for different values of $G$ at various mooring spring constant $\mu$. The other parameters are $T=0.4$, $h/H=0.4$, $\beta=30^\circ$ and $k_0h=1.25$.} \label{f4}  
	\end{center}
\end{figure}

The modulus of scattering coefficient along the different directions from the single fishing cage is plotted in Fig.~\ref{f4} for different porous-effect parameter $G$ at various spring constant $\mu$. It is observed that the scattering coefficient peaks along the incident wave direction [i.e., $\theta_g=30^\circ$]. There exist two sidebands on either side of the central band, which indicate that the wave scattering decreases periodically on either side of incident wave direction. On increasing the mooring spring constant, the membrane stretches more and the wave scattering decreases as a consequence of less oscillation by a moored cylindrical membrane. On increasing the porous-effect parameter, the wave scattering decreases owing to an increasing wave energy dissipation. However, the deviations are quite significant for smaller and moderate values of $\mu$. In the case of larger $\mu$ [Fig.~\ref{f4}(c)], the effect of $G$ becomes negligible and the sideband dampens, which is due to less scattering and dissipation by membrane  having very high stiffness.

\begin{figure}[ht!]
	\begin{center}
		\subfigure[$\mu=10^5$N/m]{\includegraphics*[width=5.4cm,height=4.4cm]{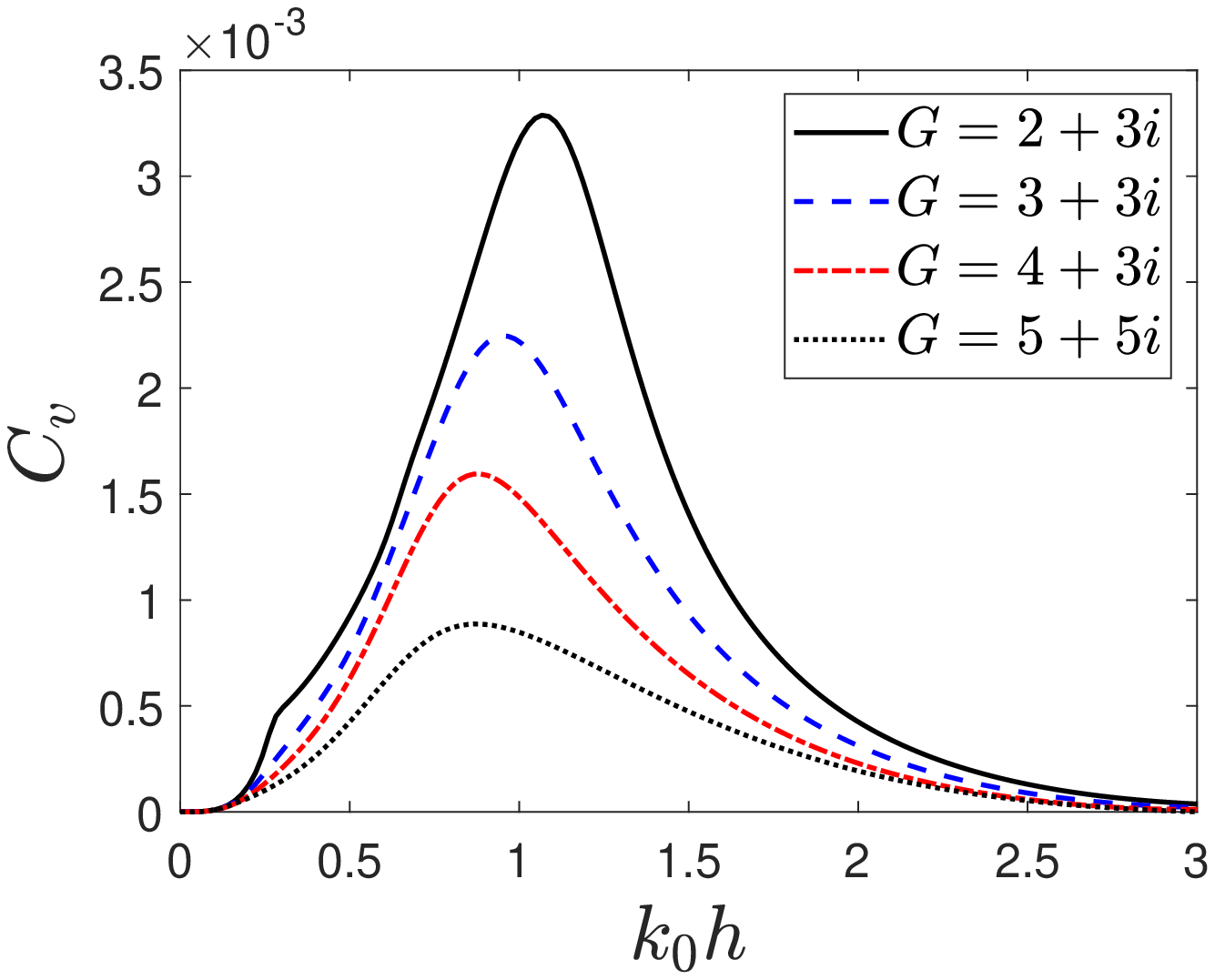}}
		\subfigure[$\mu=10^6$N/m]{\includegraphics*[width=5.4cm,height=4.4cm]{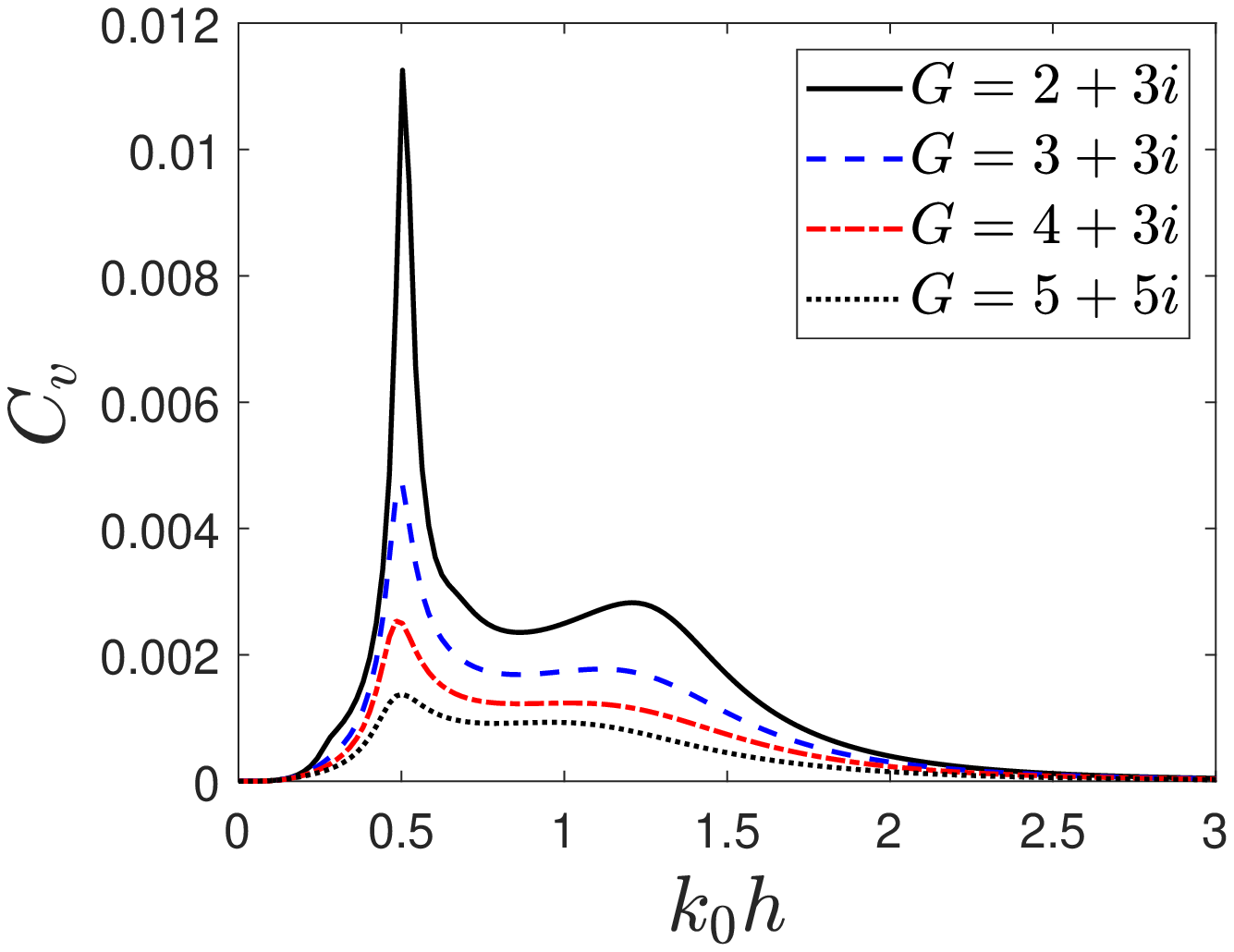}}
		\subfigure[$\mu=10^{7}$N/m]{\includegraphics*[width=5.4cm,height=4.4cm]{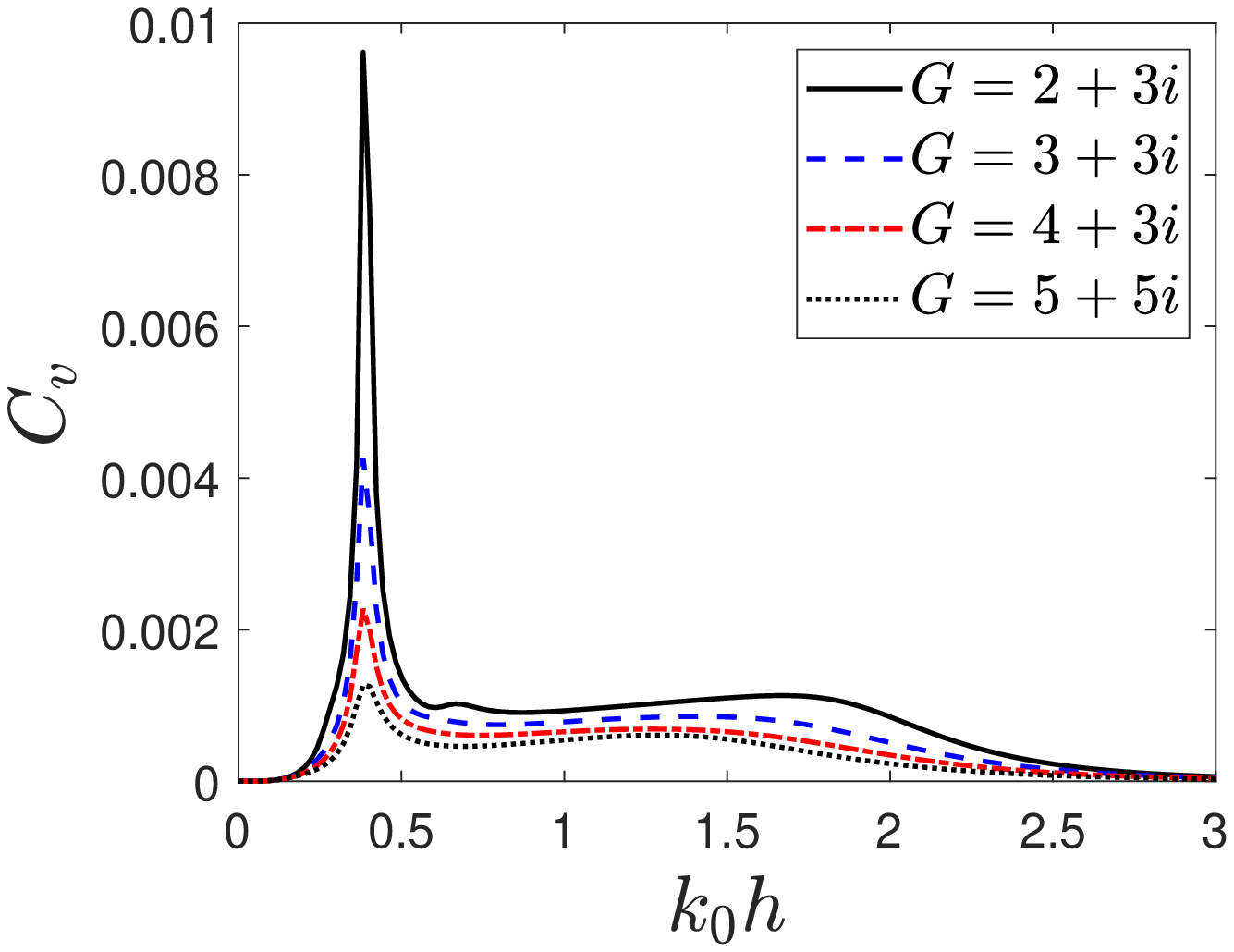}}
		\caption{Vertical wave force acting on the fishing cage against the wavenumber for different values of $G$ at various mooring spring constant $\mu$. The other parameters are $T=0.4$, $h/H=0.4$, $\beta=30^\circ$ and $k_0h=1.25$.} \label{f5} 
	\end{center}
\end{figure}

In Figs.~\ref{f5}(a), (b) and (c), the vertical wave force acting on the fishing cage is plotted as a function of non-dimensional wavenumber $k_0h$ for different values of $G$ at $\mu=10^5$N/m, $\mu=10^6$N/m and $\mu=10^7$N/m, respectively. The vertical wave force increases and attains maximum, then decreases for increasing values of $k_0h$. The maximum wave force happens at a certain wavenumber as a result of more wave reflection from the cage, in which the dissipation decreases. This specific wavenumber shifts based on the cage parameter. With an increase in the mooring spring constant, the magnitude of wave force increases initially; then, there is a moderate decrease for increasing $\mu$. It is interpreted that there exists an optimum value of mooring spring constant [i.e., $\mu=10^6$N/m], where the wave force attains maximum value resulting in more reflection. Further, there is a sharp rise in the wave force for increasing $\mu$, which is due to increasing membrane stiffness.

\subsection{Dual fishing cage}

\begin{figure}[h!]
	\begin{center}
		\subfigure[$R_{12}=H$]{\includegraphics*[width=8cm]{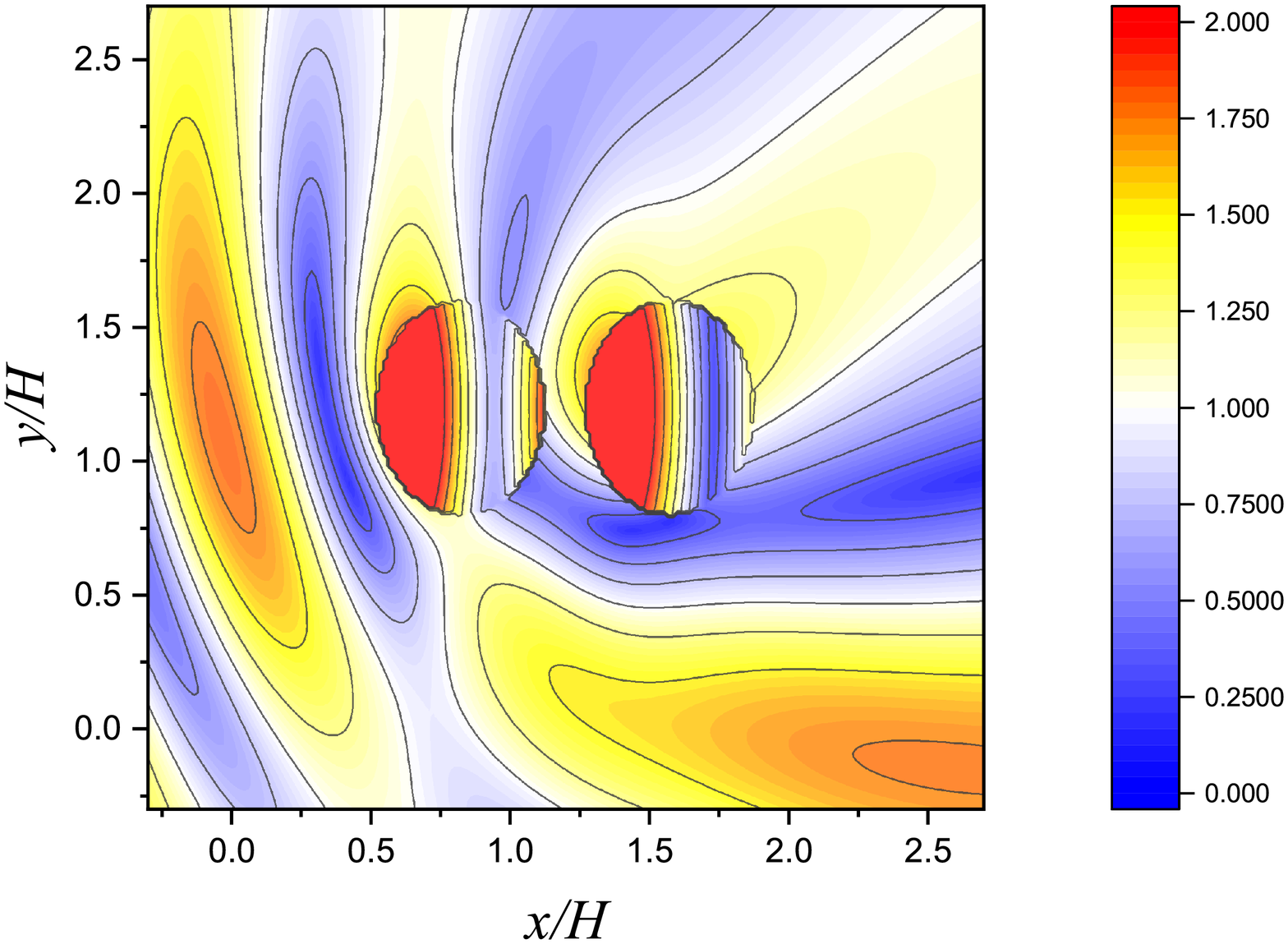}}
		\subfigure[$R_{12}=2H$]{\includegraphics*[width=8cm]{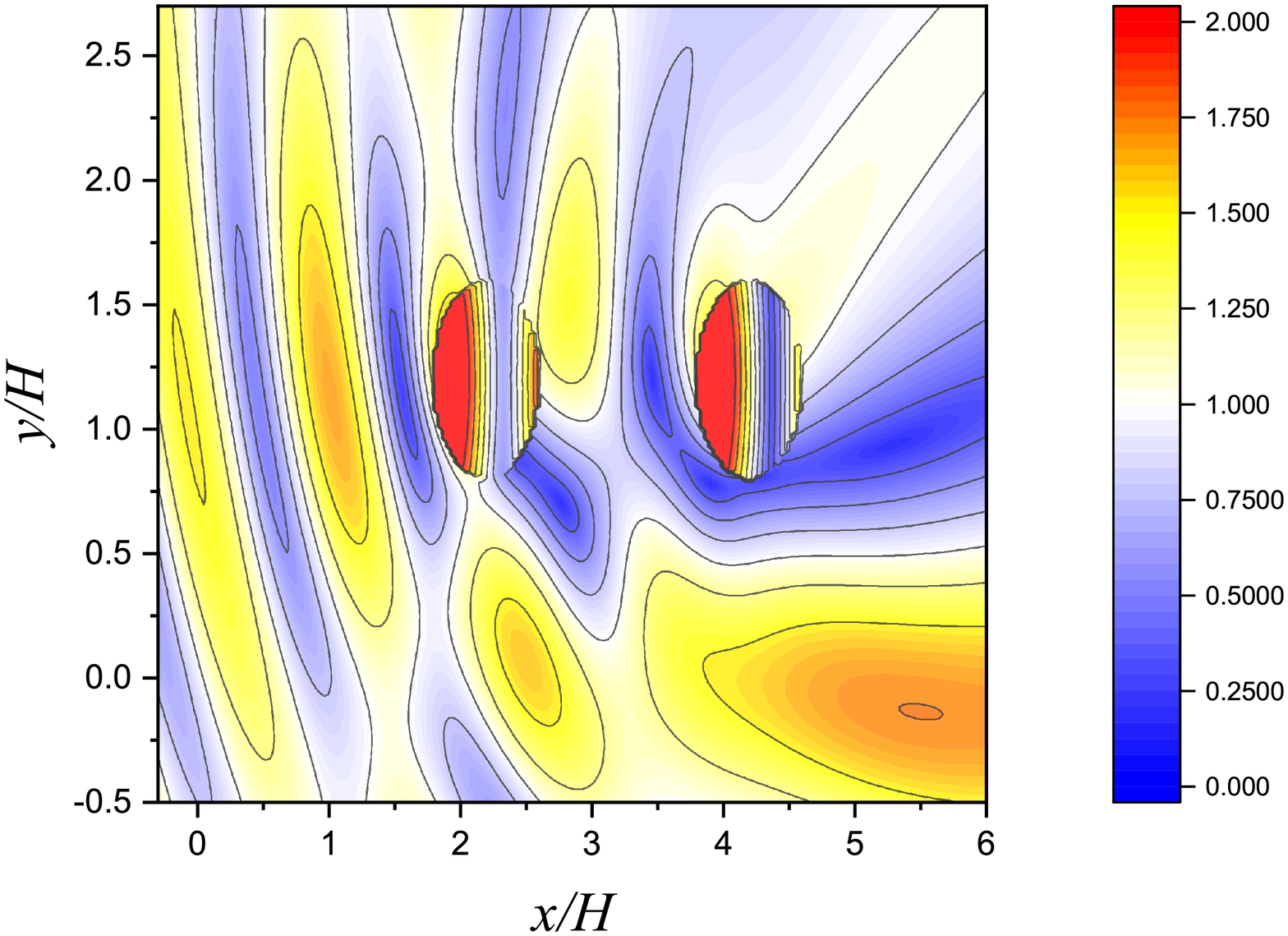}}
		\caption{Surface amplitude distributions $|\eta|$ of the dual fishing cage system for different spacing parameter $R_{12}$. The other parameters are $G=3+3\mathsf{i}$, $T=0.4$, $h/H=0.4$, $\beta=30^\circ$ and $\mu=10^6$N/m.} \label{f6} 
	\end{center}
\end{figure}
In Fig.~\ref{f6}, the effect of spacing on the amplitude of surface elevation is plotted. It is noticed the wave interacts the dual system with phase angle $\beta=30^\circ$ from both the figures . On interacting the first cage, the wave amplitude inside the cage increases initially, then there is an energy loss that happens inside the cage due to destructive interference between the incoming waves and inner scattered waves. The same phenomenon of destructive interference occurs inside the second cage. However, the constructive interference occurs between the cages on keeping the both cages close to each other (i.e. $R_{12}=H$). This results in the high amplitude waves in the region confined between the two cages as observed in Fig.~\ref{f6}(a). For larger spacing between the cages (i.e. $R_{12}=2H$), there is a loss in wave energy resulting in the damping of wave amplitudes between the cages as observed in Fig.~\ref{f6}(b). Moreover, on comparing both the figures, the wave propagation in the lee-ward side of the structure decreases while increasing the spacing between the cages. This, in turn, reduces the damages to the fishing cage system.

\begin{figure}[h!]
	\begin{center}
		\subfigure[$T=0.1$]{\includegraphics*[width=5.4cm,height=4.4cm]{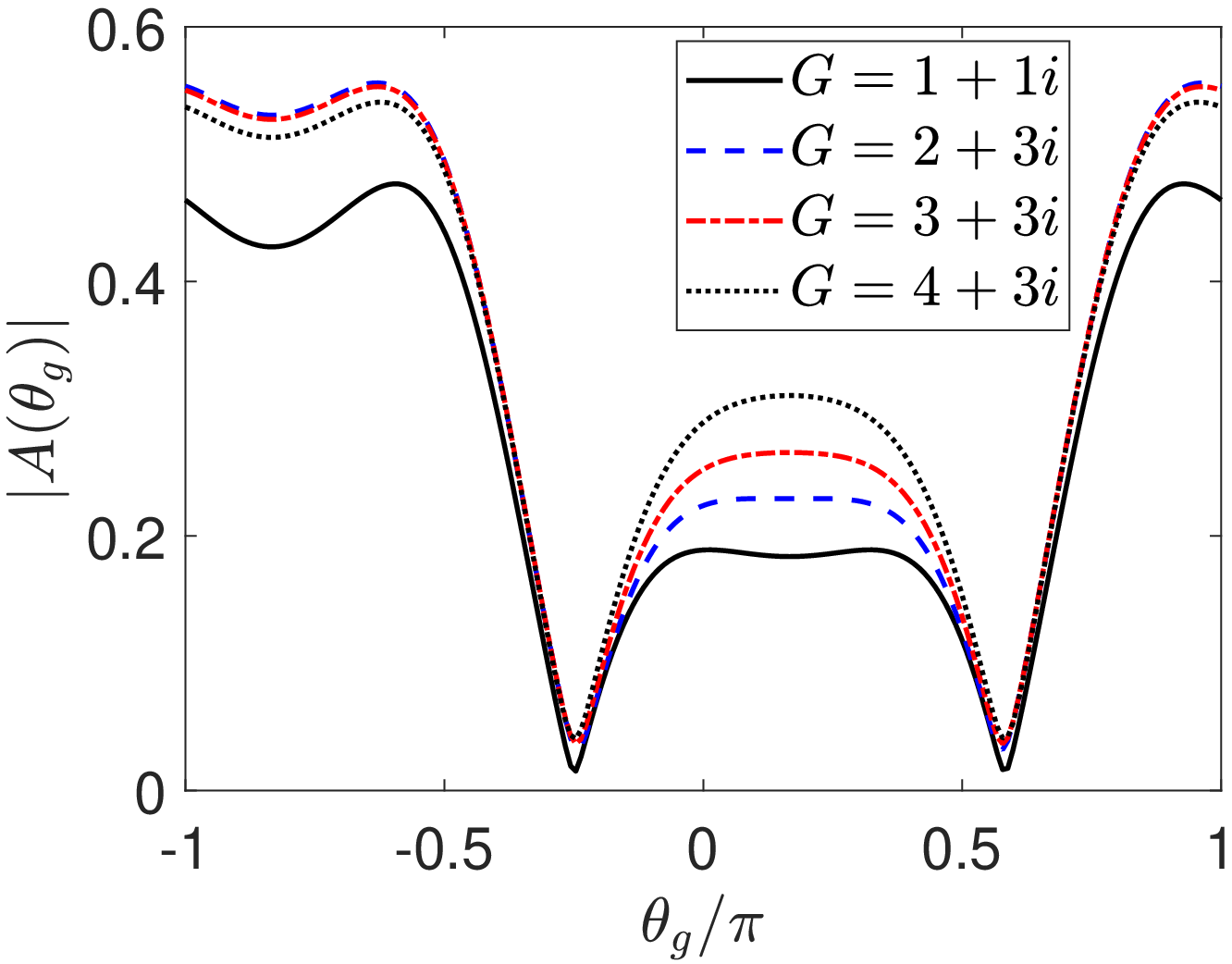}}
		\subfigure[$T=0.2$]{\includegraphics*[width=5.4cm,height=4.4cm]{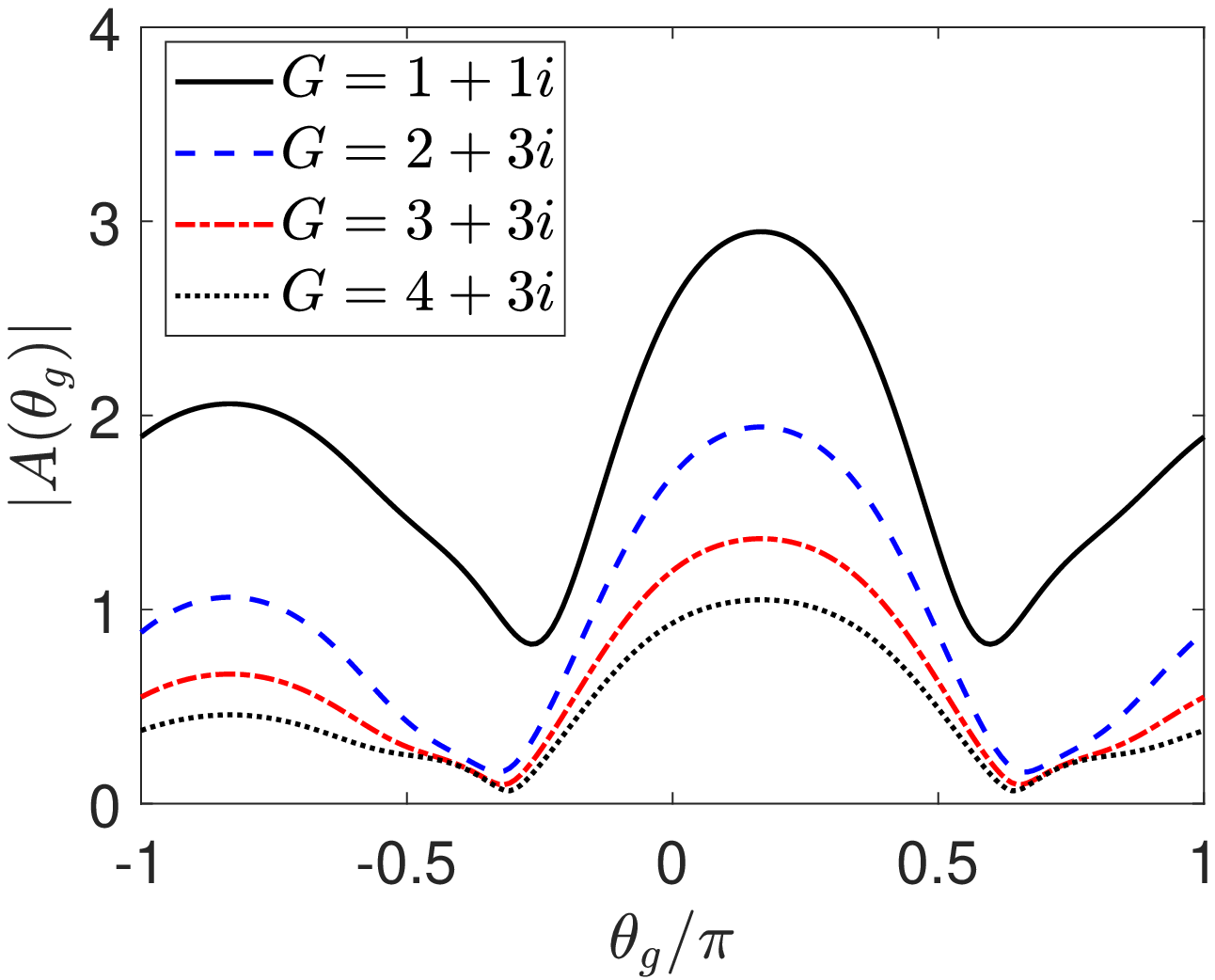}}
		\subfigure[$T=0.4$]{\includegraphics*[width=5.4cm,height=4.4cm]{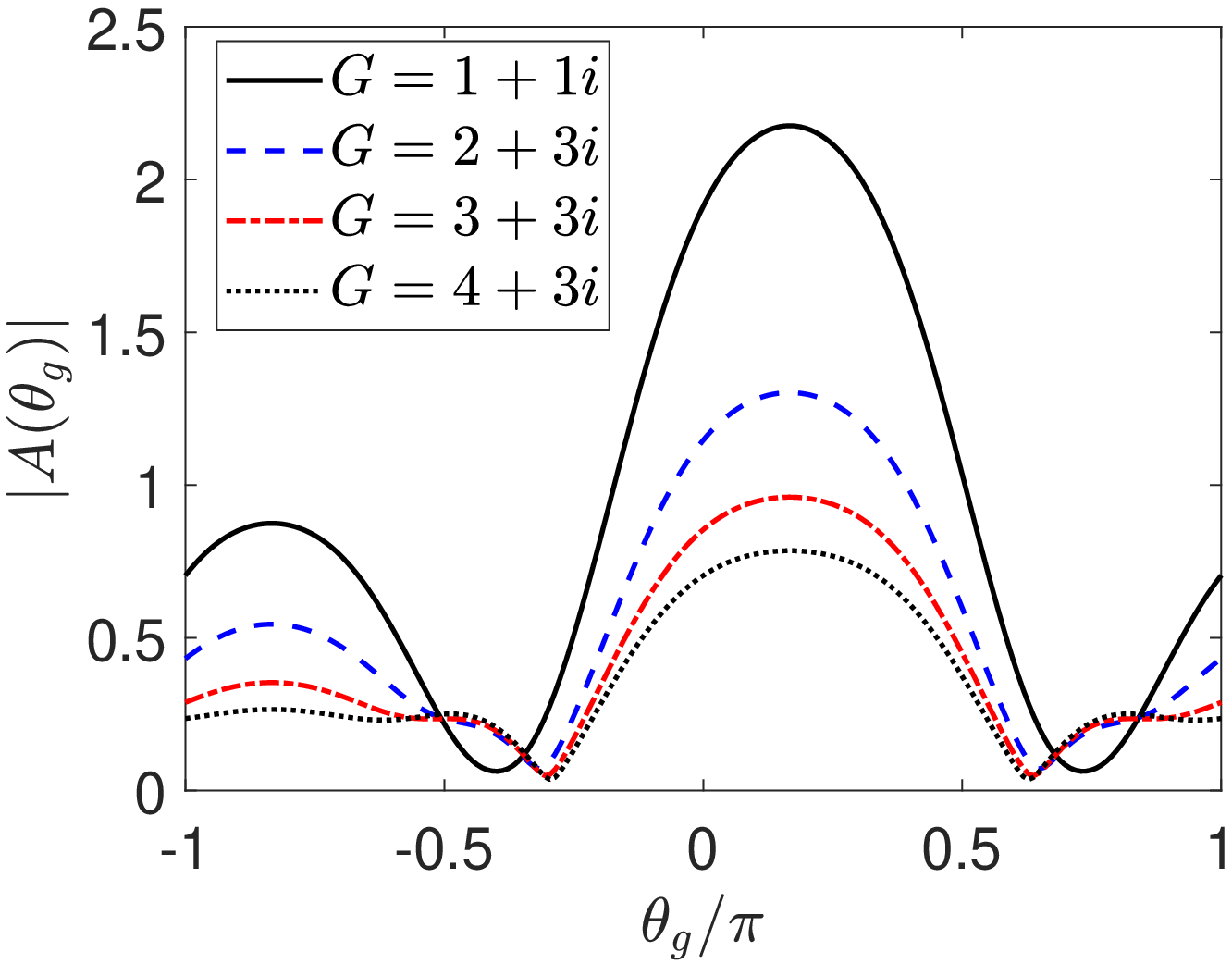}}
		\caption{Modulus of scattering coefficient $|A(\theta_g)|$ for different values of $G$ at various tensile force $T$. The other parameters are $h/H=0.4$, $\beta=30^\circ$, $k_0h=1.25$, $R_{12}=H$ and $\mu=10^6$N/m.} \label{f7} 
	\end{center}
\end{figure}

The modulus of scattering coefficient along the different directions from the dual fishing cage system is plotted in Fig.~\ref{f7} for different porous-effect parameter $G$ at various tensile force $T$. It is noticed that the central band peaks at the incident wave direction for $\theta_g=30^\circ$ in all figures. For smaller tensile force [Fig.~\ref{f7}(a)], the central band reduces and scatters less wave as compared to the sidebands, which scatters more waves. As the value of $T$ increases, the central band's increases, and the sideband decreases owing to increasing restoring force developed inside the membrane. Moreover, there exists an optimum value of $T$ [i.e., $T=0.2$] at which the maximum waves are scattered, after which it decreases slowly for increasing $T$. On increasing the value of $G$, the wave scattering from the cage system increases as a consequence of impedance offered by the cages for membrane having less tensile force. For $T\geq0.2$, the waves scatter from the dual cage system decreases for increasing porous-effect parameter, where the impedance decreases and wave energy dissipation increases. 

\begin{figure}[ht!]
	\begin{center}
		\subfigure[$T=0.1$]{\includegraphics*[width=5.4cm,height=4.4cm]{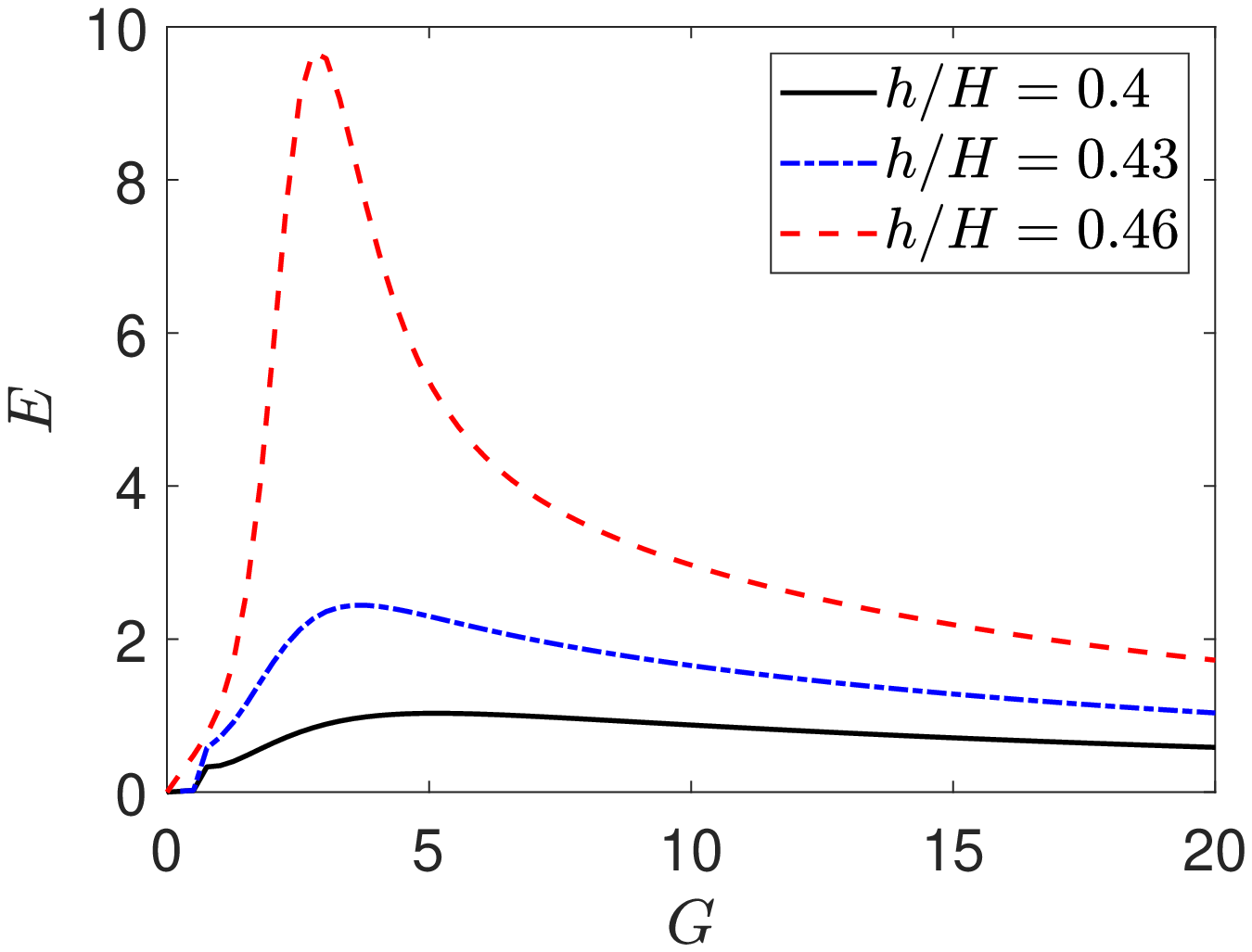}}
		\subfigure[$T=0.2$]{\includegraphics*[width=5.4cm,height=4.4cm]{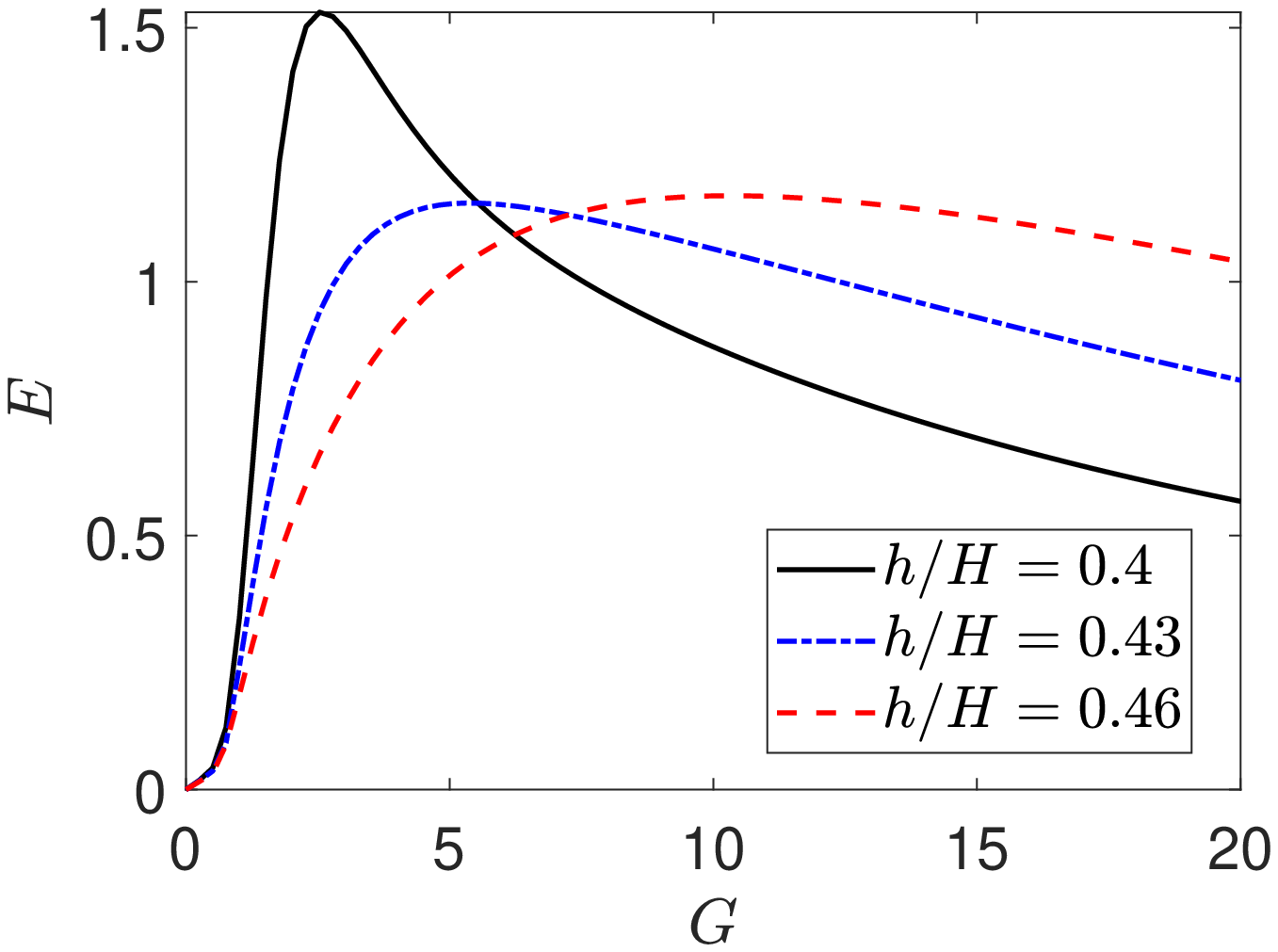}}
		\subfigure[$T=0.4$]{\includegraphics*[width=5.4cm,height=4.4cm]{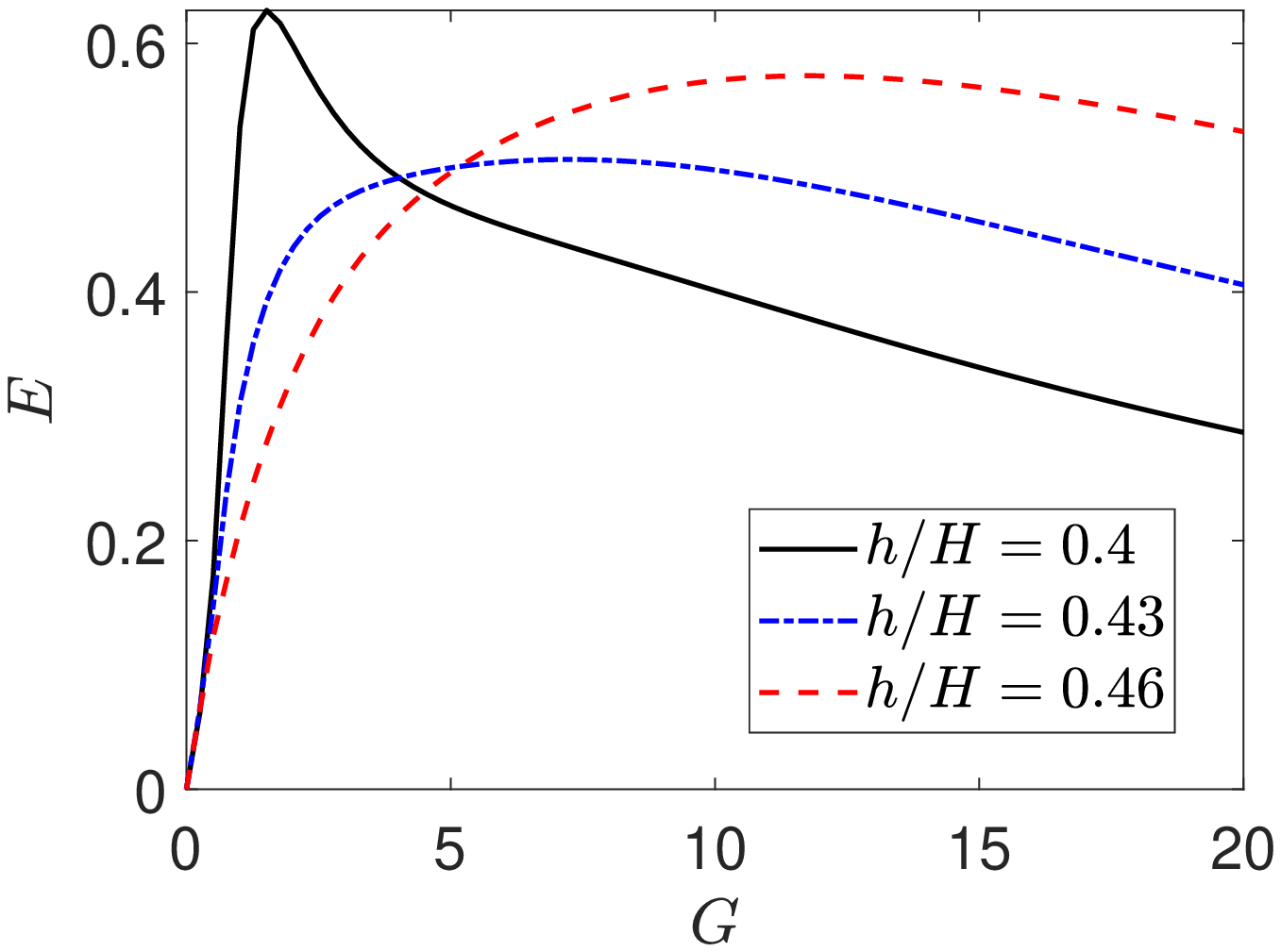}}
		\caption{Power dissipation $E$ against the real part of $G$ for different values of $h/H$ at various tensile force $T$.  The other parameters are $G=3+3\mathsf{i}$, $\beta=30^\circ$, $k_0h=1.25$, $R_{12}=H$ and $\mu=10^6$N/m.} \label{f8} 
	\end{center}
\end{figure}

The energy dissipation by the dual fishing cages against the real values of $G$ is plotted for various $h/H$ at varying $T$ in Fig.\ref{f8}, respectively. Initially, the power dissipation increases with respect to $G$, then it attains maximum and befalls gradually for larger values of $G$. It is noticed that for $T=0.1$, the energy dissipation increases for an increasing $h/H$ as a result of increasing energy dissipation due to increased surface area of the cage. In the case of $T\geq 0.2$, the energy dissipation behaves differently for $G<6$ and $G>6$. In the case of $G<6$, the energy dissipation decreases for increasing submergence depth $h/H$, where more scattering occurs as compared to the dissipation. On the other hand, for $G>6$, the energy dissipation increases for increasing $h/H$ resulting in less wave scattering as compared to the energy dissipation. Further, the membrane with less tension dissipates more energy for larger depth ratio owing to the increased energy dissipation.

\subsection{Multiple fishing cage}
\begin{figure}[ht!]
	\begin{center}
		\subfigure[]{\includegraphics*[width=8cm]{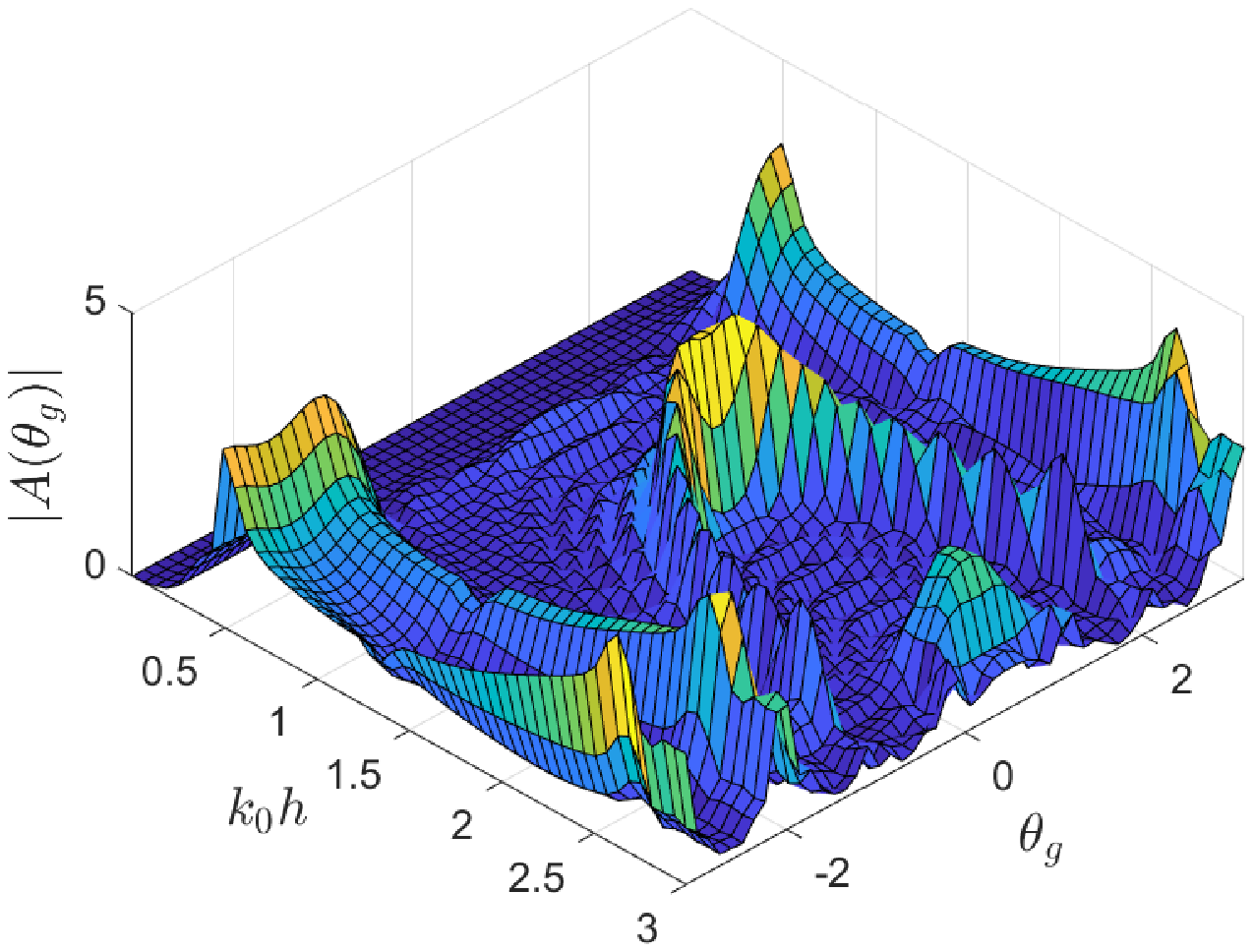}}
		\subfigure[]{\includegraphics*[width=8cm]{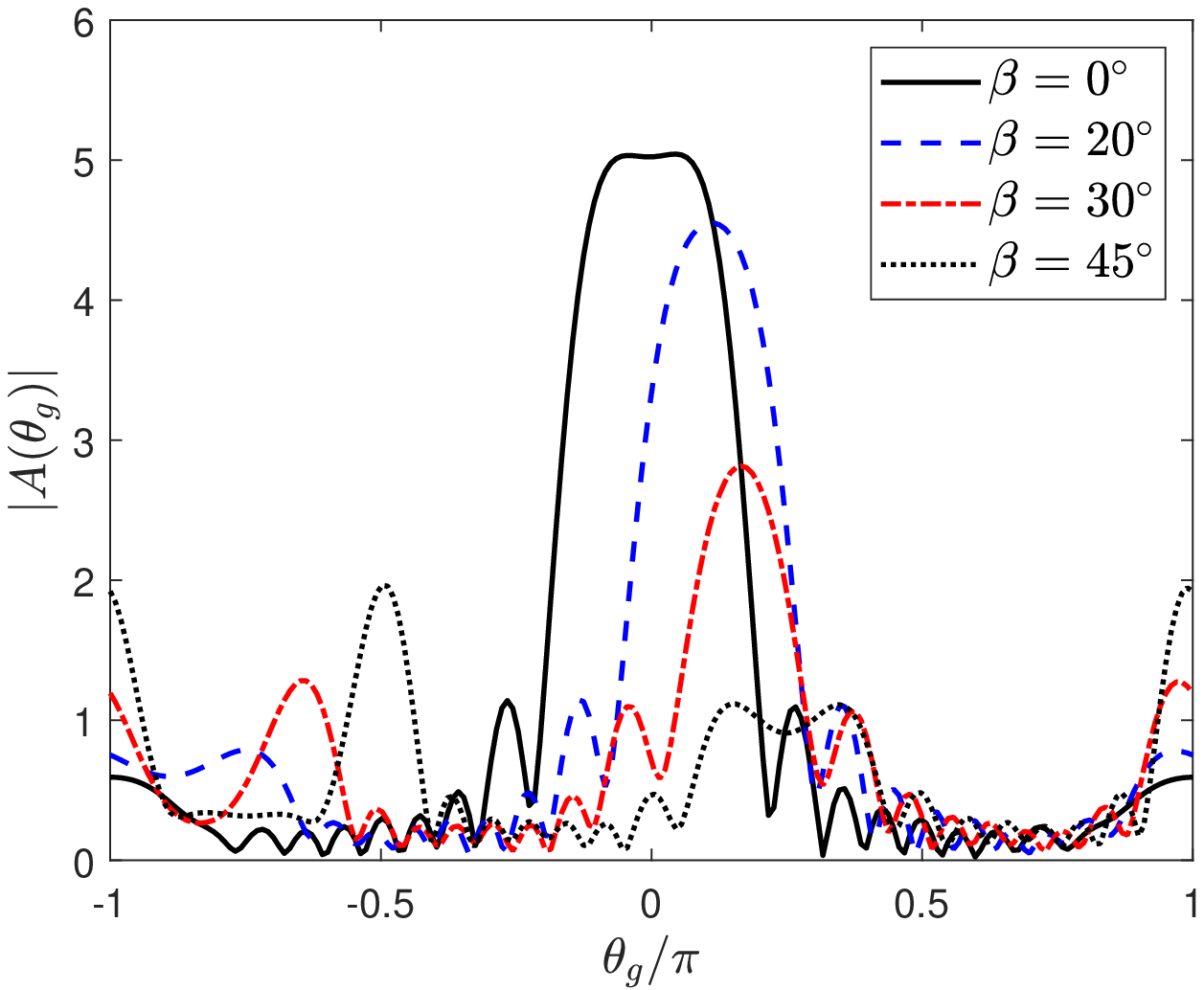}}
		\subfigure[]{\includegraphics*[width=12.2cm,height=8cm]{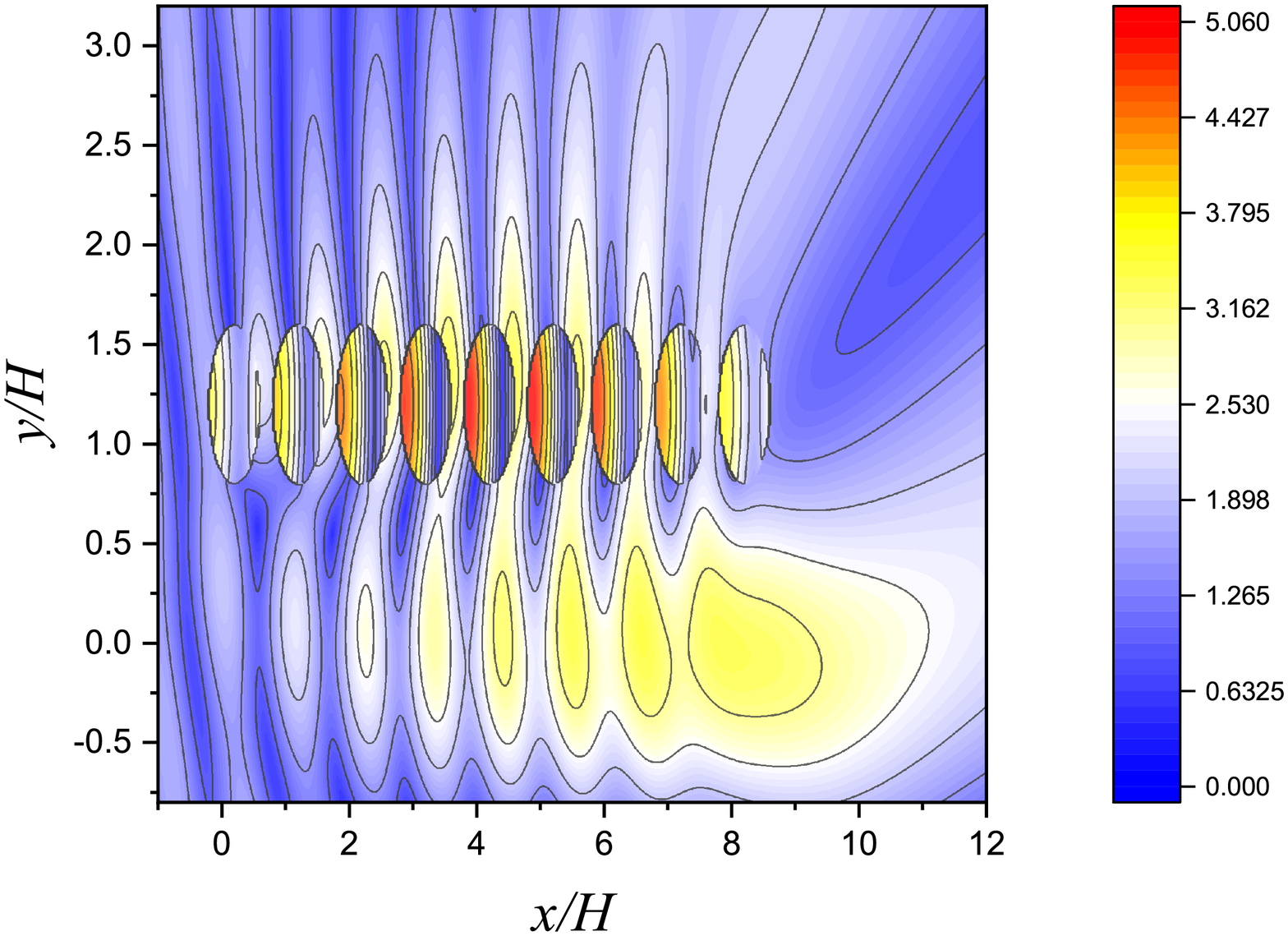}}
		\caption{Surface wave interaction with an inline arrangement of the multiple fishing showing the (a) scattering coefficient for varying $(k_0h,\theta_g)$ at $\beta=30^\circ$, (b) scattering coefficient for varying $\theta_g$ at various $\beta$ with $k_0h=1.25$, and (c) amplitude of surface elevation with $k_0h=1.25$ and $\beta=30^\circ$.  The other parameters are $h/H=0.4$ and $\mu=10^6$N/m.} \label{f9} 
	\end{center}
\end{figure}
In Fig.~\ref{f9}, the scattering coefficient and flow distributions around the multiple fishing cage system having inline arrangement are discussed. In Fig,~\ref{f9}(a), the surface plot depicting the scattering coefficients for varying values of $(k_0h,\theta_g)$ is investigated. In the inline arrangement, the incident wave is directed with the phase angle $\beta=30^\circ$ towards the system. It is observed that the wave scattering is less for smaller and larger wavenumbers along the direction of the incident wavefield (i.e., $\theta_g=30^\circ$), where most of the incoming waves pass through the system with less scattering. Around $1<k_0h<1.5$, the major portions of waves belonging to this bandwidth range are scattered more, resulting in more reflection from the system. For $k_0h=1.25$, the effect of phase angle on the scattering coefficient is plotted in Fig.~\ref{f9}(b). It is noticed that the scattering peaks at an incident wave direction, and it follows the oscillatory pattern along the other direction due to the periodic arrangement of the cage system. However, the wave scattering decreases on either side of the central band due to the inline arrangement, which allows most of the incoming waves through the cage system. Further, the wave scattering decreases for an increasing value of $\beta$ due to the less screening of waves by the other cages for higher phase angle. The corresponding flow distributions around the inline system is demonstrated in Fig.~\ref{f9}(c), where the phase angle of the incident wave is considered as $\beta=30^\circ$. In general, the wave amplitude of the surface waves decreases downstream of the cage system due to the wave dissipation by the porous cage system. Further, the wave amplitudes around the cage's circumference increases as a result of scattering between the consequent cages, and accumulated waves inside the cages are dissipated by the vertical porous structure. This reduces the transmission on the lee-side of the cage system. 

\begin{figure}[h!]
	\begin{center}
		\subfigure[]{\includegraphics*[width=8cm]{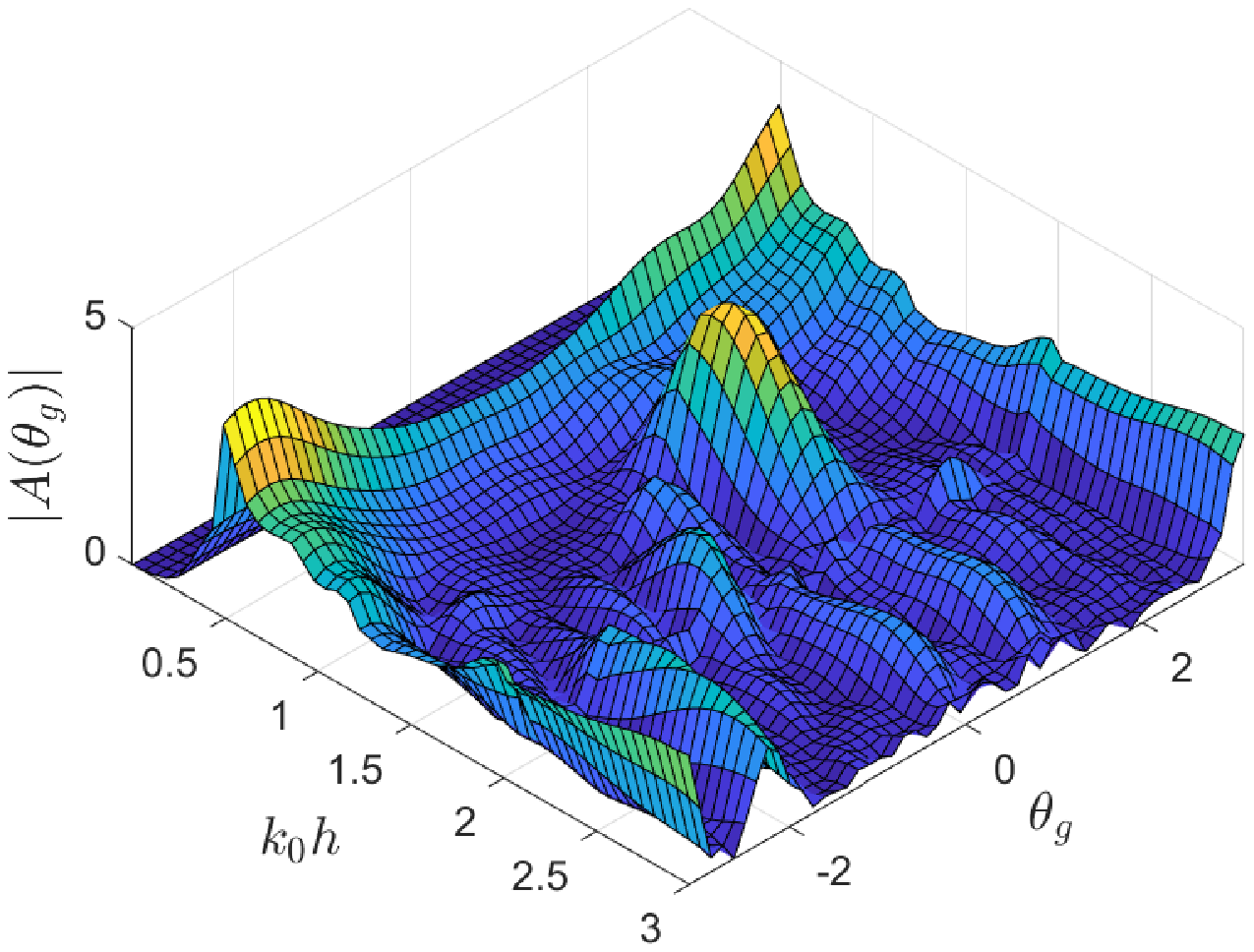}}
		\subfigure[]{\includegraphics*[width=8cm]{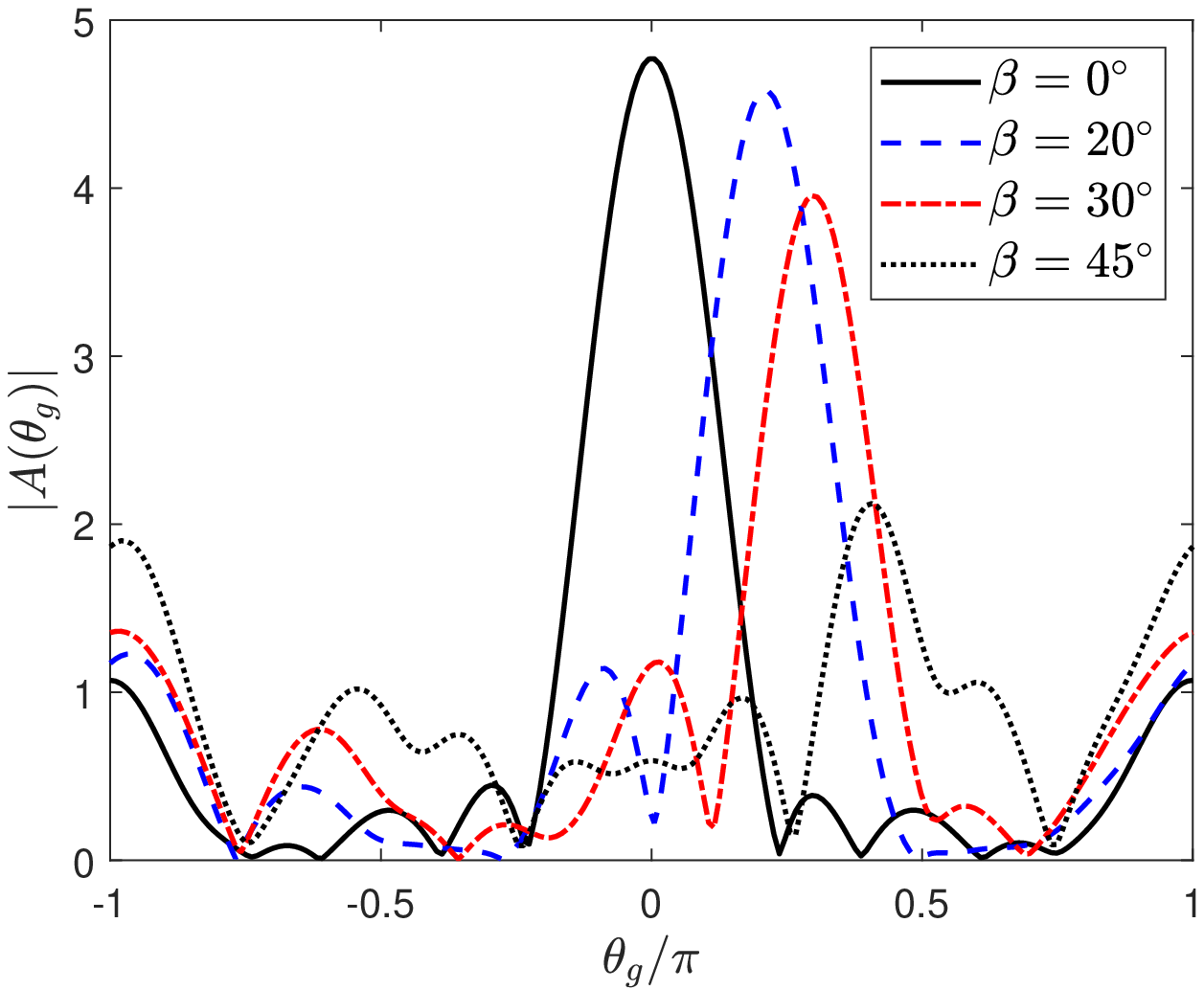}}
		\subfigure[]{\includegraphics*[width=7.8cm]{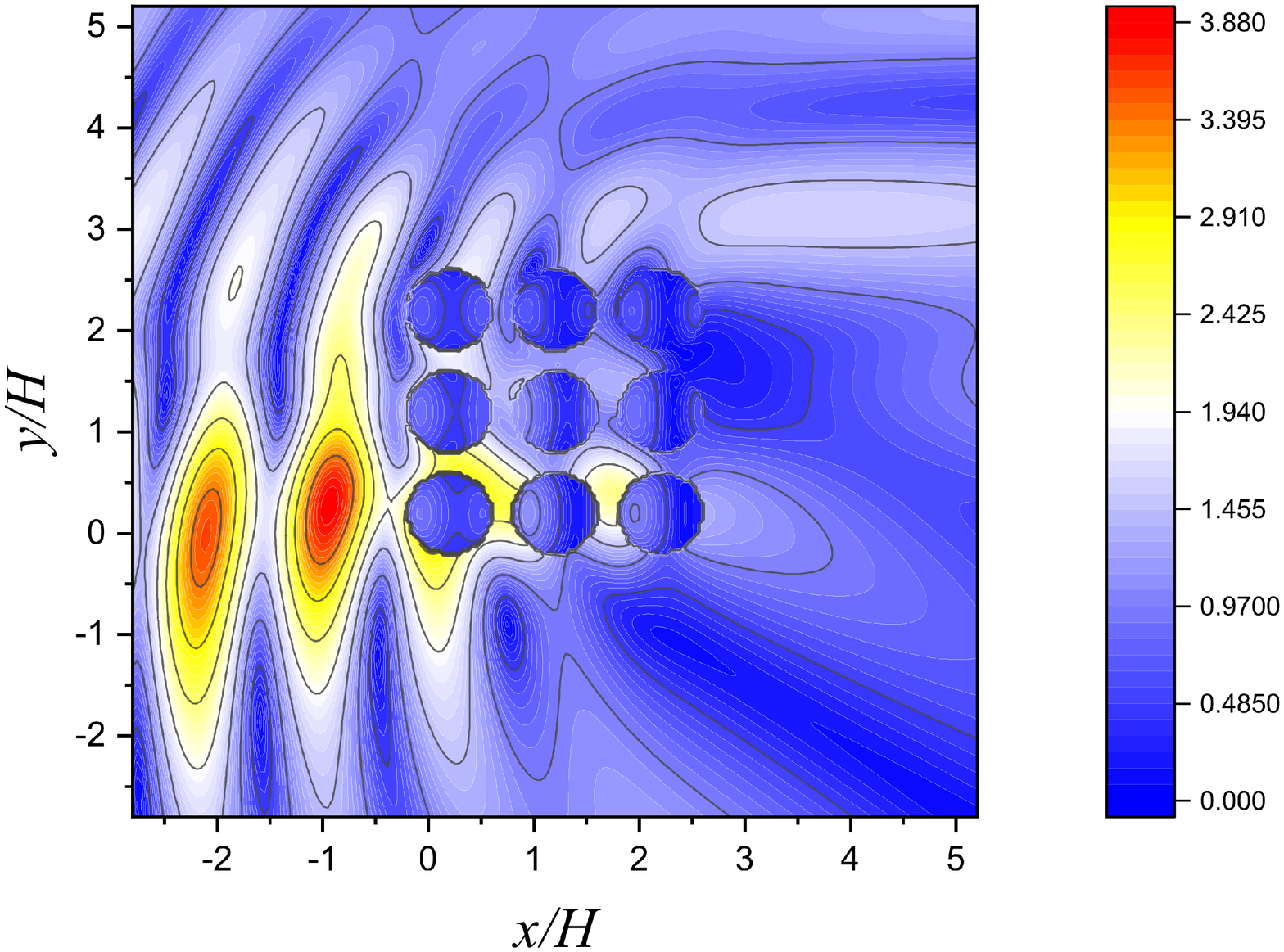}}
		\caption{Surface wave interaction with a square arrangement of the multiple fishing showing the (a) scattering coefficient for varying $(k_0h,\theta_g)$ at $\beta=30^\circ$, (b) scattering coefficient for varying $\theta_g$ at various $\beta$ with $k_0h=1.25$, and (c) amplitude of surface elevation with $k_0h=1.25$ and $\beta=30^\circ$. The other parameters are $h/H=0.4$ and $\mu=10^6$N/m.} \label{f10} 
	\end{center}
\end{figure}

Fig.~\ref{f10} exhibits the scattering coefficients and flow distributions around the multiple fishing cage system arranged in a square formation. Unlike inline arrangement, there exists a wave scattering for smaller wavenumbers in all directions. Then, the scattering coefficients befall and attain a maximum between $1<k_0h<1.5$, where more waves are scattered. Further, the wave scattering considerably increases in this arrangement as compared to the previous arrangement (Fig.~\ref{f9}(a)) due to wave interaction with the column of cages. By fixing $k_0h=1.25$, the response of phase angle on the scattering coefficients is illustrated along a different direction. It is interesting to note that the scattering coefficient increases for a certain optimum angle of $\beta=20^\circ$, where most of the waves are scattered. Moreover, the oscillatory patterns are reduced as compared to other arrangements owing to the simultaneous scattering of the incident wave by a column of cylinders. For $\beta>20^\circ$, the wave scattering decreases due to the less screening of waves by the cages along incident wave direction. The flow distributions corresponding to the square arrangement is plotted in Fig.~\ref{f10}(c), where $\beta=30^\circ$ is fixed. The constructive interference occurs in between the cages in the first column and it gradually decreases as the wave passes through the adjacent column of cages. It is observed that the high amplitude waves are damped by an array of fishing cage due to the wave energy dissipation. However, the dissipated wave amplitudes are further trapped inside the system resulting in less wave propagation on the lee-side of the cage system. The damages to the cage system are greatly avoided in such type of arrangement as a consequence of multiple scattering between the cages.

\subsection{Flow field}  

\begin{figure}[h! ]
	\begin{center}
		\subfigure[$N=1$]{\includegraphics*[width=8cm]{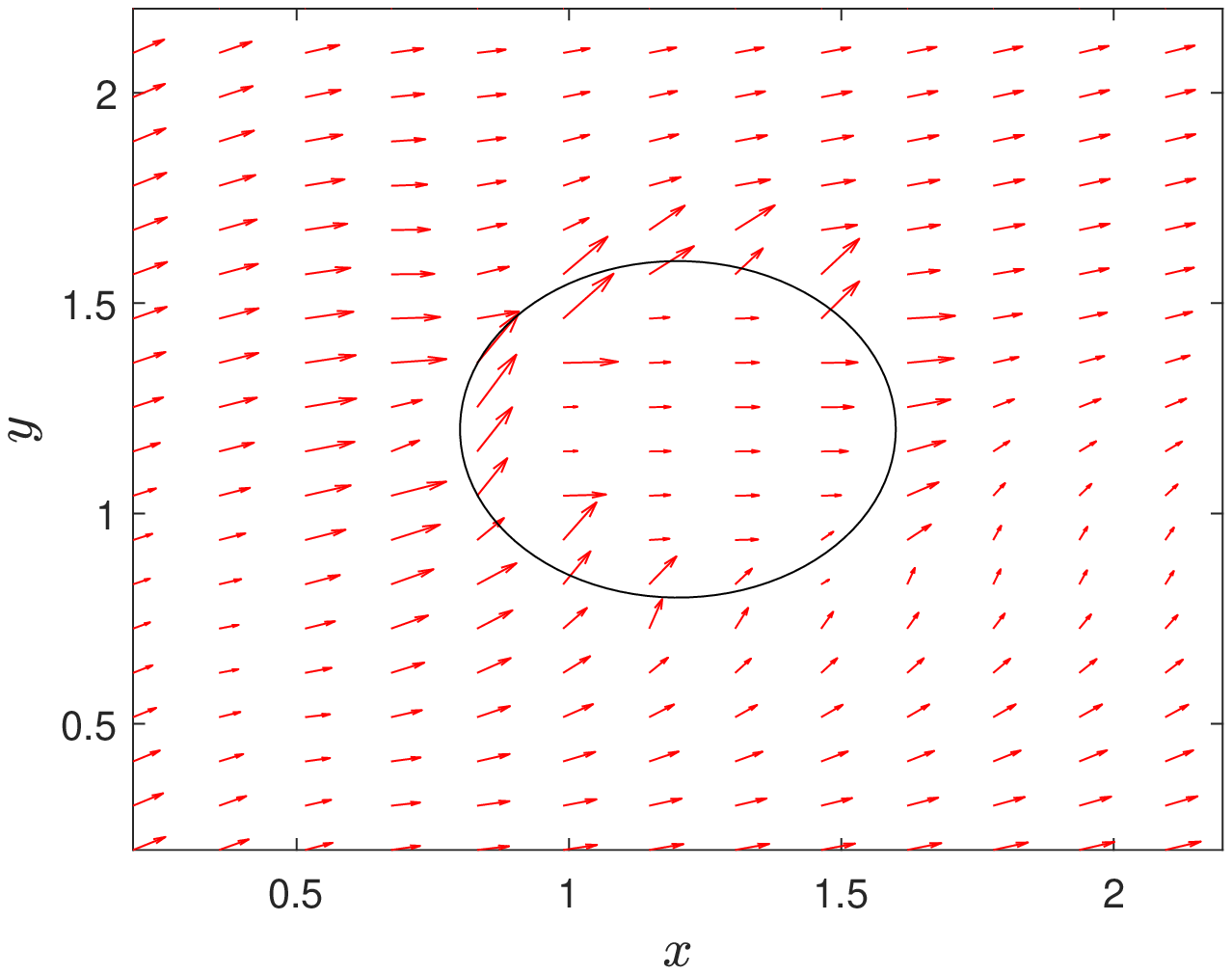}}
		\subfigure[$N=2$]{\includegraphics*[width=8cm]{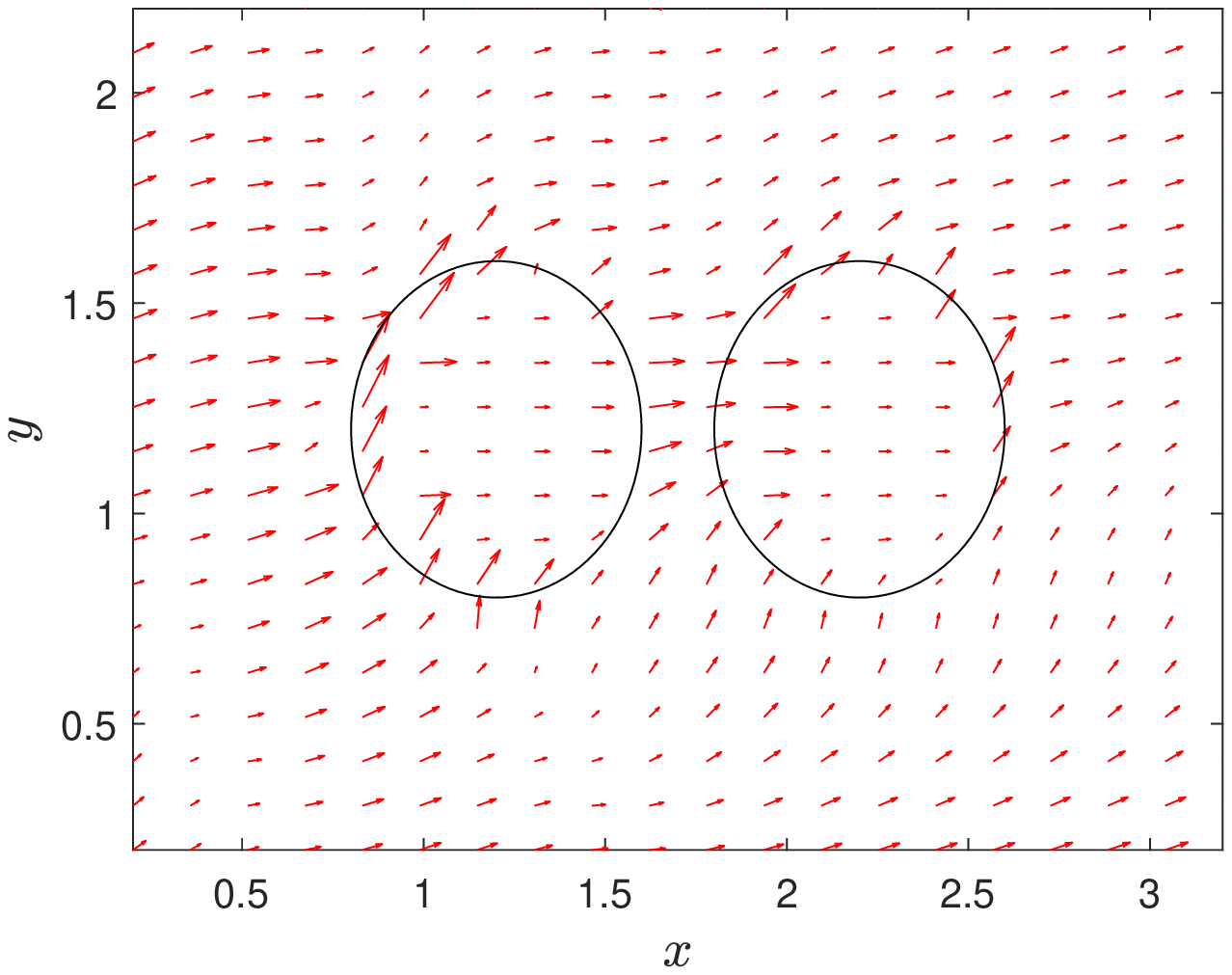}}
		\subfigure[$N=9$]{\includegraphics*[width=8cm]{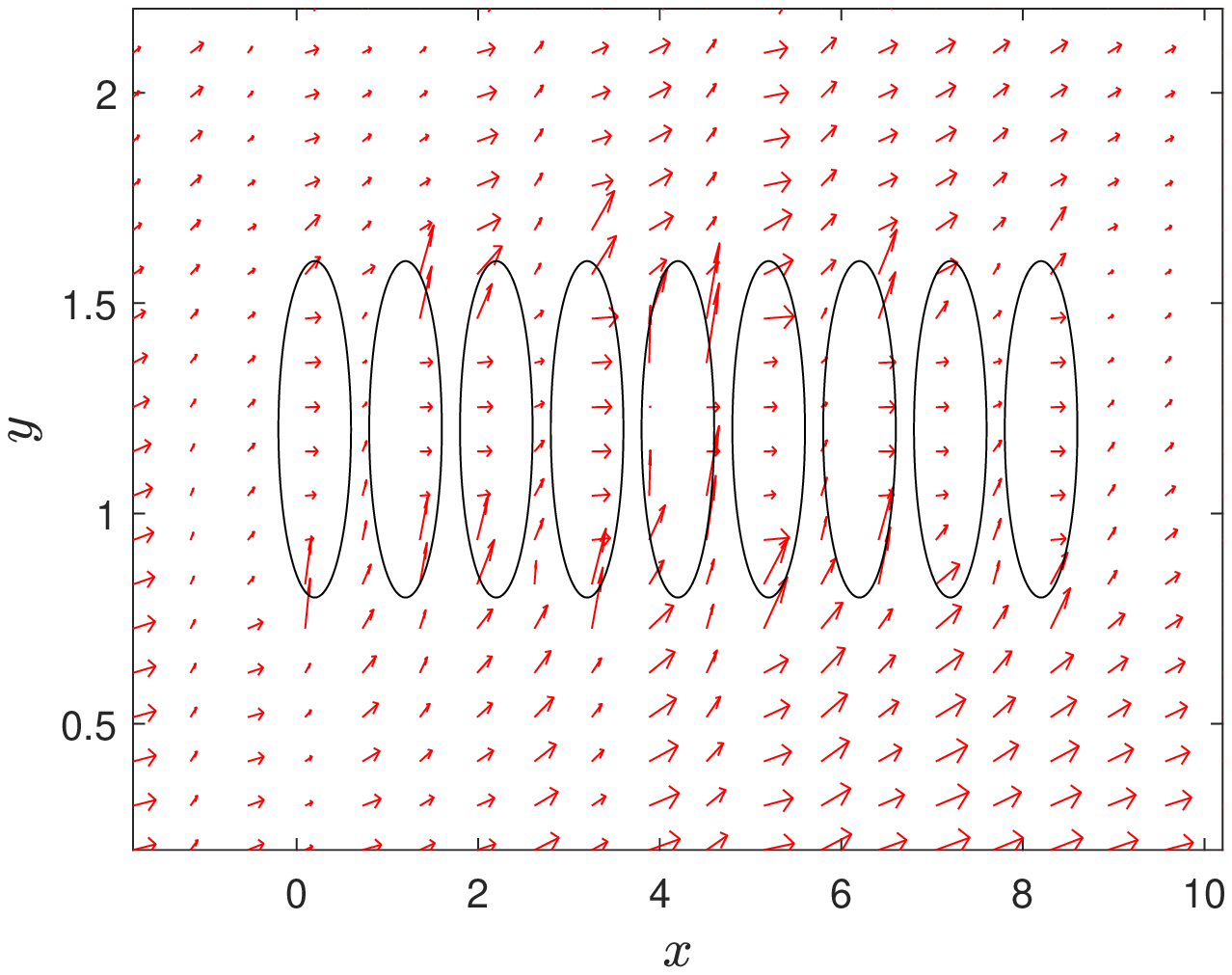}}
		\subfigure[$N=9$]{\includegraphics*[width=8cm]{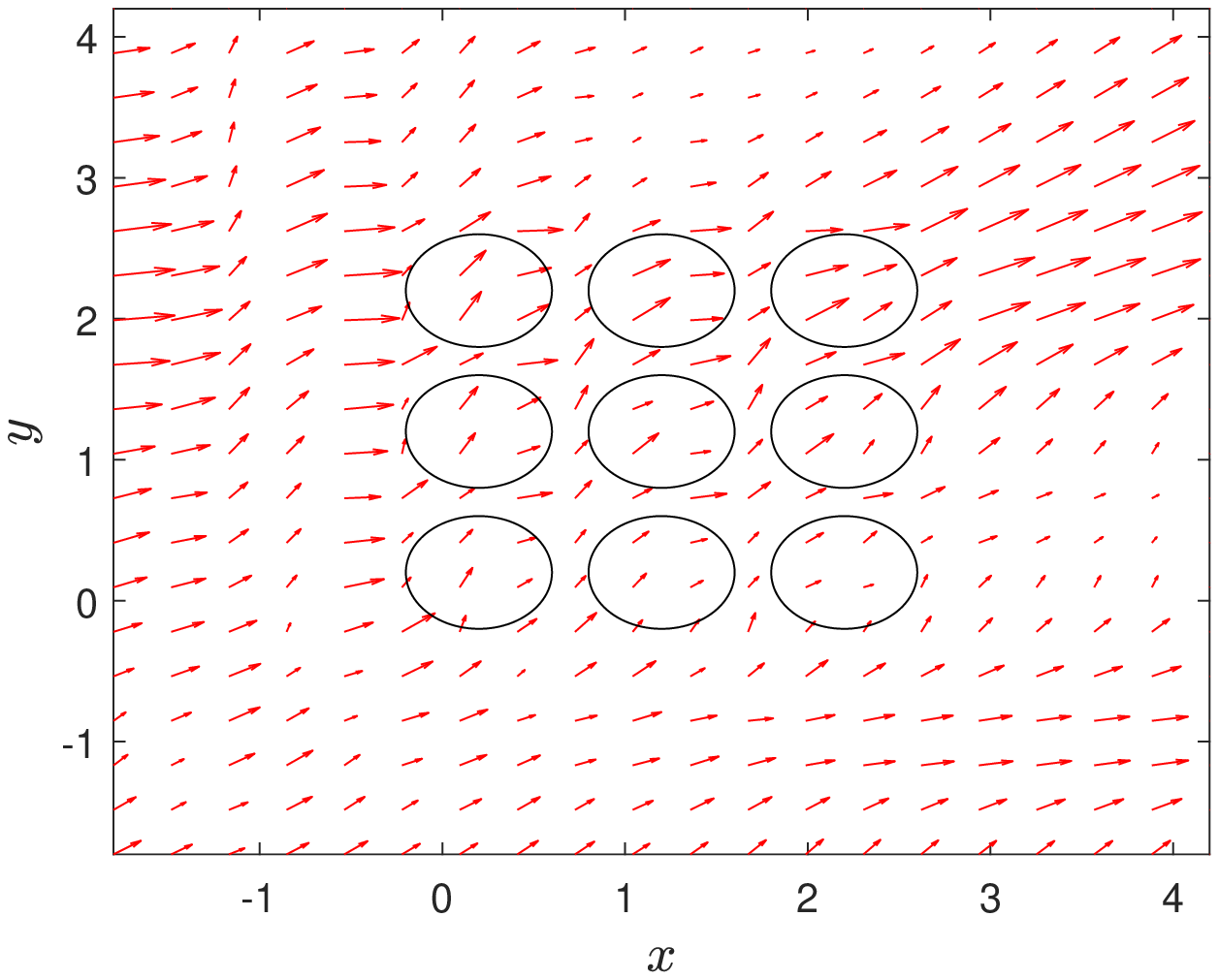}}
		\caption{Flow field around the (a) single fishing cage, (b) dual fishing cage, (c) multiple fishing cage having inline arrangement ($N=9$) and (d) multiple fishing cage having square arrangement ($N=9$). The other parameters are $G=3+3\mathsf{i}$, $T=0.4$, $h/H=0.4$, $\beta=30^\circ$, $k_0h=1.25$ and $\mu=10^6$N/m.} \label{f11} 
	\end{center}
\end{figure}

In Fig.~\ref{f11}, the flow field around the fishing cage system is plotted for different values of $N$. It is noticed that the incident surface wave with phase angle $\beta=30^\circ$ is passing through the cage system in all figures. For $N=1$ (Fig.~\ref{f11}(a)), there are more flows around the circumference of the cage facing the wind-ward side due to the wave interaction. Then, it is passing through the cage, which is due to the wave reflection and dissipation, resulting in less wave transmission in the corresponding lee-ward side. This is true from the figure \ref{f11}(a), where the less transmission on the lee-ward side is shown by well-spaced arrows denoting the flow.  In the case of a dual fishing cage (Fig.~\ref{f11}(b)), there is an interaction of scattered waves in between the cages resulting in the large accumulation of flows. However, the flow field inside the second cage decreases, resulting in less transmission on the lee-side of the cage system. The flow field through the multiple fishing cages ($N=9$) arranged in inline configuration [Fig.~\ref{f11}(c)] shows that the wave passage gradually decreases as it propagates through the cages, which reduces the wave transmission in the lee-side of cages. Moreover, from Fig.~\ref{f11}(d), the wave transmission through the first column of cages decreases for square configuration. However, there is scattering occurs between the internal cages, and more waves are transmitted on the lee-side of the cage system.

\subsection{Time simulation of solution}
\begin{figure}[h!]
\begin{center}
\subfigure[]{\label{f12a}{\includegraphics[width=8cm]{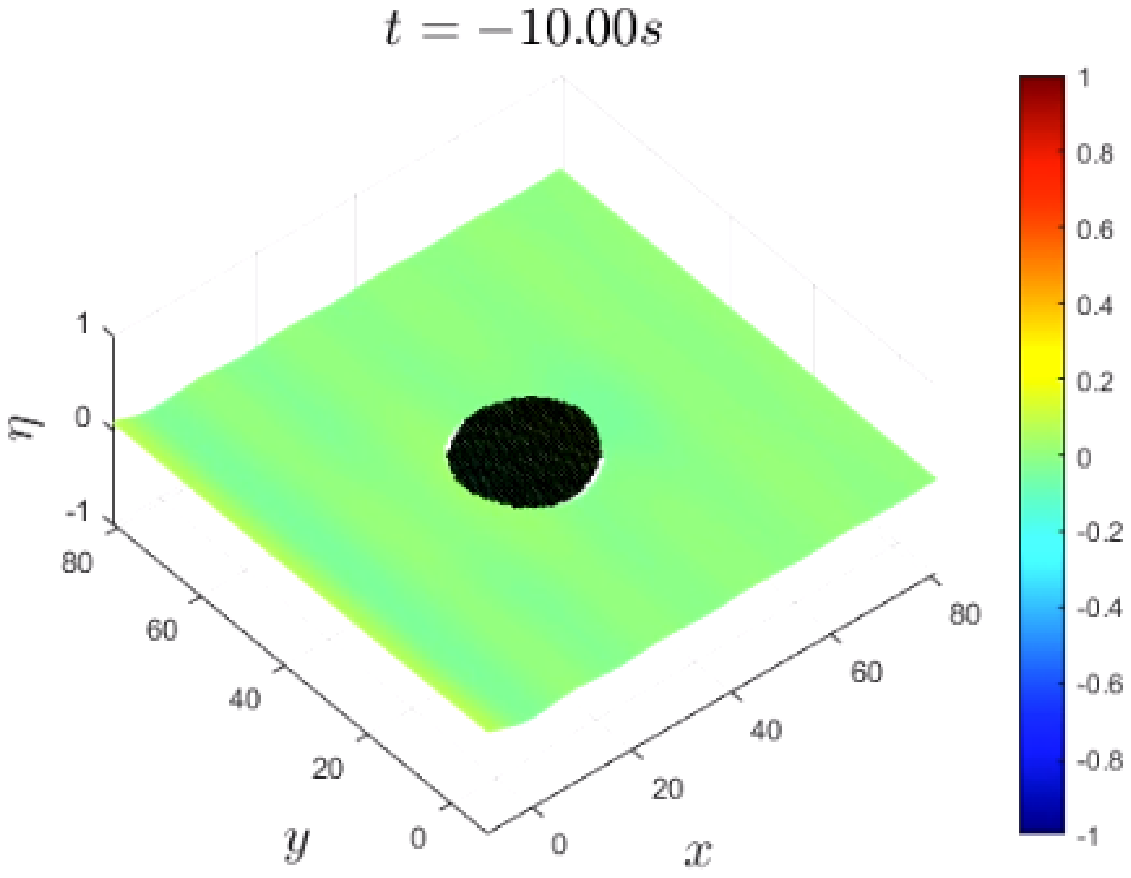}}}
\subfigure[]{\label{f12b}{\includegraphics[width=8cm]{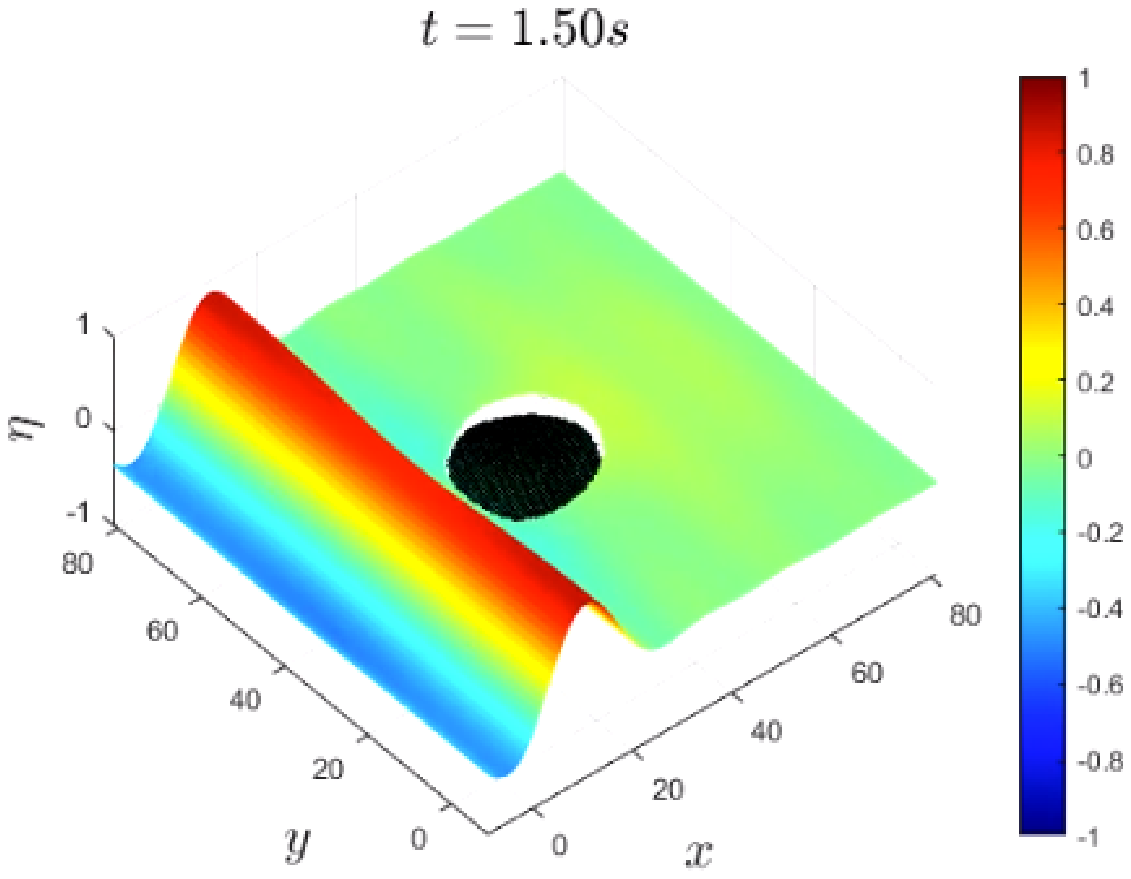}}}
\subfigure[]{\label{f12c}{\includegraphics[width=8cm]{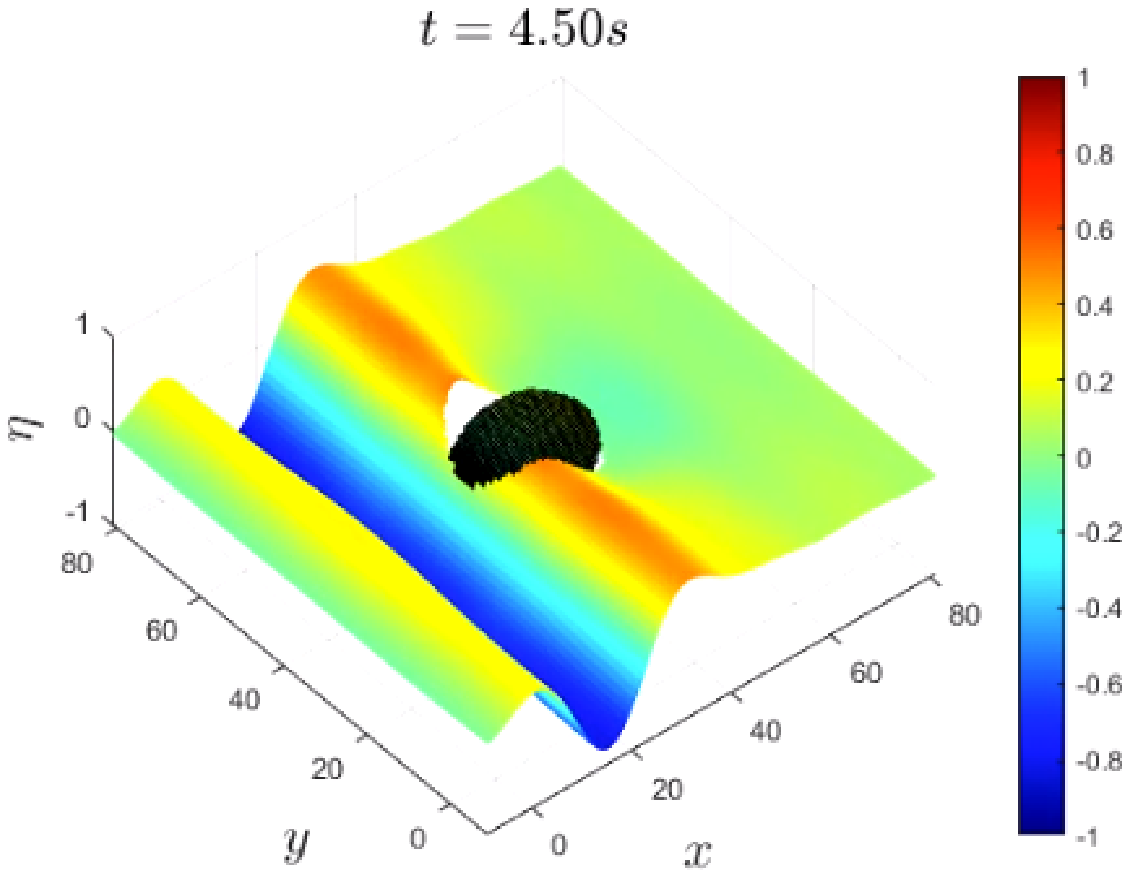}}}
\subfigure[]{\label{f12d}{\includegraphics[width=8cm]{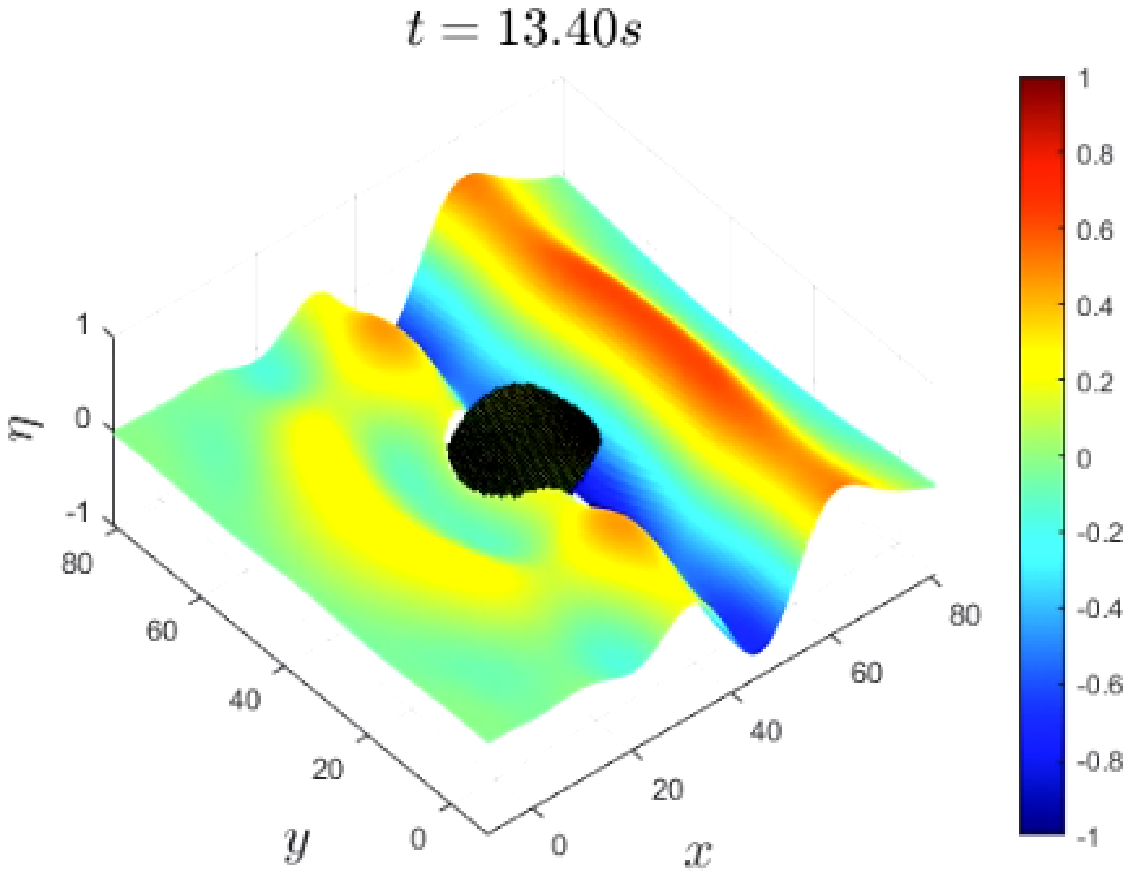}}}
\caption{Wave scattering by dual fishing cages ($N=1$). The other parameters are $G=3+3\mathsf{i}$, $T=0.4$, $h/H=0.4$, $\beta=30^\circ$, $k_0h=1.25$ and $\mu=10^6$N/m. Multimedia view:} \label{f12}
\end{center}
\end{figure}
The simulation of a surface elevation ($\eta$) can be done using the following transformation rule, which can be expressed as
\begin{align}
    \eta(x,y,0,t)=\Re\biggl[2A\sqrt{\pi s}\int^{\infty}_{-\infty}\eta(x,y,0)\,\mathrm{exp}\{-s(\bar{k}-\bar{k}_c)^2-\mathsf{i}\omega t\} d\bar{k}\bigg],
\end{align}
where $\Re$ denotes the real part. Further, the parameters such as $s$ and $\bar{k}_c$ are known as spreading function and centre frequency, respectively. For the purpose of simulation, the values are fixed as $s=2.5$ and $k_c=1/H$. 

\begin{figure}[h!]
\begin{center}
\subfigure[]{\label{f13a}{\includegraphics[width=8cm]{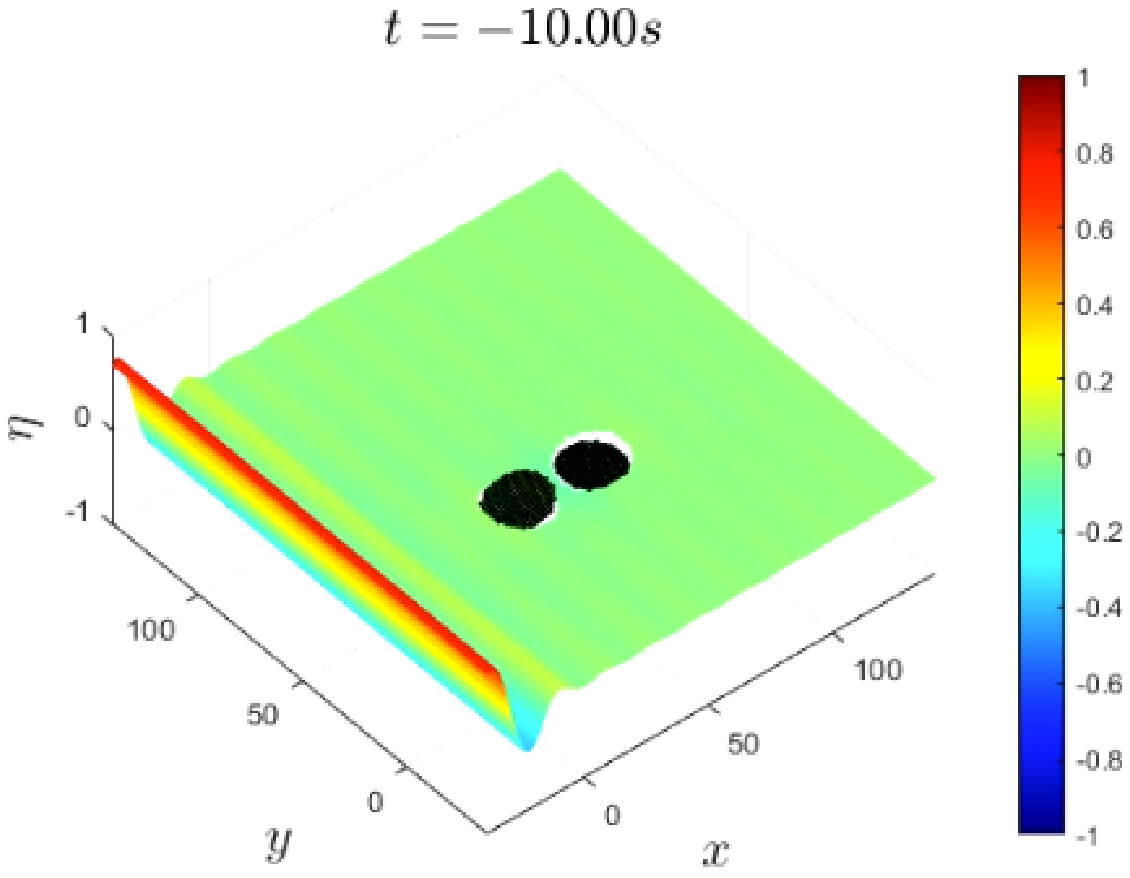}}}
\subfigure[]{\label{f13b}{\includegraphics[width=8cm]{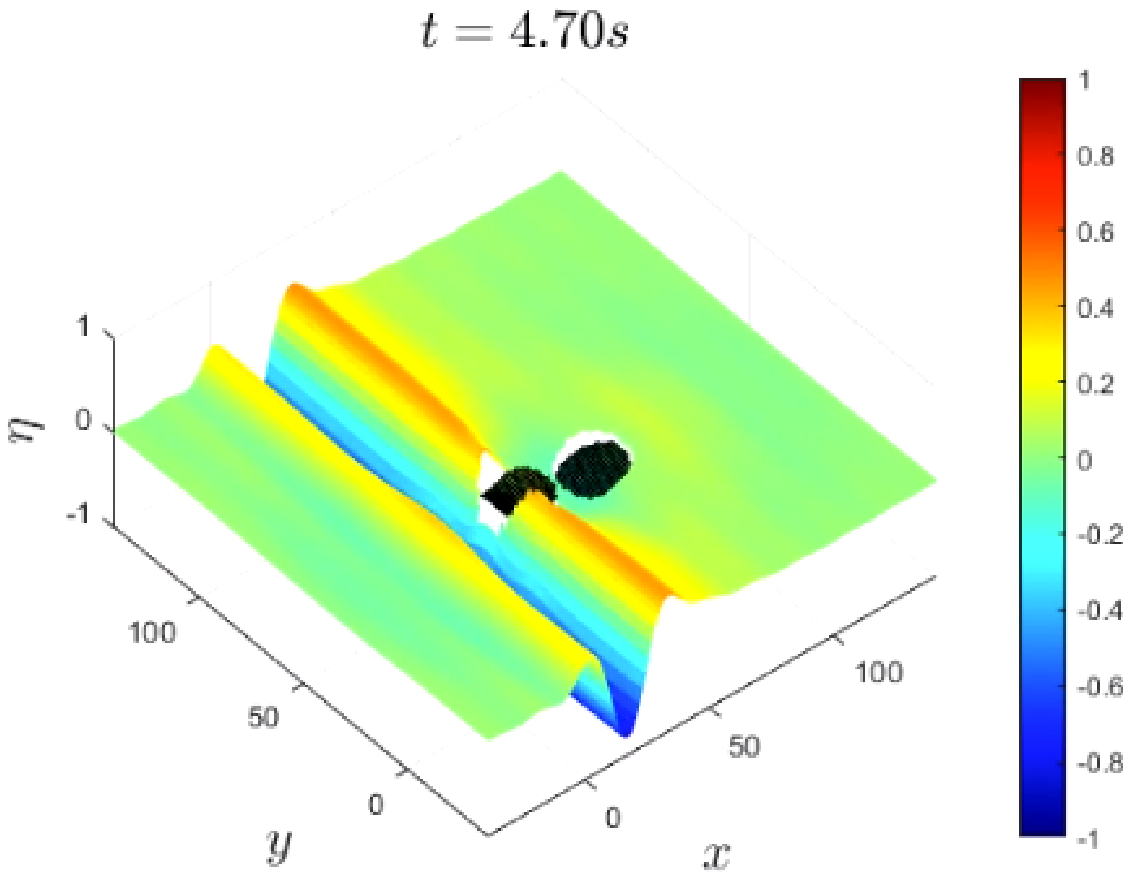}}}
\subfigure[]{\label{f13c}{\includegraphics[width=8cm]{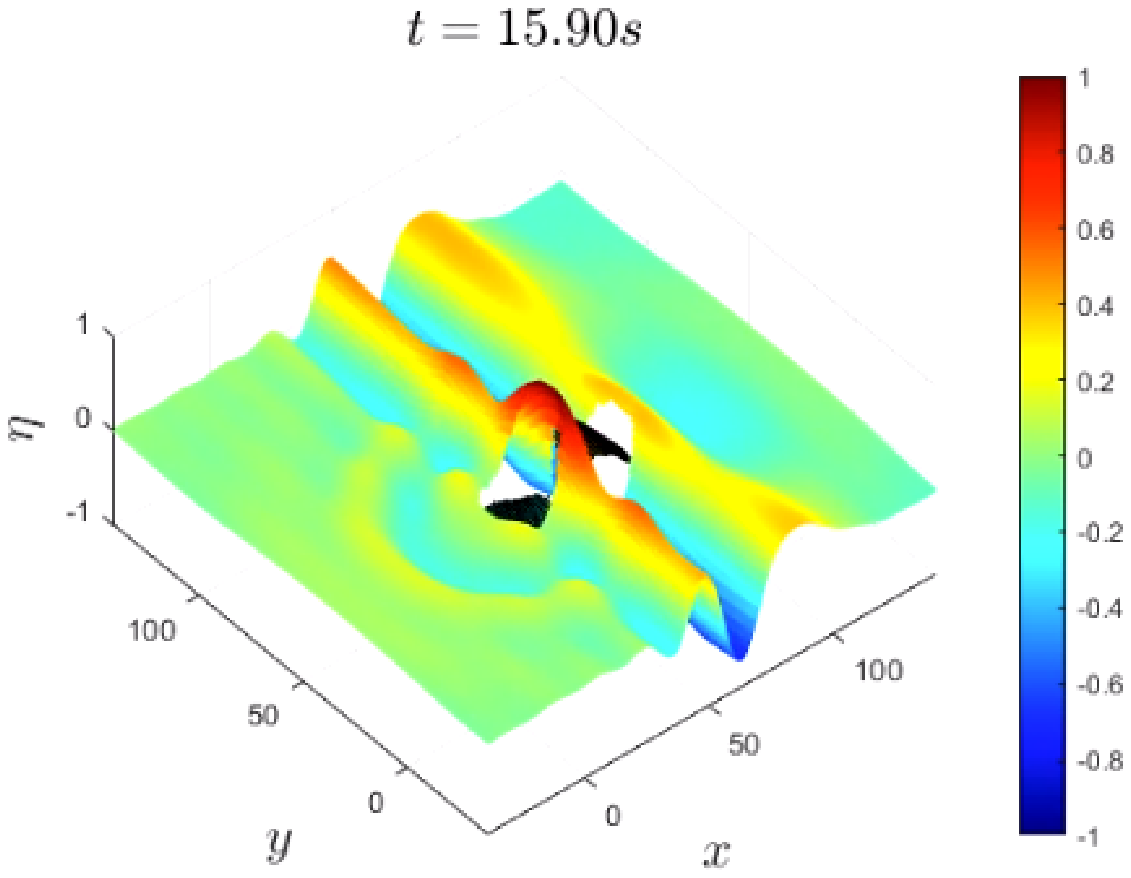}}}
\subfigure[]{\label{f13c}{\includegraphics[width=8cm]{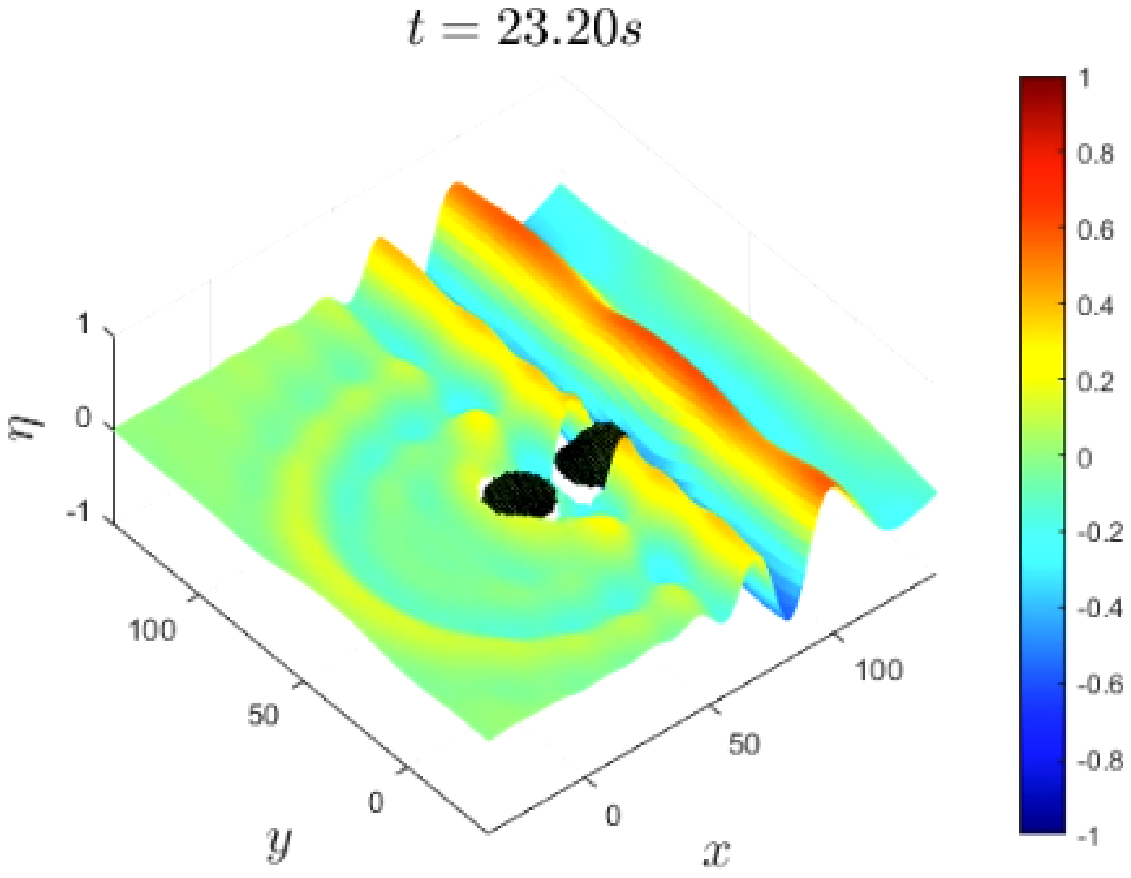}}}
\caption{Wave scattering by dual fishing cages ($N=2$). The other parameters are $G=3+3\mathsf{i}$, $T=0.4$, $h/H=0.4$, $\beta=30^\circ$, $k_0h=1.25$ and $\mu=10^6$N/m. Multimedia view:} \label{f13}
\end{center}
\end{figure}

\begin{figure}[h!]
\begin{center}
\subfigure[]{\label{f14a}{\includegraphics[width=8cm]{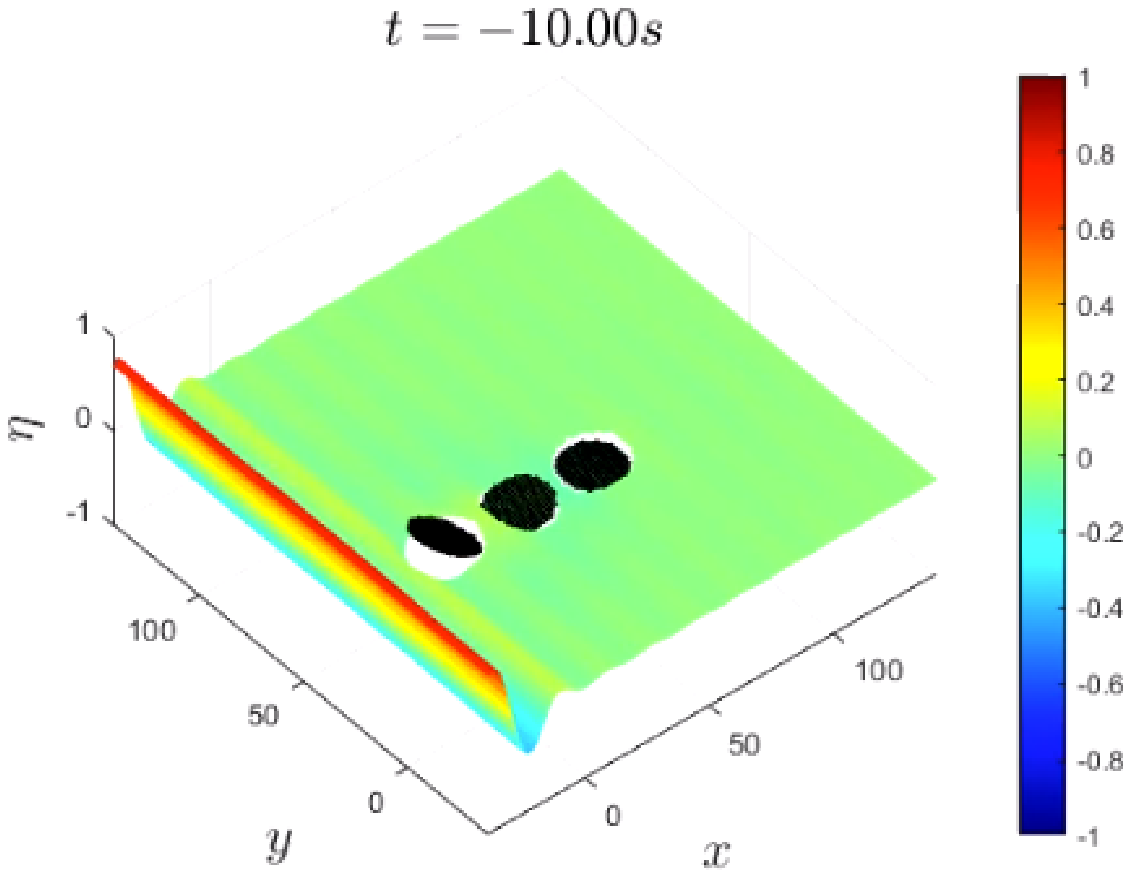}}}
\subfigure[]{\label{f14b}{\includegraphics[width=8cm]{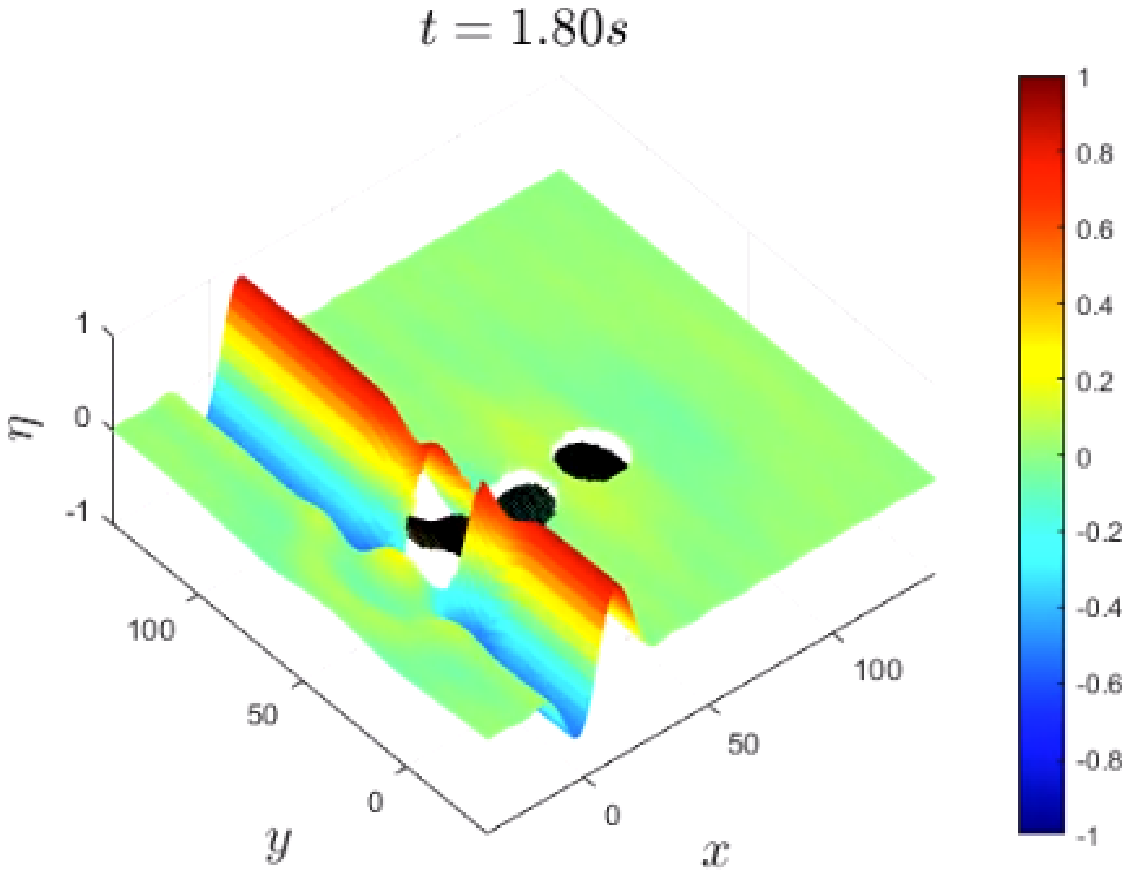}}}
\subfigure[]{\label{f14c}{\includegraphics[width=8cm]{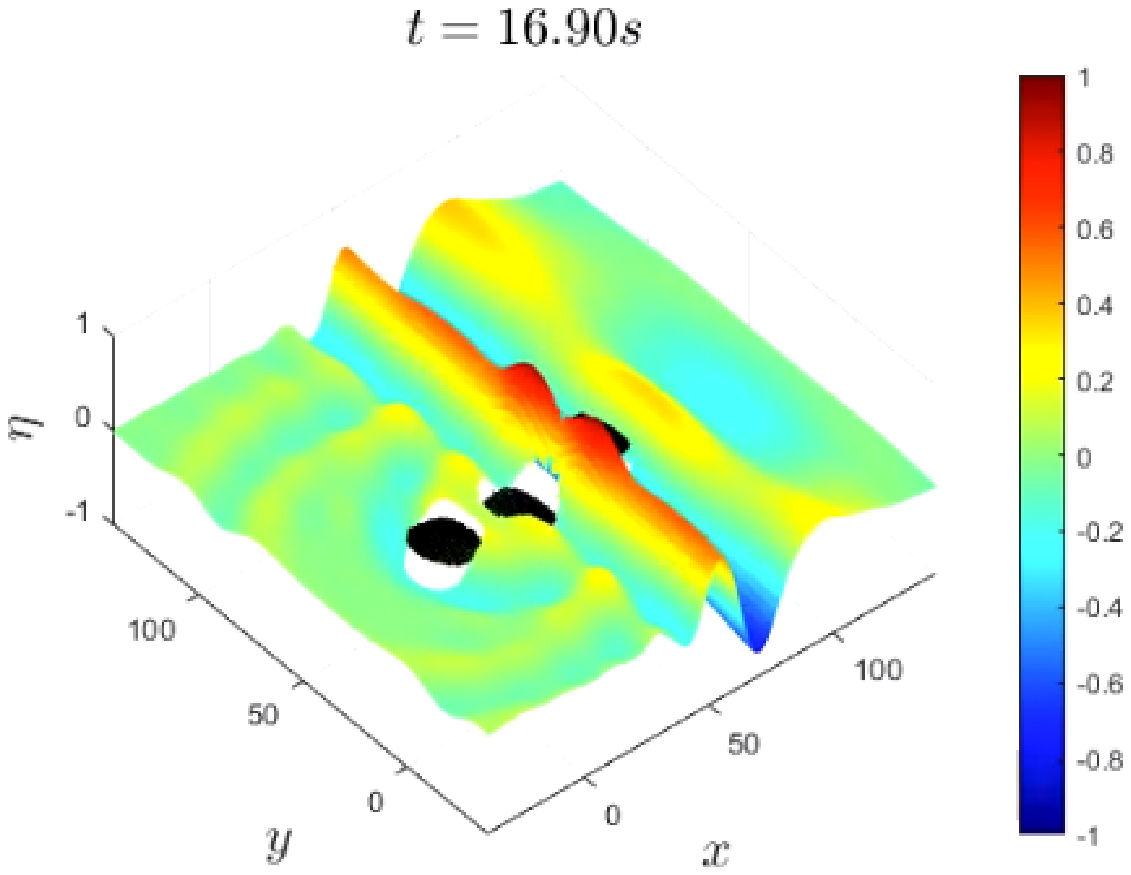}}}
\subfigure[]{\label{f14d}{\includegraphics[width=8cm]{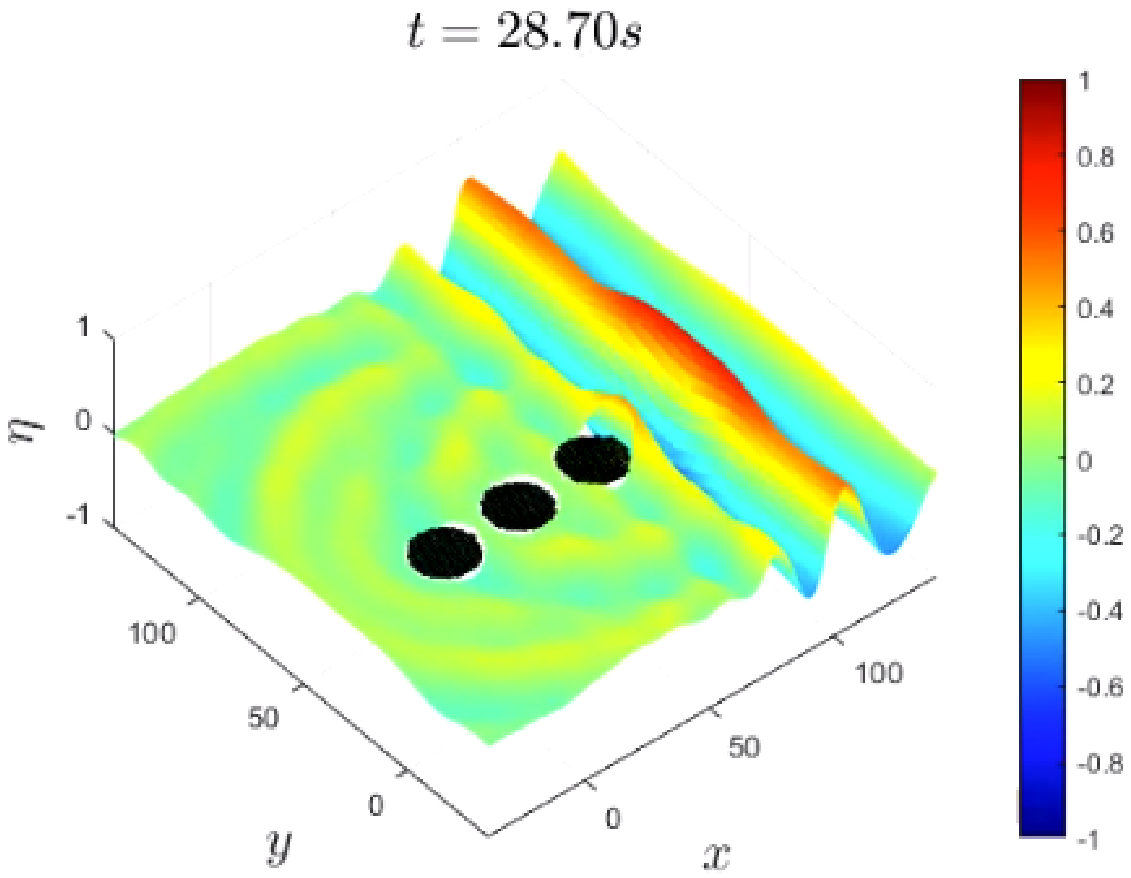}}}
\caption{Wave scattering by triple fishing cages ($N=3$). The other parameters are $G=3+3\mathsf{i}$, $T=0.4$, $h/H=0.4$, $\beta=30^\circ$, $k_0h=1.25$ and $\mu=10^6$N/m. Multimedia view:} \label{f14}
\end{center}
\end{figure}

The time simulation of the wave scattering by the single, double and triple fishing cage system having the inline arrangement are shown as static figure in Figs.~\ref{f12a}, \ref{f12b} and \ref{f12c}, respectively.
The movies corresponding to the static figures~\ref{f12a}, \ref{f12b} and \ref{f12c} are given in the multimedia file. It is noticed that the amplitude of wave interacting with the single fishing cage as in Fig.~\ref{f12} decreases as it passes through the cage. This happens due to the partial reflection and dissipation from the single fishing cage. In the case of dual fishing cage system as in Fig.~\ref{f13}, there is a multiple scattering occurs between the cages. This consequently reduces the scattered wave amplitude in both wind-ward and lee-ward sides of the cage system. For triple cage system as in Fig.~\ref{f13}, at certain wavenumber, the constructive interference occurs resulting in the increased wave amplitude in between the second and third cage in the Fig.~\ref{f13}. In general, the wave amplitude in the triple cage system decreases significantly when compared to the single and dual cage system due to the increased rate of energy dissipation and multiple scattering between the cages. Moreover, the wave amplitudes inside the cages are consequently less in Fig.~\ref{f14} when compared to Figs.~\ref{f13} and \ref{f14}. This shows that the fluid circulation becomes smoother inside the cages with an increase in the number of fish cages.  

\section{Conclusions}\label{concl}
The manuscript deals with the surface wave scattering by the system of multiple fishing cages by employing the linear water wave theory and the small-amplitude wave response. The general velocity potential in the form of the Fourier-Bessel series is obtained while solving the Laplace equation. Further, the matched eigenfunction expansion and least-square approximation are implemented for solving the system of the linear equation obtained from the interface and edge conditions. The numerical results, such as hydrodynamic wave loads, far-field wave amplitudes, and power dissipation, are investigated for different wave and cage parameters. Moreover, the effect of spatial arrangements on the far-field wave amplitudes and flow distributions are investigated. The following observations are made from the present study:
\begin{itemize}
	\item  In the single fishing cage, the wave amplitude decreases on the lee-side of cage due to the wave energy energy dissipation by the porous structure. The energy dissipation increases gradually for the waves having larger wavenumbers resulting in the less wave amplitude on the lee-side of the cage.
	\item The wave scattering from the single fishing cage decreases when it connected with the mooring spring of larger spring constant. This ultimately reduces the hydrodynamic wave loads acting on the cage. Thus, the cage system connected with the high stiffness spring reduces the damages caused by hydrodynamic wave loading acting on the cage.
	\item For the dual fishing cage, there is a constructive interference on placing the two cages close enough and it decreases as the spacing between the cages enlarges. This results in decreasing wave amplitude in the confined region between the cages. However, the wave amplitude inside the cage decreases as a consequence of destructive interference between an incoming wave and waves scattered from an inner region of a cage.
	\item The wave scattered from the dual fishing cage increases as compared to the single cage, which increases for moderate values of tensile force. Further, the membrane with less tension dissipates more energy for larger depth ratio owing to the increased energy dissipation. Thus, this can also be employed as the power absorber.
	\item The deployment of the multiple fishing cage in different spatial arrangements result in the oscillatory pattern in the scattering coefficients. In the case of inline deployment, the periodic optima in the scattering coefficients are more and approach zero for certain cage parameters resulting in the Fabry-Perot effect at certain directions in the far-field. In the case of a square arrangements of multiple cages, this effect becomes weaker.  However, the damages to the cage system are greatly avoided in square arrangement as a consequence of multiple scattering between the cages and it further enables the smooth water circulation inside the cage owing to the wave amplitude reduction. 
\end{itemize}

\section*{Acknowledgment}
HB gratefully acknowledges the financial support from SERB, Department of Science and Technology, Government of India through “CRG” project, Award No. CRG/2018/004521. 

\section*{Data Availability}
The data that supports the findings of this study are available
within the article, highlighted in the related figure captions and corresponding discussions.


\bibliography{REF}  
\end{document}